\documentclass[reprint, superscriptaddress, nofootinbib, longbibliography, amsmath, amssymb, aps, pra]{revtex4-1}

\usepackage{color}
\usepackage{graphicx}% Include figure files
\usepackage{dcolumn}% Align table columns on decimal point
\usepackage{bm}% bold math
\usepackage{verbatim}
\usepackage{subfigure}
\usepackage{hyperref}
\usepackage{mathrsfs}

\DeclareRobustCommand{\gobblefive}[5]{}
\newcommand*{\SkipTocEntry}{\addtocontents{toc}{\gobblefive}}

\newcommand{\ddo}{\hat{\rho}}

\newcommand{\ket}[1]{\left| #1 \right\rangle}
\newcommand{\bra}[1]{\left\langle #1 \right|}

\usepackage{amssymb}
\usepackage{amsmath}

\newcommand{\qq}{\mathbf{q}}

\begin{document}

\title{Quantum Indistinguishability by Path Identity:\\The awakening of a sleeping beauty}% Force line breaks with \\

\author{Armin Hochrainer}
\affiliation{Vienna Center for Quantum Science \& Technology (VCQ), Faculty of Physics, University of Vienna, Austria.\\
Institute for Quantum Optics and Quantum Information (IQOQI) Vienna, Austrian Academy of Sciences, Austria.}
\author{Mayukh Lahiri}
\affiliation{Vienna Center for Quantum Science \& Technology (VCQ), Faculty of Physics, University of Vienna, Austria.\\
Institute for Quantum Optics and Quantum Information (IQOQI) Vienna, Austrian Academy of Sciences, Austria.}
\affiliation{Department of
Physics, Oklahoma State University, Stillwater, Oklahoma, USA}
\author{Manuel Erhard}
\affiliation{Vienna Center for Quantum Science \& Technology (VCQ), Faculty of Physics, University of Vienna, Austria.\\
Institute for Quantum Optics and Quantum Information (IQOQI) Vienna, Austrian Academy of Sciences, Austria.}
\author{Mario Krenn}
\affiliation{Vienna Center for Quantum Science \& Technology (VCQ), Faculty of Physics, University of Vienna, Austria.\\
Institute for Quantum Optics and Quantum Information (IQOQI) Vienna, Austrian Academy of Sciences, Austria.}
\affiliation{Department of Chemistry \& Computer Science, University of Toronto, Canada.\\
Vector Institute for Artificial Intelligence, Toronto, Canada.}
\author{Anton Zeilinger}
\email{anton.zeilinger@univie.ac.at}
\affiliation{Vienna Center for Quantum Science \& Technology (VCQ), Faculty of Physics, University of Vienna, Austria.\\
Institute for Quantum Optics and Quantum Information (IQOQI) Vienna, Austrian Academy of Sciences, Austria.}

\date{\today}% It is always \today, today,

\begin{abstract}
Two photon-pair creation processes can be arranged such that the paths of the emitted photons are identical. Thereby the path information is not erased but is never born in the first place. In addition to its implications for fundamental physics, this concept has recently led to a series of discoveries in the fields of imaging, spectroscopy, and quantum information science. Here we present the idea of path identity and provide a comprehensive review of the recent developments.
\end{abstract}

\maketitle

\tableofcontents

%\makeatletter
%\let\toc@pre\relax
%\let\toc@post\relax
%\makeatother 

\section{Introduction}
One of the fundamental principles of quantum mechanics is the following: If an event can occur in more than one alternatives, and there is no way to distinguish between the alternatives, interference occurs. Discussing double-slit experiments, Richard Feynman said that this phenomenon \textit{has in it the heart of quantum mechanics. In reality, it contains the only mystery} \cite{feynman_quantum_1965}. This simple but profound principle can be used to explain many other of the basic quantum mechanics experiments.

In the early 1990s, the group of Leonard Mandel has pushed the concept of indistinguishability to a new level. Instead of considering different alternatives of single (or multiple) photons, they have created alternatives of the origin of a photon pair itself. By cleverly overlapping (or \textit{identifying}) one of the photon's paths (an ingenious idea suggested by Jeff Ou), there is no information anywhere in the universe about the origin of the second photon. Thus -- to apply Feynman's principle -- the second photon is in a superposition of being created in either of the crystals. Zou, Wang and Mandel (ZWM) have exploited this idea in a remarkable way \cite{zou1991induced}: They were able to measure phase shifts introduced in photons that they never detected. %This concept has since been called \textit{Path Identity}.

\begin{figure}[t]
\centering
\includegraphics[width=1\linewidth]{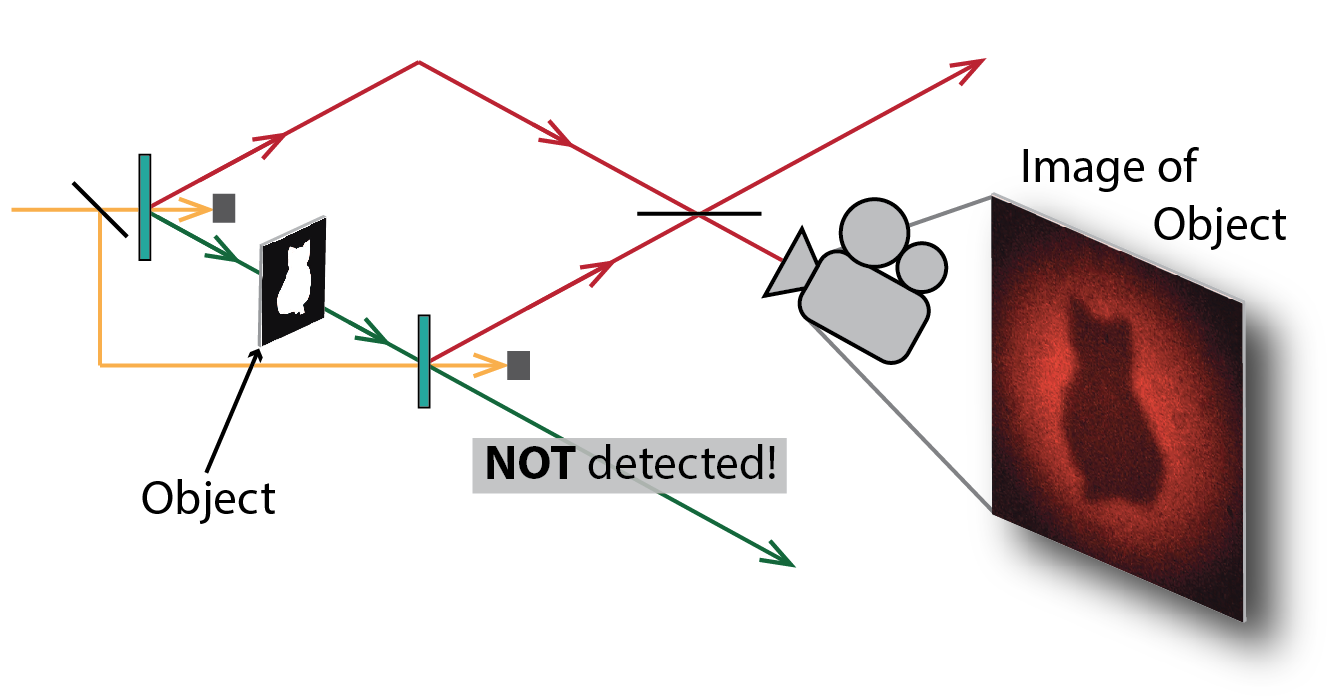}
\caption{Quantum imaging without detecting the photons that interact with the object. The yellow beam coherently pumps two nonlinear crystals, such that one of them creates a photon pair. The photon pair is in a superposition of being created in the first or the second crystal. When the green photon path is identified, the red photon is in a superposition of being in the upper or lower path, which leads to interference at a beam splitter. By introducing an object (a picture of a cat) into the identified green path, the origins become partially distinguishable, which can be observed in the resulting interference pattern. Remarkably, the image of the cat is constructed without detecting any of the green photons. It is important to mention that this concept is entirely different from conventional \textit{Ghost imaging}, where both photons need to be detected, and which can be performed classically \cite{bennink2004quantum}.}
\label{fig:MandelCat}
\end{figure}

Historically it is interesting that after some activities over a period of roughly ten years, investigating this phenomenon has nearly stopped around the year 2000. Then, in 2014, the field has been revived, when it was shown that one can extend the scope even further, by imaging an object without ever detecting photons that interacted with the object itself -- see Fig. \ref{fig:MandelCat}. An important addition is that the detected light can have an entirely different wavelength than the light interacting with the object. A multitude of critical applications has been discovered since then in quantum imaging, quantum spectroscopy, quantum information science and other fields, which are interesting for basic research as well as for practical tasks with potential impact on industrial technologies. 

As a testament to the fundamental importance of the concept, Path Identity has recently been identified as one of the core concepts that should be used in high school education for quantum physics to understand the idea of a \textit{photon} better \cite{malgieri2017test}.

In this review, we will focus on the developments during the last few years, which enormously widened the scope of ZWM's experiment and Ou's idea. In Section II, we demonstrate the concept of \textit{Path Identity} with three defining examples from the early 1990s, followed by a detailed technical account of the ZWM experiment and its importance in the historical context. After laying the foundations, we explain how general properties of single photons (Section IV) or even correlated photon systems (Section VI) can be obtained without ever detecting the photon itself. The applications to modern quantum information science will be discussed based on Single Photon and Multi-Photon entanglement generation (Section VI) with a connection to the mathematical field of Graph Theory (Section VII). A novel type of multi-photon quantum interference is discussed subsequently (Section VIII). Concepts related to Path Identity are discussed (Section IX), followed by conclusions and currently open questions that would be interesting to be investigated in the future (Section X).

In our review, we focus on true single photon or entangled photon quantum phenomena. For a detailed account of so-called \textit{non-linear} interferometers in the high-gain regime of non-linear optics, see e.g. \cite{chekhova2016nonlinear}.

\section{The Principles of Path Identity}
In the following chapter, we explain the ideas of three experiments from the early 1990s, which define the principles of \textit{path identity}. Detailed investigations and extensions of the works in the last 25 years (and in particular of the last flourishing five years) will be the main content of the rest of this review.  
\begin{figure*}[!ht]
\centering
\includegraphics[width=\textwidth]{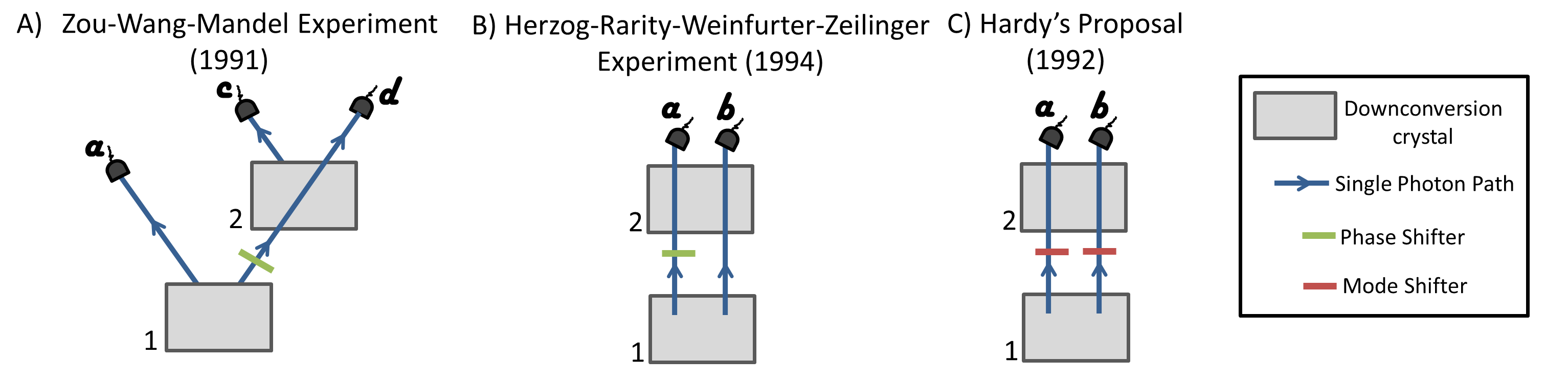}
\caption{Three historic experiments initiated the research in Path Identity. A) The experiment by Zou, Wang and Mandel has first shown that identifying the path of photons can induce coherence in an unexpected way \cite{zou1991induced,wang1991induced}. B) Herzog, Rarity, Weinfurter and Zeilinger identified that two-photon interference phenomena occur when it is fundamentally indistinguishable in which crystal the photon pair has been created \cite{herzog1994frustrated}. C) Hardy proposed the mode shifting and identification two paths as a method for the generation of quantum entanglement \cite{hardy1992source}. The source has become a cornerstone of photonic quantum entanglement experiments \cite{kwiat1995new, pan2012multiphoton, zhong201812}. The review concerns these ideas and their generalizations and applications during the last quarter of a century.}
\label{fig:Intro_3experiments}
\end{figure*}

\subsection{The Zou-Wang-Mandel Experiment: Induced coherence without induced emission}\label{ZWM-Mario}
In 1991, Zou, Wang and Mandel (ZWM) demonstrated an experiment where they induced coherence between two photonic beams without interacting with any of them \cite{zou1991induced,wang1991induced}. They used two non-linear crystals which produce photon pairs. In the experiment, one photon pair is in a superposition of being created in crystal 1 (creating photons in path $a$ and $b$) and crystal 2 (photons in $c$ and $d$). The striking idea (originally proposed by Zhe-Yu Ou) was to overlap one of the paths from each crystal (Fig. \ref{fig:Intro_3experiments}A), which can be written as the \textit{Path Identity} 
\begin{align}\label{PI-eq}
\ket{b} \to \ket{d}. 
\end{align}
Thereby, the \textit{which-crystal information} of the final photon in path $d$ is removed. Importantly, the information of the photon's origin is not erased but \textit{has never been created in the first place}. The resulting state can be written as 
\begin{eqnarray}
\ket{\psi}&=&\frac{1}{\sqrt{2}}\left(\ket{a,d}+\ket{c,d} \right)\nonumber\\
&=&\frac{1}{\sqrt{2}}\left(\ket{a} + \ket{c} \right)\ket{d},
\end{eqnarray} 
which shows that one photon is in a superposition of being in path $a$ or path $c$. If one adds a phase shifter of $\phi$ between the two crystals, the state is
\begin{eqnarray}
\ket{\psi}&=&\frac{1}{\sqrt{2}}\left(\ket{a,d}+e^{i\phi}\ket{c,d}\right)\nonumber\\
&=&\frac{1}{\sqrt{2}}\ket{d}\left(\ket{a} +e^{i\phi} \ket{c} \right).
\end{eqnarray}
Now the phase is encoded between the photon's path $a$ and $c$, which never interacted with the phase shifter in the first place. The phase can be extracted if one superposed paths $a$ and $c$ by a beam splitter. A detailed analysis of this \textit{mind boggling} experiment \cite{greenberger1993multiparticle} is shown in chapter \ref{subsec:theory-ZWM}.

It took more than 20 years until it has been recognized that this type of interference can be exploited for quantum imaging \cite{lemos2014quantum} and spectroscopy \cite{kalashnikov2016infrared} in a way that uses the potential to probe in one wavelength and measure in another wavelength. We come back to these core insights in chapter \ref{section:objectrecon}.

\subsection{Frustrated Down-Conversion: Interference in Photon Pair Creation}\label{Sec:HerzogIntro}
Shortly after the demonstration of ZWM interference, an experiment showed quantum interference that occurs when both paths of the photon pairs are identified \cite{herzog1994frustrated}. 

Similarly, as in the ZWM-Experiment, the experimental setup (depicted in Figure \ref{fig:Intro_3experiments}B) consists of two crystals which are pumped coherently and in such a way, that only one of the crystals creates a photon pair. A photon pair now could be created in the first or the second crystal. A phase plate can shift the relative phase between the two processes, which leads to the final state of
\begin{eqnarray}
\ket{\psi}&=&\frac{1}{\sqrt{2}}\left(\ket{a,b} + e^{i \phi} \ket{a,b} \right)\nonumber\\
&=&\frac{1}{\sqrt{2}}\ket{a,b}\left(1 + e^{i \phi}\right)
\end{eqnarray}
One finds that by changing the phase between the two crystals, one can enhance or suppress the production of the photon pairs. This is a remarkable interference effect because while a single crystal produces a constant number of photon pairs, adding a second crystal could lead to complete zero output, on the single-photon level.

Recent generalizations to multi-photonic systems show, among others, connections to graph theory and quantum computation, which we will explain in chapter \ref{section:multiphotonInterf}.

\subsection{Entanglement by Path Identity}\label{Sec:EbPIIntro}

In the two experiments mentioned above, paths of indistinguishable photons are identified, and phases between photon-pair sources are altered, which led to a new kind of single-photon interference.

Let us now ask what would happen if instead of the phase the modes themselves (e.g., the polarization of the photons) are changed, such that the photons are no longer indistinguishable. A conceptual sketch can be seen in Figure \ref{fig:Intro_3experiments}C.

If one pumped only crystal 1, then one always gets a vertically polarized photon pair in detector a and b, $\ket{\psi_1} = \ket{a_V,a_V}$. When pumping crystal 2, one always obtains two photons with horizontal polarization, $\ket{\psi_2} = \ket{a_H,a_H}$. If the two crystals are pumped coherently, then a pair is generated that is a coherent superposition of the two possibilities, i. e.
\begin{align}
\ket{\psi}=\frac{1}{\sqrt{2}}\left(\ket{a_H,b_H} + \ket{a_V,b_V} \right)
\end{align}
This state describes two photons that are entangled in their polarization. 

An entanglement source of this kind was first described by Lucien Hardy in 1992 \cite{hardy1992source}, only a few months after the ZWM experiment. Hardy described it as a deterministic, collinear emitting source of polarization-entangled photon pairs that can be used for the definitive violation of Bell's inequality. The implementation of this source was achieved by Kwiat et al. in 1995 \cite{kwiat1995new}, and the design of the source is still in use today - especially in the generation of the most complex photonic entanglement states such as a 12-photon entangled system \cite{zhong201812}, or an 18-qubit entangled photon state of several degrees of freedom \cite{wang201818}.

Despite the frequent use of this source, it took 25 years that the concept of path identification was generalized to vast classes of entanglement, such as any high-dimensional two-photon systems as well as vast types of high-dimensional many-body systems \cite{krenn2017entanglement}. We come back to that in chapter \ref{section:EbPI}.

\section{The Zou-Wang-Mandel Experiment}\label{Sec:detail-ZWM}

We now describe the ZWM experiment, first conceptually, then in detail. Afterwards, we present several of its fundamental conclusions for quantum physics in general.

\subsection{Description of the Experiment}\label{Sec:descrip-ZWM}
\begin{figure}[hbp]  \centering
    \includegraphics[width=0.9\linewidth]{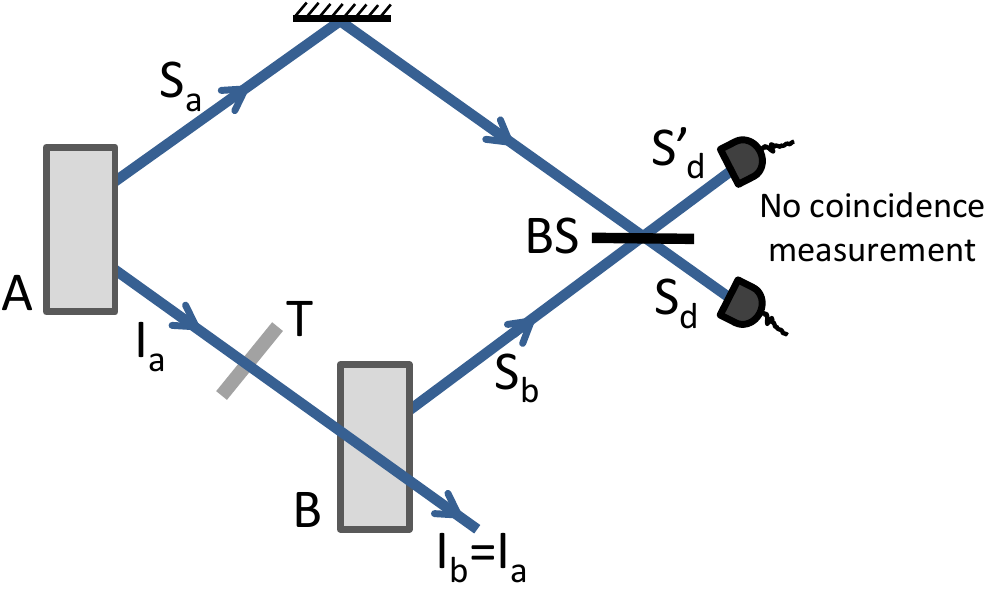}
    \qquad \caption{Zou-Wang-Mandel experiment. $A$ and $B$ are two
        identical photon-pair sources. $A$ and $B$ emit can a photon pair in paths ($S_a$, $I_a$) and ($S_b$, $I_b$) respectively. Paths $I_a$ and $I_b$ are made identical. An object with amplitude transmission coefficient $T$ is placed on $I_a$ between $A$ and $B$. A single-photon interference pattern is observed at both outputs of the beam splitter (BS) if the $S_a$ and $S_b$ are superposed. The visibility is proportional to the modulus of the amplitude transmission coefficient }
    \label{fig:Mand-setup}
\end{figure}
\par
The concept of Path Identity can be best illustrated with an absorptive element between the two crystals, which influences the which-crystal information, and therefore the interference visibility.

The experiment is illustrated in Fig. \ref{fig:Mand-setup}. Two bi-photon sources are denoted by $A$ and $B$. These two sources are two identical nonlinear crystals pumped by two \emph{mutually coherent} laser beams (not shown in the Figure). Two photons belonging to a pair are denoted by $S$ and $I$ (which stands for Signal and Idler, for historical reasons). Source $A$ emits the $S$ and $I$ along paths $S_a$ and $I_a$ respectively. Likewise, $B$ emits the photons along paths $S_b$ and $I_b$. The two crystals are pumped weakly, such that in most cases only maximally one crystal produces a pair of photons at a time.
\par
The paths $S_a$ and $S_b$ are superposed by a beam splitter (BS) and single-photon counting rate is measured at one or both outputs of BS. If the concept of path identity is applied, i.e. if $I_a$ is sent through source $B$ and the aligned with $I_b$ interference occurs in the ideal scenario. However, in practice, the pump and the down-converted light have finite spectral width, i.e. finite coherence time. Therefore, further conditions relating to the arm lengths of the interferometer are required to be satisfied. If the pump light has much longer coherence length than the down-converted light, the following condition must be satisfied \cite{wang1991induced}: 
\begin{align}\label{coh-length-cond-1}
\left|l_{S_a}-l_{S_b}-l_I\right|\ll L_{\text{dc}}, 
\end{align}
where $l_{S_a}$ is the optical path length along $S_a$ from $A$ to BS; $l_{S_b}$ is the optical path length along $S_b$ from $B$ to BS; $l_I$ is the optical path length along $I_a$ from $A$ to $B$; and $L_{\text{dc}}$ is the coherence length of the down-converted light. If this condition is met, there exists no information that allows one to determine from which crystal a signal photon arrived at the detector. If beam $I_a$ is blocked between $A$ and $B$ or condition (\ref{coh-length-cond-1}) is not met, the interference is lost. In this case, it is possible to determine from which crystal a signal photon arrived by detecting $I$ photon in coincidence. 
\par
We extend the analysis by placing an attenuator with amplitude transmission coefficient $T$ is placed on $I_a$ between sources $A$ and $B$. The visibility of the interference patterns turns out to be linearly proportional to $|T|$. This can be understood by the quantum mechanical rules of addition of probability.
An idler photon passes through the attenuator with a certain probability. The transmission probability is equal to $|T|^2$. When the idler photon passes through the object, we have a complete path identity. In this case, there exists no information about the origin of a single photon detected after the beam splitter, and thus interference occurs. If the photon does not pass through the object, the path identity is broken, and the signal photon does not contribute to the interference pattern. Therefore, there are three alternative ways through which a signal photon can arrive at the detector: 1) the signal photon has been emitted by crystal B; 2) the signal photon is emitted by crystal A, and the partner idler photon is transmitted through the object, and 3) the signal photon is emitted by crystal A, and the partner idler photon is blocked. Note that alternative 1 and 2 are indistinguishable and therefore, their probability adds. Alternative 3 is distinguishable from the rest, and thus its probability adds with the combined probability of the other two. If one determines the total probability in this way, one finds that the visibility is proportional to $|T|$.
\par
In the next section, we provide a more rigorous analysis of the experiment.

\subsection{Brief Theoretical Analysis}\label{subsec:theory-ZWM}
\par
The original analysis presented by Zou, Wang, and Mandel is based on quantum field theory \cite{zou1991induced,wang1991induced}. Here, we present an equivalent but simpler treatment. For the sake of simplicity, we do not consider the multimode nature of optical fields in this section. Multimode nature of the field will be considered later when we discuss imaging and spectroscopy experiments. We also slightly modify the notations used in Sec. \ref{ZWM-Mario}.
\par
We denote the photon pair states produced individually at $A$ and $B$
by $\ket{S_a}\ket{I_a}$ and $\ket{S_b}\ket{I_b}$ respectively. In the experimental condition, these two sources emit coherently, and they rarely produce more than one photon pair jointly. The resulting state, therefore, becomes
\begin{align}\label{ini-state-ZWM}
\ket{\psi_0}=\frac{1}{\sqrt{2}}\left(\ket{S_a}\ket{I_a}+
e^{i\phi} \ket{S_b}\ket{I_b}\right),
\end{align}
where $\phi$ is an arbitrary phase, and we have assumed that both sources have equal emission probability.
\par
Suppose that the two outputs
of the beam splitter are denoted by $S_d$ and $S_d'$ (Fig. \ref{fig:Mand-setup}).
The transformation of the states due to the beam splitter is given
by 
\begin{subequations}\label{BS-1}
    \begin{align}
    &\ket{S_a} \to \frac{1}{\sqrt{2}}\left(\ket{S_d}+i\ket{S_d'}\right), \label{BS-1-1} \\
    &\ket{S_b} \to
    \frac{e^{i\phi_S}}{\sqrt{2}}\left(\ket{S_d'}+i\ket{S_d}\right),
    \label{BS-1-2}
    \end{align}
\end{subequations}
where $\phi_S$ is the phase difference due to different propagation
distances along paths $S_a$ and $S_b$. 
\par
As mentioned before, the crucial part of the experiment is to make the paths $I_a$ and $I_b$ identical. This is done by sending $I_a$ through $B$ and then perfectly aligning with $I_b$. The quality of this alignment (path identity) can be theoretically modelled by placing an attenuator on beam $I_a$ between the sources $A$ and $B$. In our case, the attenuator is an object with complex amplitude transmission coefficient $T=|T|\exp[i\Phi_T]$. An idler photon passes through the object with probability $|T|^2$. Therefore, the object can be treated as a beam splitter and the path identity condition can be expressed as
\begin{align}\label{PI-T-ZWM}
\ket{I_a} \to \exp[i\theta_I](T\ket{I_b}+R\ket{l}),
\end{align}
where $|T|^2+|R|^2=1$ and $\ket{l}$ represents a photon that is lost or absorbed. We note that Eq. (\ref{PI-T-ZWM}) essentially reduces to Eq. (\ref{PI-eq}) when $|T|=1$.
\par
Applying the transformations
(\ref{BS-1}) and (\ref{PI-T-ZWM}) to Eq. (\ref{ini-state-ZWM}), we find that the state $\ket{\psi_0}$ transforms to
\begin{align}\label{final-state-ZWM}
&\ket{\psi_f}= \frac{1}{2}
\big(e^{i\theta_I}T\ket{I_b}+ie^{i(\phi+\phi_S)}\ket{I_b} +e^{i\theta_I}R\ket{l}
\big)\ket{S_d}
\nonumber \\ & +\frac{1}{2}
\big(ie^{i\theta_I}T\ket{I_b}+e^{i(\phi+\phi_S)}\ket{I_b} +ie^{i\theta_I}R\ket{l}
\big)\ket{S_d'}.
\end{align}
In order to determine the photon counting rates at outputs $S_d$ and $S_d'$, we carry out the following steps: 1) we determine the density operator $\ddo_f=\ket{\psi_f}\bra{\psi_f}$; 2) we trace over $\ket{I_b}$ and $\ket{l}$ to obtain the reduced density operator that represents the state of the single-photon at the outputs of the beam splitter (BS). The coefficients associated with $\ket{S_d}\bra{S_d}$ and $\ket{S_d'}\bra{S_d'}$ in the expression of the reduced density operator are proportional to the photon counting rates at $S_d$ and $S_d'$ respectively. The photon counting rates are found to be given by
\begin{subequations}\label{ph-ct-rt-ZWM-T}
    \begin{align}
    &R_d \propto \frac{1}{2}(1-|T|\sin (\phi_{ab}+\phi_S-\theta_I-\Phi_T)), \label{ph-ct-rt-ZWM:a} \\
    &R_{d'}\propto \frac{1}{2}(1+|T|\sin (\phi_{ab}+\phi_S-\theta_I-\Phi_T)). \label{ph-ct-rt-ZWM:a}
    \end{align}
\end{subequations}
\par
We note that both the amplitude and phase of the object's transmission coefficient appears in the interference pattern. This is a striking phenomenon because the interference pattern is obtained by detecting photons that have never interacted with the object. 

In the first decade after the experiment by Zou-Wang-Mandel experiment in 1991, a manifold of theoretical and experimental investigations have been reported. A few notable and exciting results contain the following: Studies of the interference's time delays and coherence times demonstrated elegantly the fundamental importance of information in quantum mechanics \cite{zou1992new, wang1992time, zou1992observation, zou1993control, grayson1993interference, barbosa1994forced, wang1999propagation}. Unlike \textit{quantum erasure experiments}, where information is deleted to observe interference, in the ZWM experiment, the which-way information is never born. This fundamental contrast is analyzed in more detail in several papers \cite{kwiat1994three, ryff1995changing, ou1997nonlocal}. The ZWM experiment has also inspired new mathematical methods for describing SPDC processes \cite{luis19962, casado1997parametric, mista2000quantum}, has been used for experimentally falsifying certain variations of de Broglie's deterministic pilot wave interpretation of quantum mechanics \cite{zou1992can}, and has inspired proposals to detect effects in quantum cosmology \cite{yurtsever2004signalling}.

Since it is possible to extract the information about the object from the recorded interference pattern, the concept of path identity also inspired an entirely new class of imaging, spectroscopy, and tomography experiments. In all these experiments, one does not need to detect the photon that interacts with an object.

\subsection{Non-classicality of the Zou-Wang-Mandel experiment}\label{subsec:nonclss-ZWM}
\begin{figure}[htbp] \centering
    \subfigure[] {
        \label{ZWM-data}
        \includegraphics[width=0.45\textwidth]{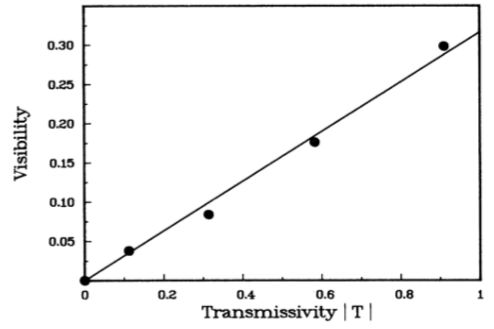}
    } \hskip 0.2cm
    \subfigure[] {
        \label{WM-plots}
        \includegraphics[width=0.45\textwidth]{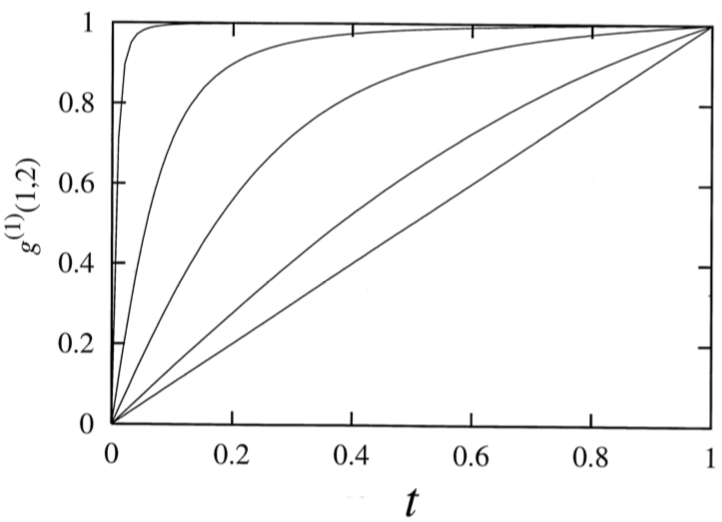}
    }  \caption{Non-classicallity of the Zou-Wang-Mandel experiment. (a) [Figure adapted from Ref. \cite{zou1991induced}]. Experimental data shows that the visibility of the single-photon interference pattern is linearly proportional to $|T|$. (b) [Figure adapted from Ref. \cite{wiseman2000induced}.] $g^{(1)}(1,2)$ refers to visibility and $t$ is the same as $|T|$ in the text and Fig. (a). The theoretical curves show that when the effect of stimulated emission is prominent, the linear dependence is no longer observed.} \label{fig:non-class-ZWM}
\end{figure}
The Zou-Wang-Mandel experiment does not have any classical explanation. We now briefly touch upon the quantitative evidence that supports this fact. We find from Eq. (\ref{ph-ct-rt-ZWM-T}) that the visibility of the interference pattern is linearly proportional to the modulus of the amplitude transmission coefficient of the attenuator. This linear dependence was experimentally verified by Zou, Wang, and Mandel (Fig. \ref{ZWM-data}). The physical reason for this linear dependence is the fact that the effect of stimulated emission is negligible in this experiment. 
\par
When beam $I_a$ is sent through source $B$ and then aligned with beam $I_b$, it is natural that stimulated emission would take place at $B$. However, in the experiment, $A$ and $B$ are two low-gain nonlinear crystals which are weakly pumped by laser light. In this case, the probability of down-conversion is so low that when emission occurs at $B$, beam $I_a$ is rarely occupied by a photon. Therefore, spontaneous emission dominates over stimulated emission and the quantum state, jointly produced by $A$ and $B$, becomes a linear superposition of two-photon states. Zou, Wang, and Mandel showed that such a quantum state can explain the linear dependence shown in Fig. \ref{ZWM-data}.
\par
Non-classicality of this experiment was also verified by Wiseman and M\o lmer independently \cite{wiseman2000induced}. They considered a situation in which the crystal gain can be arbitrarily increased. They showed that for high crystal gain, i.e. when the effect of stimulated emission becomes prominent, the above-mentioned linear dependence is no longer observed (Fig. \ref{WM-plots}). For very high gain, the problem can be treated by classical nonlinear optics, but in that case, the linear dependence is no longer found. 
\par
The non-classicality of the Zou-Wang-Mandel experiment has also drawn attention recently. A good description has been presented in Ref. \cite{kolobov2017controlling}. It has also been shown that if the laser pump is replaced by a single-photon pump, the possibility of stimulated emission becomes strictly forbidden, and even then the interference occurs \cite{lahiri2017can}.
\par
The ZWM experiment's non-classicality makes it an important milestone in the history of quantum photonic experiments as we elaborate more in the next section.

\subsection{Quantum nature of photonic experiments}
The classical (non-quantum) theory clearly distinguishes between particles and waves: a classical particle does not display interference effect whereas a classical wave does. One of the great successes of quantum mechanics lies in predicting interference phenomena displayed by systems which were previously understood as classical particles. Well-known examples are neutron  \cite{zeilinger1986complementarity,rauch2015neutron} or macro-molecule interference \cite{fein2019quantum}. Quantum theory is also applicable to light, and therefore every optical interference effects can also be understood as quantum mechanical effects. However, the scientific community usually does not perceive such a view as appropriate. An optical phenomenon is called truly quantum mechanical when the classical electromagnetic theory of light is unable to explain it, and the only explanation comes from quantum mechanics. In this paper, we use the term \textit{quantum mechanical} in this sense. Therefore, the intensity modulation observed in Young's double-slit experiment or in a Mach-Zehnder interferometer is a classical effect irrespective of the state of light used in the experiment.
\par
It turns out that most optical phenomena that involve measurement of intensity or single-photon counting rate display no quantum mechanical effect. Although the photoelectric effect suggests the quantum nature of light, it has been shown that a highly successful quantitative theory of photoelectric detection can be developed without considering the quantum nature of light \cite{mandel1964theory}. The essence of this work lies in the semi-classical theory of light-matter interaction, in which light is treated by classical electromagnetic theory and the atoms by quantum mechanics. It was first experimentally demonstrated by Clauser that the semi-classical theory of photoelectric detection provides incorrect results if one measure \emph{coincidence counts} at two detectors illuminated by light generated by the cascade $9~^1P_1\to 7~^3S_1\to 6~^3P_1$ in excited $^{202}$Hg atoms \cite{clauser1974experimental}. It has turned out in later investigations that most quantum mechanical effects in the optical domain involve coincidence detection of at least two photons (i.e., intensity correlation). Several such phenomena has been studied in the field of quantum optics; three notable ones are photon antibunching \cite{kimble1977photon}, two-photon interference \cite{horne1986einstein,ghosh1987observation,hong1987measurement,horne1989two}, and Bell test experiments \cite{freedman1972experimental,aspect1981experimental,weihs1998violation}. 
\par
The experiment by Zou, Wang, and Mandel \cite{wang1991induced,zou1991induced} turns out to be the first demonstration of the quantum nature of an optical interference effect that relies only on the detection of single photons. In the language of optical coherence theory, such interference effects can be called lowest-order correlation effects\footnote{The terminology is not uniform. Such correlation effects have been called both ``first-order'' \cite{glauber1963quantum} and ``second-order'' \cite{mandel1995optical} in the literature. We use the term ``lowest-order'' to avoid confusion.}. Therefore, the concept of path identity is also significant for optical coherence theory. We elaborate more on this point below. 

\subsection{Implications for Optical Coherence Theory}\label{Sec:imp-opt-coh}
\SkipTocEntry\subsubsection*{Path Identity and the Degree of Coherence}\label{subsec:doc-PI}
\par
Optical coherence theory studies phenomena that are a manifestation of statistical fluctuations present in optical fields \cite{born_wolf,mandel1995optical}. For example, when two light beams (or equivalently light from two point sources) are superposed, the corresponding fields add linearly. If the field corresponding to one beam is fully uncorrelated with that of the other beam, no interference occurs. Interference requires correlation between the fields that are superposed. In fact, the visibility of the interference pattern depends on the amount of correlation. Mathematically, a typical interference pattern can be expressed in the general form \cite{born_wolf}
\begin{align}\label{inten-law-coh}
R=R_1+R_2+2\sqrt{R_1R_2}~|\gamma_{12}|\cos\phi,
\end{align}
where $R_1$ and $R_2$ are individual intensity contributions from the interfering fields, $\phi$ is a phase, and $|\gamma_{12}|$ is the modulus of the degree of coherence, which is a measure of the correlation between the two superposed fields. Clearly, the visibility is linearly proportional to $|\gamma_{12}|$. The degree of coherence, therefore, provides a quantitative measure of the ability of light to interfere. 
\par
It is important to appreciate that such a description of interference is not in contradiction with quantum mechanics. This fact can be understood as follows: when two beams are superposed, the absence of ``which-way information'' implies that the superposed fields are maximally correlated and intensities of the corresponding beams are equal \cite{wootters1979complementarity,greenberger1988simultaneous,mandel1991coherence,jaeger1993complementarity,englert1996fringe}. In the other extreme case, when which-way information is fully available, the superposed fields are completely uncorrelated. This fundamental connection between the two apparently distinct interpretations has been elegantly demonstrated by the ZWM experiment.
\par
According to the classical understanding, a change in correlation (the degree of coherence) between the two superposed fields requires a direct interaction with at least one of them. However, according to quantum mechanics, one only needs to ensure that the which-way information does not exist and such an act may not require direct interaction with the fields. The path identity employed in the ZWM experiment allows one to control the correlation between the fields without interacting with any of them. In their experiment, $|\gamma_{12}|$ is equal to the modulus of the amplitude transfer coefficient ($|T|$) of the object that never interacts with beams $S_a$ and $S_b$ (see Fig. \ref{fig:Mand-setup}). The ZWM experiment is therefore a landmark experiment that highlights the quantum mechanical aspect of the degree of coherence. 

\SkipTocEntry\subsubsection*{Path Identity and Optical Polarization}\label{subsec:par-pol-PI}
\begin{figure}[t] \centering
    \subfigure[] {
        \label{figa:PI-pol-setup}
        \includegraphics[width=0.45\textwidth]{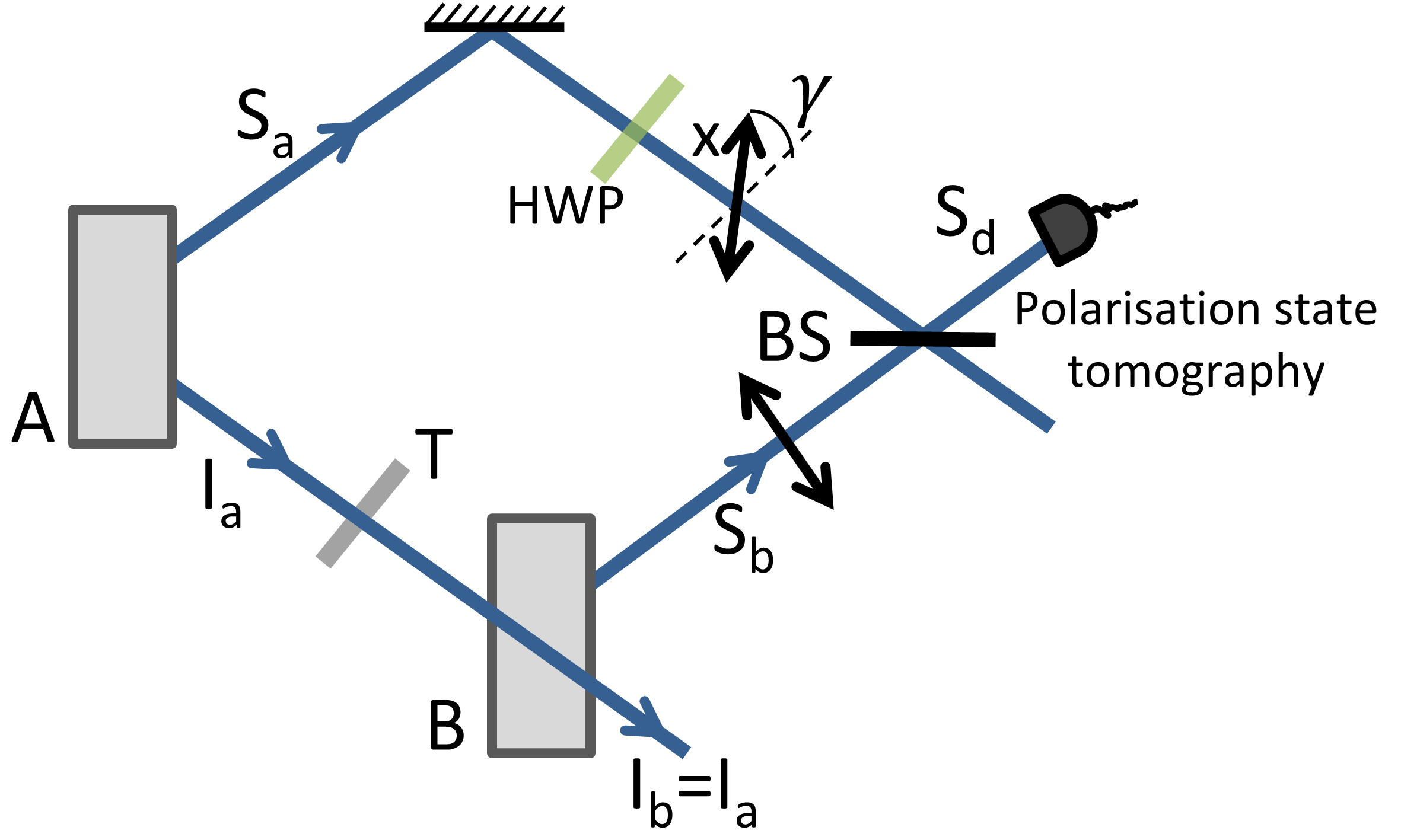}
    } \hskip 0.5cm
    \subfigure[] {
        \label{figb:dop-T-plot}
        \includegraphics[width=0.45\textwidth]{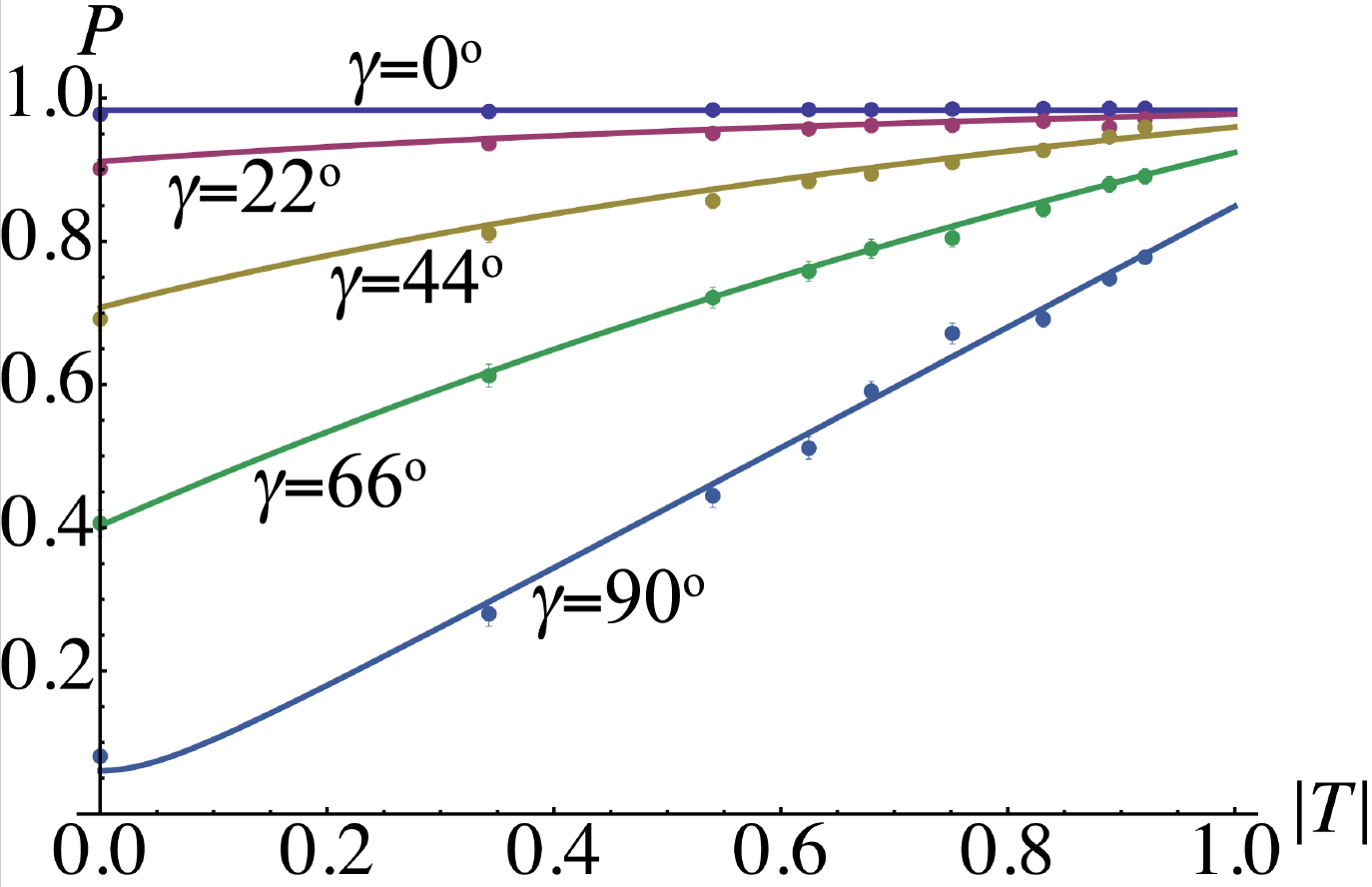}
    } \caption{Degree of polarization and path identity: (a) Schematics of the experiment. (b) Experimentally observed dependence of the degree of polarization on $|T|$ for various values of $\gamma$. Solid lines represent theoretical curves considering experimental imperfections. (Adapted from \cite{lahiri2017partial})} \label{fig:dop-PI}
\end{figure}
Polarization properties of a light beam at a point can also be understood as an effect of correlation between the components of electric field vectors at that point \cite{pancharatnam1956generalized,wolf1959coherence}. The degree of polarization ($P$) quantifies how polarized a light beam is at a point. For example, $P=1$ implies fully polarized light and $P=0$ implies unpolarized light. In the intermediate case, $0<P<1$, the light is partially polarized.
\par
Suppose that we choose two mutually orthogonal transverse directions, $x$ and $y$, perpendicular to the direction of beam propagation. If the intensities associated with individual field components along $x$ and $y$ are equal, the degree of polarization is the modulus of the normalized correlation function of these two field components \cite{born_wolf}. Like the degree of coherence, the degree of polarization can also be understood from the principles of quantum mechanics \cite{lahiri2011wave}. The concept of path identity allows us to design an experiment that establishes the quantum mechanical aspect of the degree of polarization. 
\par
We consider a modification of the ZWM experiment, as illustrated in Fig. \ref{figa:PI-pol-setup}. Suppose that the beams $S_a$ and $S_b$ are polarized along directions $x$ and $H$ (horizontal) respectively; both directions are transverse. The direction $x$ encloses an angle $\gamma$ with the direction $H$. One can vary the angle $\gamma$ by the use of a half-wave plate (HWP). The correlation between these two superposed field components can be controlled by varying the value of $|T|$. As a result, the polarization properties of light generated by superposition changes. Theoretical analysis shows that the degree of polarization at an output of BS is given by \cite{lahiri2017partial}
\begin{align}\label{dop-ZWM}
P=\frac{|T|+\cos\gamma}{1+|T|\cos\gamma},
\end{align}
where the interferometric phase has been set to a multiple of $2\pi$.
Figure \ref{figb:dop-T-plot} shows the experimental results that support the theoretical prediction.
The experiment thus shows that the degree of polarization of a light beam can be changed without interacting with it; an effect that is purely quantum mechanical.

\subsection{Investigation of complementarity}
\begin{figure}[b]
  \centering
  \includegraphics[width=0.5\textwidth]{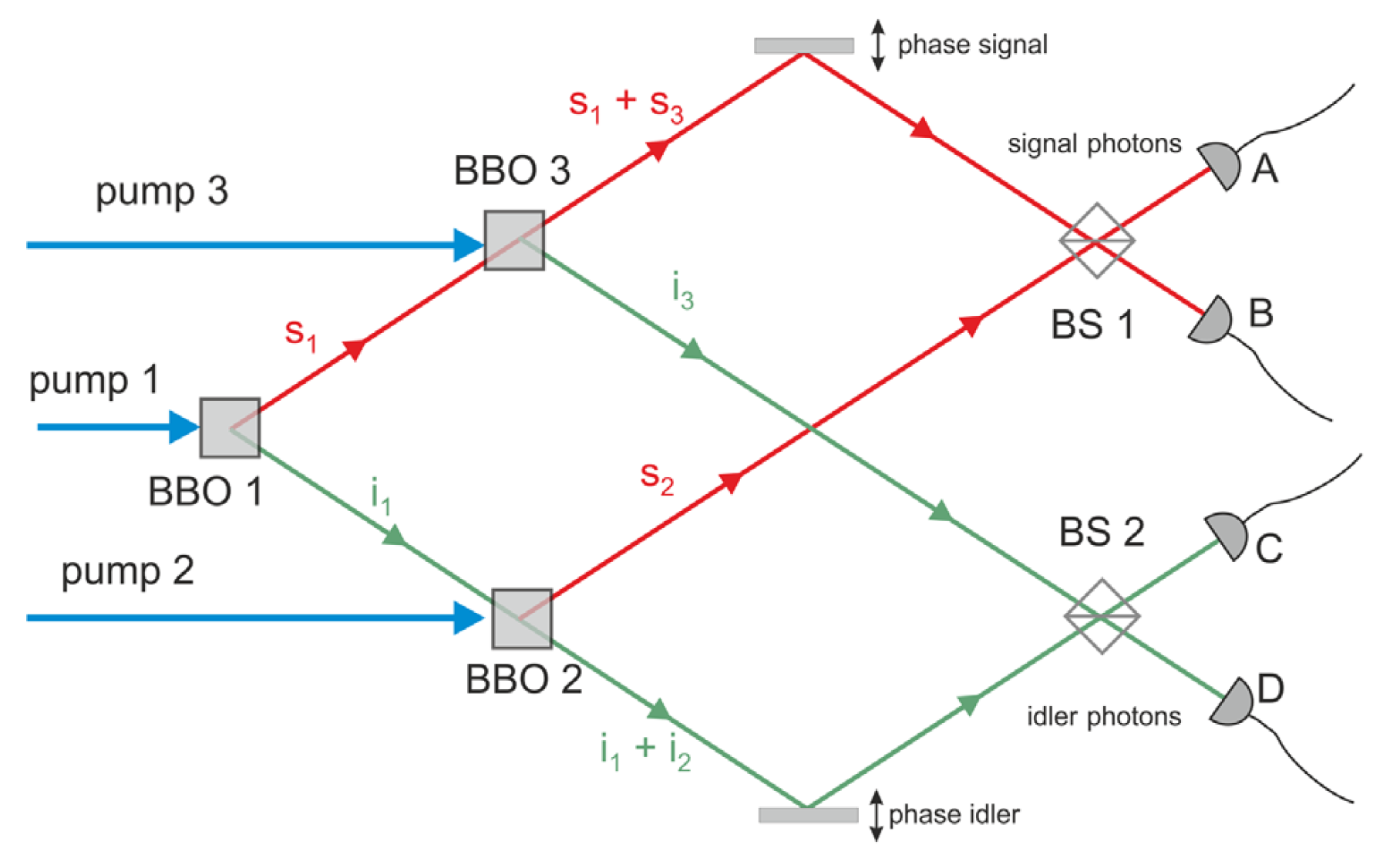}
  \caption{An extension of the ZWM experiment to three crystals allows for more general tests of the complementarity principle. Image from \cite{heuer2015induced}.
  }
\label{fig:MenzelThreeCrystals}
\end{figure}

Wave-particle complementarity has been a cornerstone in the development of quantum mechanics since it has been introduced by Niels Bohr in 1928 \cite{bohr1928quantenpostulat}. It defines a relation between the \textit{which-way} information $K$ (which is related to particle-like behaviour) and the interference visibility $V$ (which is related to wave-like behaviour). It states that 
\begin{align}
K^2 + V^2 \leq 1.
\label{complementarity}
\end{align}

In the ZWM experiment, the visibility $V$ can only be optimal when there is no which-crystal information, $K$. An extension of the ZWM to three crystals has led to more general theoretical \cite{ryff1995interference, ryff2015comment} and experimental \cite{heuer2015induced, heuer2015phase} studies of complementarity. 

The quantum state in the experiment performed in \cite{heuer2015induced} (depicted in Figure \ref{fig:MenzelThreeCrystals}), before the beam splitters, can be written as
\begin{align}
\ket{\psi}=g_1 e^{i(\phi_s+\phi_i)}\ket{s_3,i_2} + g_2 e^{i\phi_i}\ket{s_2,i_2} + g_3 e^{i\phi_s}\ket{s_3,i_3}.
\label{menzel1}
\end{align}
Here, $\phi_s$ and $\phi_i$ are the phases introduced by the path delay at the top and bottom mirror. Furthermore, the path identification $\ket{s_1} \to \ket{s_3}$ and $\ket{i_1} \to \ket{i_2}$ is used. The coefficients $g_1$, $g_2$ and $g_3$ correspond to the pump power (thus photon pair rate) of the crystals BBO 1, 2 and 3, respectively.

If $g_3$=0, $g_1$=$g_2$=$g_1$ and $\phi_i$=0, one recovers the original ZWM experiment, and the state in eq. (\ref{menzel1}) can be written as 
\begin{align}
\ket{\psi}&=e^{i\phi_s}\ket{s_3,i_2} + \ket{s_2,i_2}\nonumber\\
&=\left(e^{i\phi_s}\ket{s_3} + \ket{s_2}\right)\ket{i_2}.
\label{menzel1}
\end{align}
There, one has a single photon in perfect, equally weighted superposition between path $s_2$ and $s_3$. Varying the relative phase $\phi_s$ leads to (ideally perfect) modulation of the count rates in detector A, as well as coincidence count rates between detector A and D. The perfect interference appears because one has no information in which crystal the photon pair was created. Thus, when a photon is observed in detector A, one does not know whether the photon arrived at the BS through path $s_2$ or $s_3$. This principle indistinguishability of the two events lead to perfect visibility $V=1$.

The new, exciting situation appears, when crystal BBO 3 is pumped as well. Now, there are three possibilities for a photon to arrive at detector A: Through path $s_1$  or through path $s_2$ (created in BBO 1 or BBO 2). These two possibilities cannot be distinguished, as the idler photons are path identified. However, the photon could also be created in BBO 3. There, one has additional information about the idler path because $i_2$ and $i_3$ are not identified. This additional path information $K$ decreases the visibility $V$.

The state for this situation can be written as (with $g_1$=$g_2$=$g_3$=$1$ and $\phi_i=0$) 
\begin{align}
\ket{\psi}&=& \left(e^{i(\phi_s)}\ket{s_3,i_2} + \ket{s_2,i_2} + e^{i\phi_s}\ket{s_3,i_3}\right) \nonumber\\
&=& \left(e^{i(\phi_s}\ket{s_3} + \ket{s_2}\right)\ket{i_2} +  e^{i\phi_s}\ket{s_3,i_3}.
\label{menzel2}
\end{align}
The first two terms lead to a modulation of the interference patter in detector A, whereas the last term contributes to path information, thus leads to an incoherent background in the interference.

One can erase the information of the idler, by detecting the idler photon after the BS2 in detector D. The detection in D makes it impossible to know whether the idler photon arrived through path $i_2$ or $i_3$. If one heralds the detector clicks in A on an event in detector D (i.e. measures coincidence counts between A and D), no path information about photon in detector A exists in principle. Thus one recovers perfect interference fringes. 

The same authors generalized the experiments to stimulated emission configuration, by overlapping $i_1$ and $i_3$ with a stimulating HeNe laser \cite{heuer2015complementarity}, and transmission objects in the idler paths. They find general relations between pump powers, transmission magnitude and visibilities. These experiments demonstrate the connection between fundamental information of the photon pair's origin and visibility beautifully.

\subsection{Transmission measurement with three sources}
\begin{figure}
\centering
\includegraphics[width=0.5\textwidth]{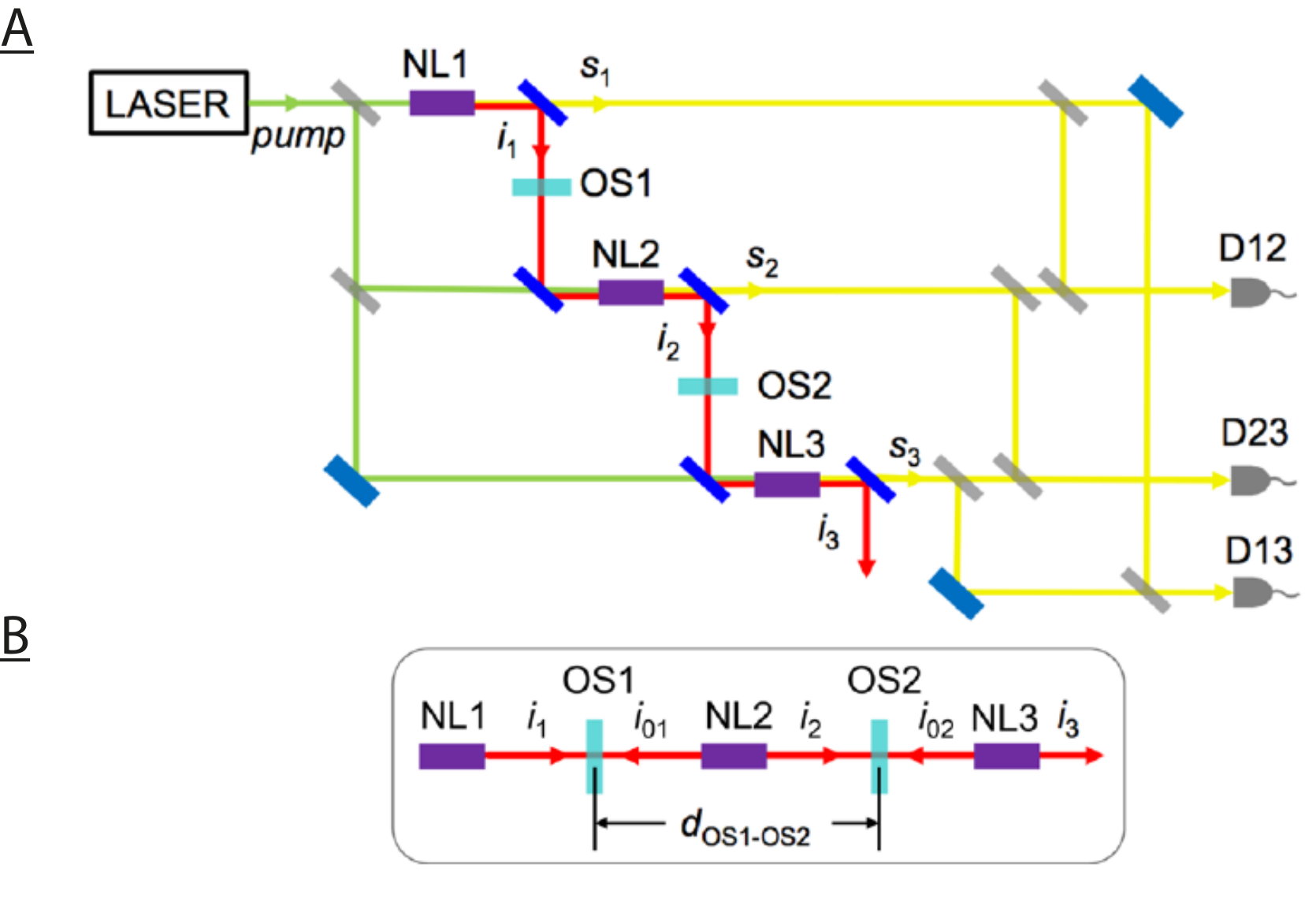}
\caption{Three crystal setup analyzed in \cite{lee2017quantum} (A). If the coherence lengths are long enough, the reflectivity of sample OS1 affects the interference pattern between NL2 and NL3. This is possible because the two samples OS1 and OS2 form an optical cavity (B). Intuitively, if the reflectivity of OS1 is high, an idler photon from NL2 has many tries to pass through OS2, because idler photons that are not transmitted by OS2 remain inside the cavity and can still be transmitted to NL3 later. This increases the effective transmission, and thus the visibility observed at D23. Figure from \cite{lee2017quantum}.}
\label{fig:cho3crystals}
\end{figure}

A variant of the previous experiment, analyzing the consequences of the existence of information using cavities has recently been demonstrated by \cite{lee2017quantum}. The experimental scheme with three sources (Fig. \ref{fig:cho3crystals}A) leads to a different peculiar feature of it of the ZWM phenomenon. If two partially reflecting objects are inserted in the idler beam, they act as a cavity for idler photons (Fig. \ref{fig:cho3crystals}B). 

This gives rise to the interesting observation that the visibility between the signal beams of the last two crystals NL2 and NL3, depends on the properties of the object OS1 placed between the first and the second crystal, in particular on its reflectivity.

In a simplified picture, this effect can be understood as follows.
The visibility observed in the original (two-crystal) ZWM experiment depends on the transmission of an object placed in the idler beam between the two crystals. If only a fraction of idler photons originating from the first crystal are transmitted through the second crystal, a photon detected after the setup is more likely to have been emitted by the second source. Thus, partial which-way information can be obtained, resulting in reduced visibility.

Here, however, idler photons that were not transmitted are not absorbed by the object but remain inside the cavity. A fraction of them is transmitted after one or more round trips. Thus, given the coherence length requirements to observe interference are still satisfied, the quality factor (i.e. the number of round trip times) of the cavity determines the actual probability of transmitting an idler photon after an arbitrary number of round trips. As the number of round trips of an idler photon depends on the reflectivity of both mirrors (objects), so does the observed visibility. In the limiting case of OS1 having perfect reflectivity, an idler photon is never lost unless it passed through OS2, even if it has imperfect transmission. In this case, at least theoretically, unit visibility is observed (regardless of the transmission of OS2).

\section{Reconstruction of Objects with Undetected Photons}\label{section:objectrecon}
In this section, we consider experiments which can reconstruct properties of objects without ever detecting the photons that interact with it. The properties can, for instance, be spatial (for quantum imaging) or in the frequency domain (for spectroscopy). A key interest in this method comes from the fact that the wavelengths of the detected photon and the probing (idler) photon can be significantly different, thus allowing for novel quantum technologies, as we discuss here.

\subsection{Quantum Imaging with undetected Photons}\label{subsec:q-img}
\begin{figure*} \centering
    \subfigure[] {
        \label{fig:img-setup}
        \includegraphics[scale=0.45]{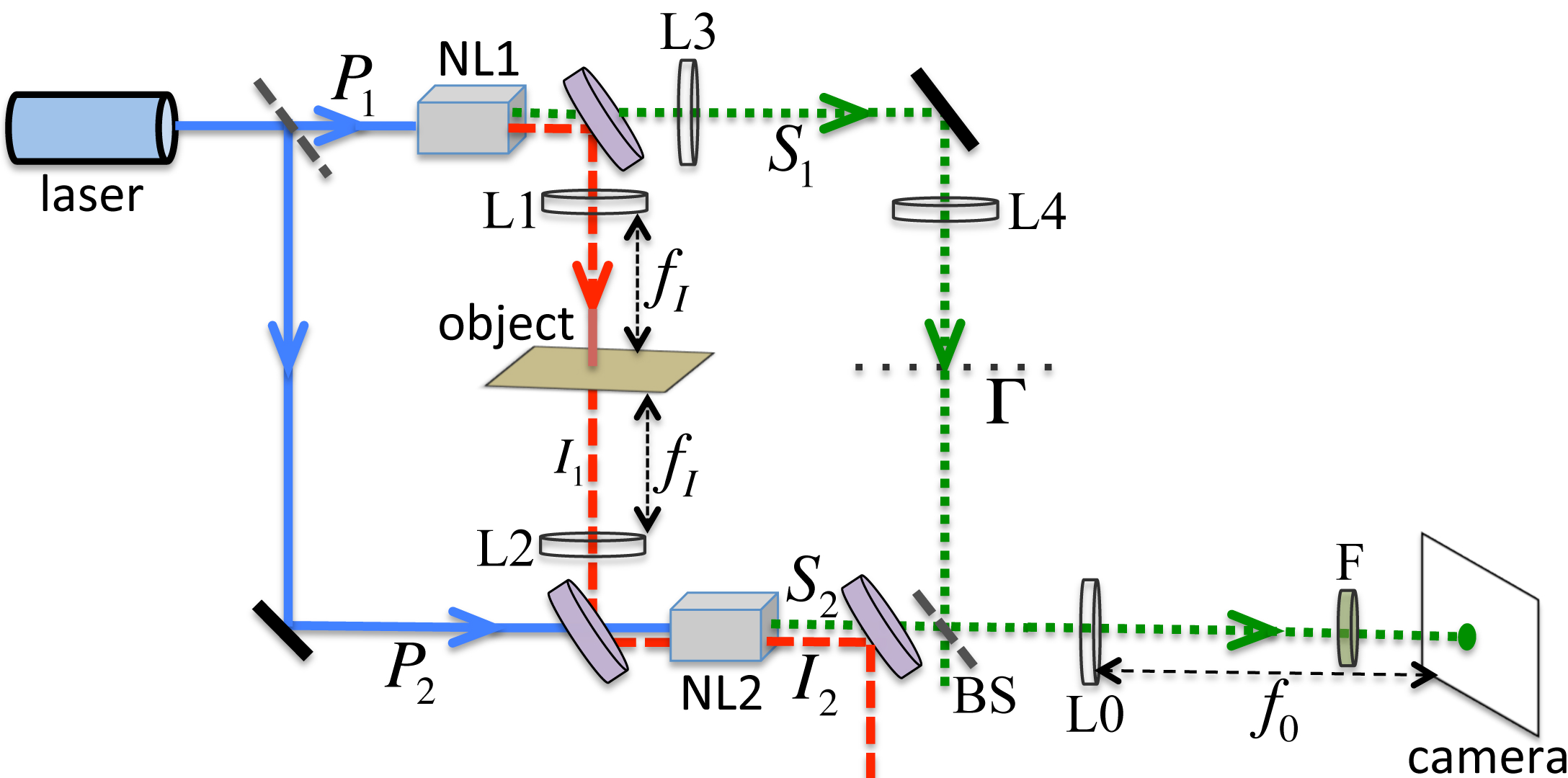}
    } \hskip 0.2cm
    \subfigure[] {
        \label{fig:4-f-obj-ed-b}
        \includegraphics[scale=0.22]{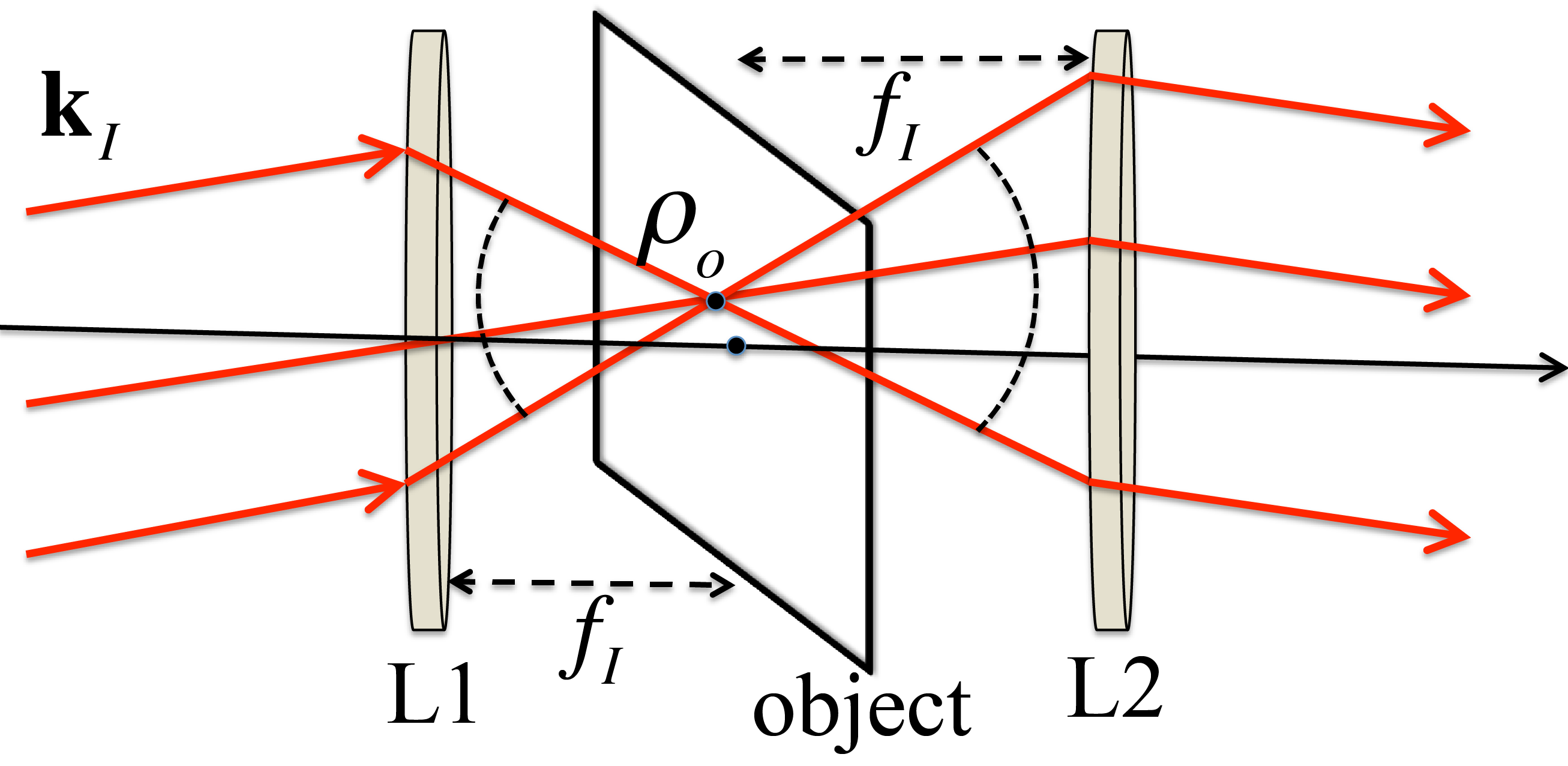}
    } \hskip 0.2cm
    \subfigure[] {
        \label{fig:det-sys}
        \includegraphics[scale=0.22]{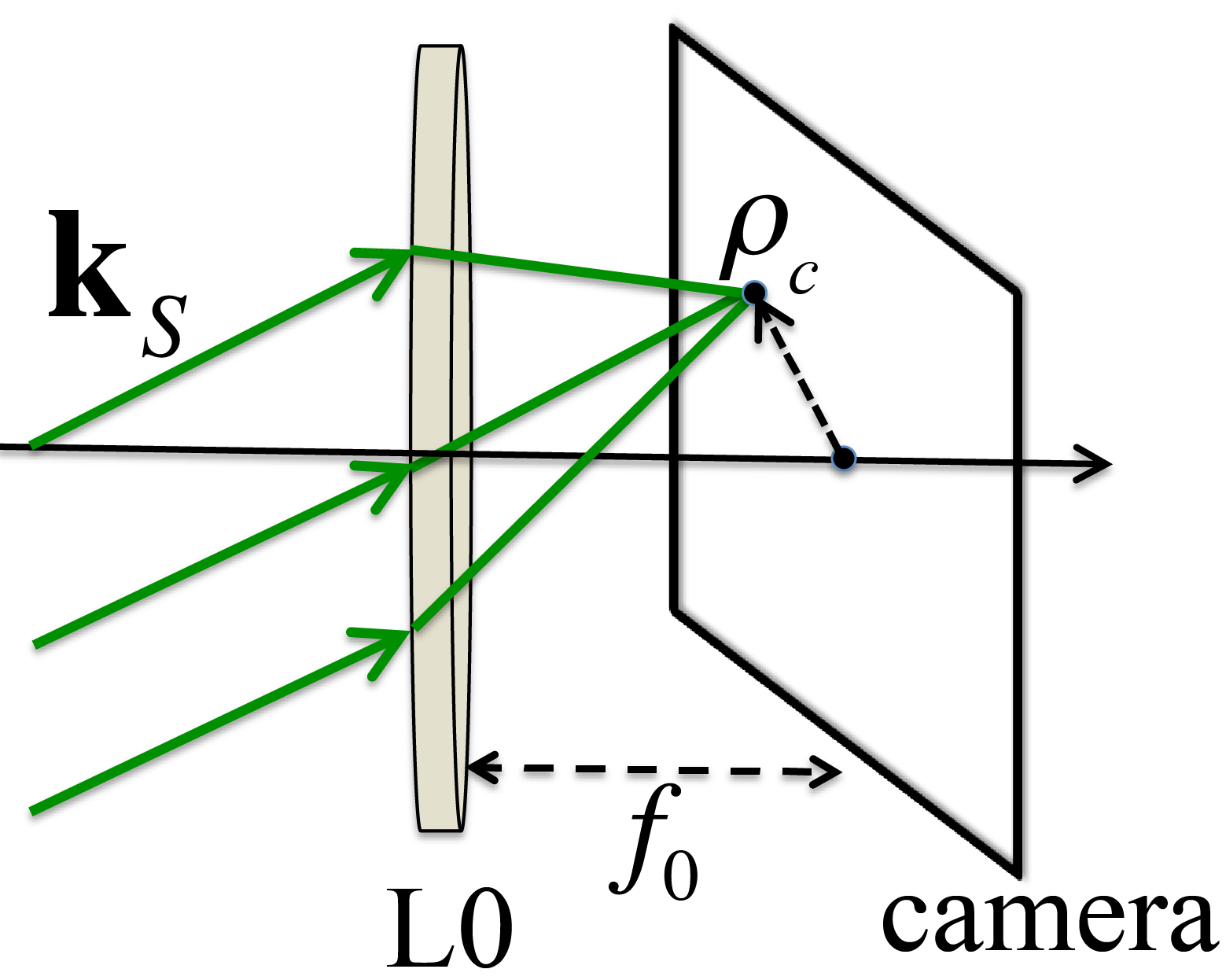}
    }  \caption{(Adapted from \cite{lahiri2015theory}): (a) Imaging setup. NL1 and NL2 are two identical non-linear crystals pumped coherently by 545 nm laser light. Each crystal can produce a signal (810 nm) and idler (1550 nm) photon by collinear spontaneous parametric down-conversion. Dichroic mirrors separate signal and idler light and sent to different paths. Alignment of $I_1$ and $I_2$ gives the path identity. The object is placed on $I_1$ between NL1 and NL2. 4-f lens systems are used in signal and idler arms. (b) 4-f lens system on idler path. An idler wave vector is focussed at a point,
        $\pmb{\rho}_{o}$, on the object plane. The emerging spherical wave from $\pmb{\rho}_{o}$ is converted back to a plane wave. (c) Detection
        system: the corresponding signal wave vector is focussed at a point, $\pmb{\rho}_{c}$, on the image plane (camera).} \label{fig:imaging-setup-details}
\end{figure*}
The imaging experiment that uses the concept of path identity was demonstrated in Ref. \cite{lemos2014quantum}. The experiment is illustrated in Fig. \ref{fig:img-setup}. 
The non-linear crystals, NL1 and NL2, are used as photon-pair sources. They are pumped coherently by a laser beam. Each of these crystals can generate a photon pair by the process of spontaneous parametric down-conversion (SPDC). In the experiment, the signal and idler have different wavelengths: the wavelengths of signal and idler are 810 nm and 1550 nm, respectively. A 4f lens system is in the path of the idler beam such that the image of NL1 falls on NL2. The object is placed in the idler path between the two non-linear crystals and at the center of the 4f system. Therefore, the object is at the Fourier plane of both crystals. An equivalent 4f system is placed in the path of the signal beam ($S_1$) generated by NL1. After a beamsplitter superposes the signal beams, a camera detects the outputs. The camera is placed at the focal plane of a positive lens, $L_0$, which is the Fourier plane of both crystals. 
\begin{figure}[htbp]
    \centering
    \includegraphics[width=9cm]{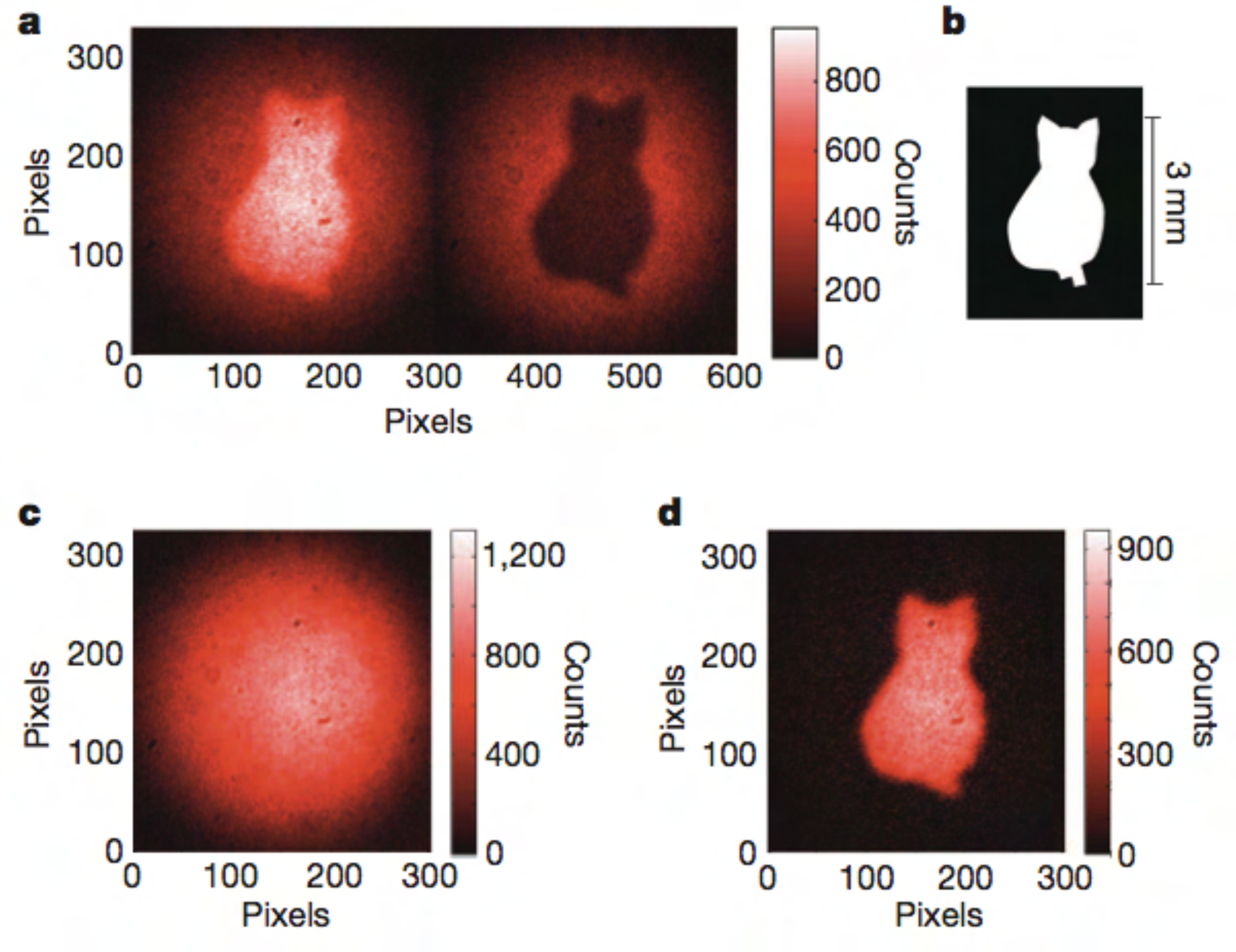}
    \caption{(adapted from \cite{lemos2014quantum}) a) The two outputs of the beam splitter. The image of the cat shows the area where interference occurs.  b) The absorptive object: the cat. c) The sum of the
        outputs of the beam splitter contains no image. d) The subtraction of the outputs leads to enhancement of
        image contrast.}\label{fig:Undetected2}
\end{figure}
\par
The 4f system in the path of the idler consists of two positive lenses (Fig. \ref{fig:4-f-obj-ed-b}). The first lens ($L_1$) focuses the plane wave to a point, $\pmb{\rho}_o$, on the object. The spherical wave emerging from this point is converted back to a plane wave by the second lens ($L_2$). This plane wave then passes through NL2. If an idler photon is emitted in this plane wave mode, its partner signal photon will be in a mode that satisfies the phase-matching condition. The associated plane wave is focused at a point, $\pmb{\rho}_c$, on the camera. It follows from the discussion of Sec. \ref{subsec:theory-ZWM} that the visibility recorded at $\pmb{\rho}_c$ will be determined by the modulus of the amplitude transmission coefficient at $\pmb{\rho}_o$. Therefore, for an absorptive object, the visibility map observed at the camera gives the image. Furthermore, as is clear from Eqs. (\ref{ph-ct-rt-ZWM-T}) the phase of the object at each point can also be constructed. There are numerous ways of constructing the image. A simple method of subtracting the two outputs of the beamsplitter was adopted in Ref. \cite{lemos2014quantum}. 
\begin{figure}[htbp]
    \centering
    \includegraphics[width=9cm]{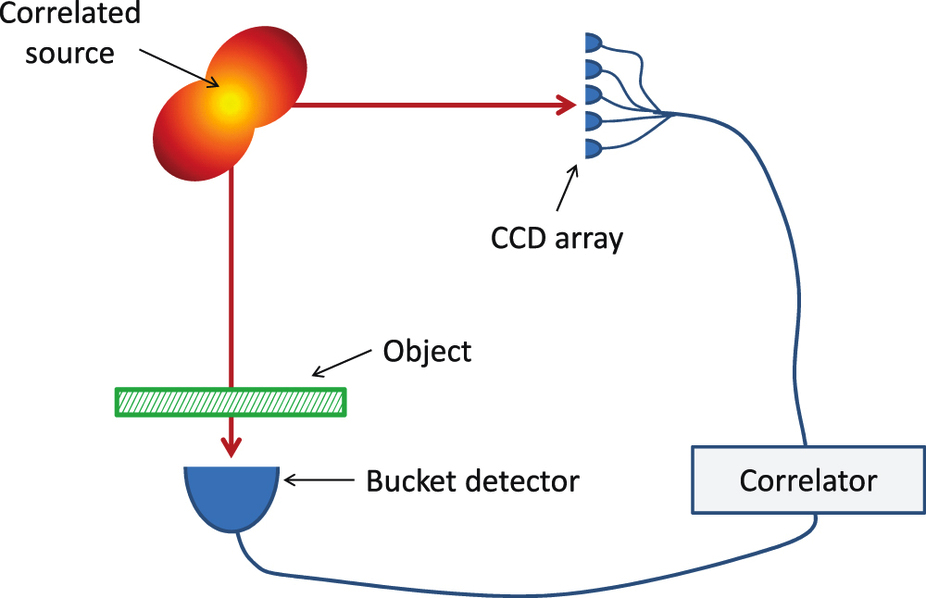}
    \caption{Ghost imaging is fundamentally different from imaging by path identity. In ghost imaging, the photons interacting with the object must be detected, and coincidence or an equivalent measurement must be performed. (Figure adapted from \cite{ragy2012nature}.) }\label{fig:GhostImaging}
\end{figure}

Figure \ref{fig:Undetected2} shows the images of an absorptive object. The object is a piece of cardboard from which the silhouette of a cat has been removed. [Fig. \ref{fig:Undetected2}(b)]. Interference patterns that contain the image are obtained at the two outputs of BS
are shown in Fig. \ref{fig:Undetected2}(a). When the two outputs are
added, the image disappears [Fig. \ref{fig:Undetected2}(c)]. This is expected from Eq. (\ref{ph-ct-rt-ZWM-T}). When one of the outputs is subtracted from the other, the image contrast is enhanced because the background gets nullified [Fig. \ref{fig:Undetected2}(d)].

The different wavelengths of signal ($\bar{\lambda}_S$) and idler
($\bar{\lambda}_I$) photons also play a role in the image
magnification. It can be shown by explicit calculation that the
magnification is given by $M=f_0 \bar{\lambda}_S/(f_I \bar{\lambda}_I)$,
where $f_0$ and $f_I$ are the focal lengths illustrated in Fig.
\ref{fig:img-setup}. A rigorous theory of the imaging experiment can be found in \cite{lahiri2015theory}, and comprehensive analysis of the image quality in terms of pump power has been conducted in \cite{kolobov2017controlling}.
\par
We would like to stress that imaging by path identity is fundamentally different from ghost imaging \cite{pittman1995optical,gatti2004ghost}. In ghost imaging, the light that interacts with the object must be detected, and coincidence or an equivalent measurement must be performed (Fig. \ref{fig:GhostImaging}). In contrast to ghost imaging, imaging by path identity does not involve the detection of the light which illuminates the object, and it does not involve any coincidence or equivalent measurement. Besides, ghost imaging can be understood purely classically \cite{bennink2004quantum}, while imaging by path identity is a genuine quantum mechanical phenomenon, as we described in Sec. \ref{subsec:nonclss-ZWM}.

\subsection{Quantum spectroscopy with undetected photons}
%Introduction why spectroscopy is important and interesting, what can we learn from it?
%Rotational and vibrational modes of molecules are in the mid- and far-infrared region
%
In the previous section, we explained how the principle of path identity is used for imaging. This effect, based on the distinguishability of the photons in the spatial degree of freedom, can be applied to distinguishability in any degree of freedom, e.g. frequency domain. In the following section, we explain how distinguishability in the frequency domain is exploited to accomplish spectroscopy with undetected photons. Today, spectroscopy is one of the most important workhorses in various fields in science and technology. Ranging from biology, chemistry, climate research, and fundamental cosmology, spectroscopy reveals key information about these diverse systems in a broad range of the electromagnetic spectrum~\cite{stewart2004infrared}. Especially interesting are the infrared and far-infrared regions of the electromagnetic spectrum. These spectral ranges are typical for vibrational and rotational modes of different molecules such as carbon dioxide, for example.

However, special optical equipment and especially detectors with low efficiencies pose significant technological obstacles. Kalashnikov et al.~\cite{kalashnikov2016infrared} used the principle of path identity in a clever way to probe the spectrum of carbon-dioxide and detected at a different wavelength. Particularly interesting is that the detection wavelength can be chosen such that it lies in the visible range where efficient detectors exist.

\begin{figure}[htbp]
    \centering
    \includegraphics[width=0.45\textwidth]{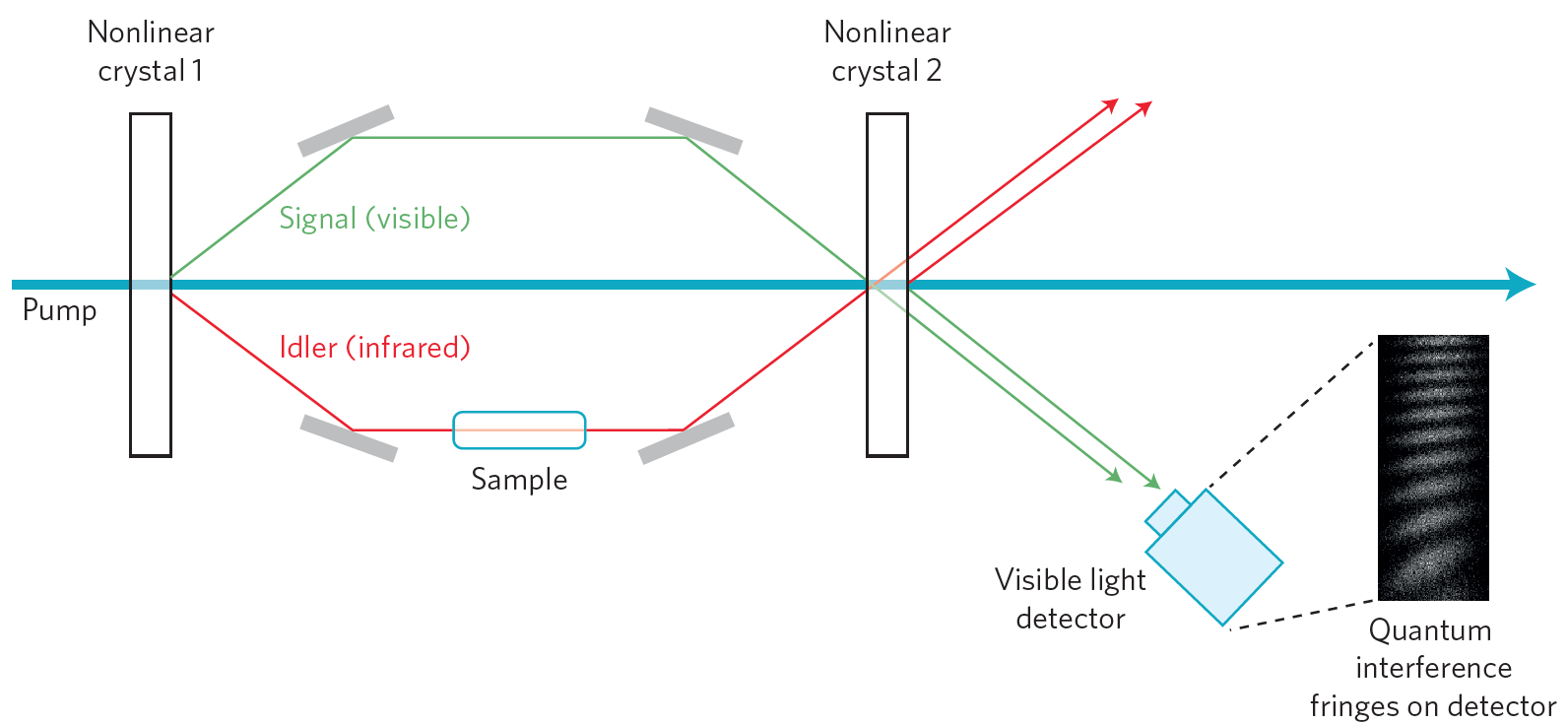}
    \caption{Principle scheme of spectroscopy with undetected photons. Two non-linear crystals probabilistically emit identical and frequency correlated photon pairs. Identifying the paths of the respective photon pairs then allows observing interference in the signal photon, for example. The acquired phase in the interference pattern depends on all three photons involved: The pump, signal and idler, which are all at different frequencies. The interference visibility is governed by the transmission of the sample placed in the idler beam. Thus at an absorption line of the sample under study, the visibility in the spectrogram detected at the camera disappears. Image is taken from Wolf \& Silberberg~\cite{wolf2016quantum}.}
    \label{fig:spectroscopy-principle}
\end{figure}

%From Mandel we know that absorption changes the interference effect -> absorption lines of atoms/molecules/air/CO2
In Sec. ~\ref{Sec:detail-ZWM} we studied the explicit dependencies of the expected intensities upon an object in the idler path, see Fig.\ref{fig:Mand-setup}. The linear relation between the observed visibility and the absorption of the idler beam suggests that this technique could be used for absorption spectroscopy. In this article, the authors explored this effect to detect infrared absorption lines of carbon dioxide molecules ($\text{CO}_2$) detecting visible light. The principle scheme is depicted in Fig.\ref{fig:spectroscopy-principle}. Two non-linear crystals made from Lithium Niobate (LiNbO3) emit non-degenerate photon pairs correlated in their frequencies. The idler photon is centred around 4161nm and can be absorbed by the $\text{CO}_2$ molecules between the two non-linear crystals, as shown in Fig.\ref{fig:spectroscopy-principle}. Overlapping both, signal and idler photon paths identically at the second non-linear crystal allows observing interference in the signal photon. Thereby the interference visibility is determined on the transmission of the $\text{CO}_2$ gas sample.
%Explain real experiment and details, equations and difference to real situation
Fig.\ref{fig:spectroscopy-exp-detail} shows the experimental apparatus in detail. In contrast to the idealized scheme in Fig.\ref{fig:spectroscopy-principle}, all rays are guided collinearly through the absorbing medium. The non-linear crystal is pumped with a 532nm laser. It produces quasi-collinear photon pairs perfectly correlated in their degenerate wavelength at $\lambda_s=610$nm and $\lambda_i=4161$nm for signal and idler photon, respectively. After the two non-linear crystals and the medium to be studied, the intensity of the signal photon is given by the following relationship~\cite{belinsky1992interference,klyshko1993ramsey}:

\begin{figure}[htbp]
    \centering
    \includegraphics[width=0.45\textwidth]{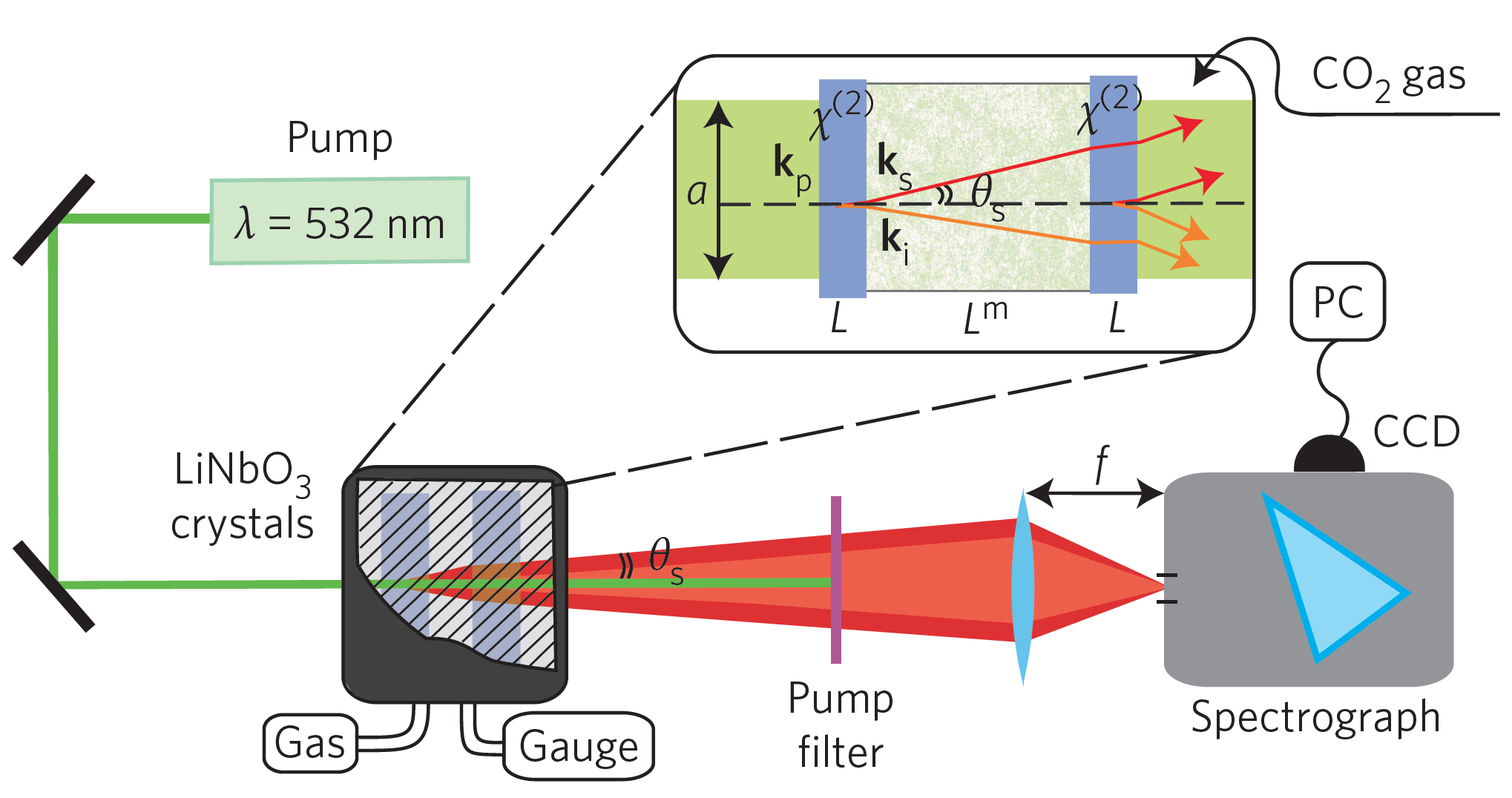}
    \caption{Experimental Details of spectroscopy with undetected photons. Two Lithium Niobate non-linear crystals are employed to create the non-degenerate but perfectly frequency correlated photon pairs. A quasi-collinear emission scheme is used to identify their respective paths and the emission angle $\theta$ of the down-conversion is small relative to the pump beam waist $a$. A vacuum chamber is used to host the two non-linear crystals and the sample of interest, here carbon dioxide $\text{CO}_2$. A lens images the down-converted photons onto a slit to select a specific angular emission spectrum. The following spectrograph enables precise determination of the signal wavelength. Finally, a two-dimensional spectrogram in the angular-wavelength dimensions is recorded by a camera. Figure adapted from Kalashnikov et al.~\cite{kalashnikov2016infrared}}
    \label{fig:spectroscopy-exp-detail}
\end{figure}

\begin{figure*}[htbp]
    \centering
    \includegraphics[width=0.95\textwidth]{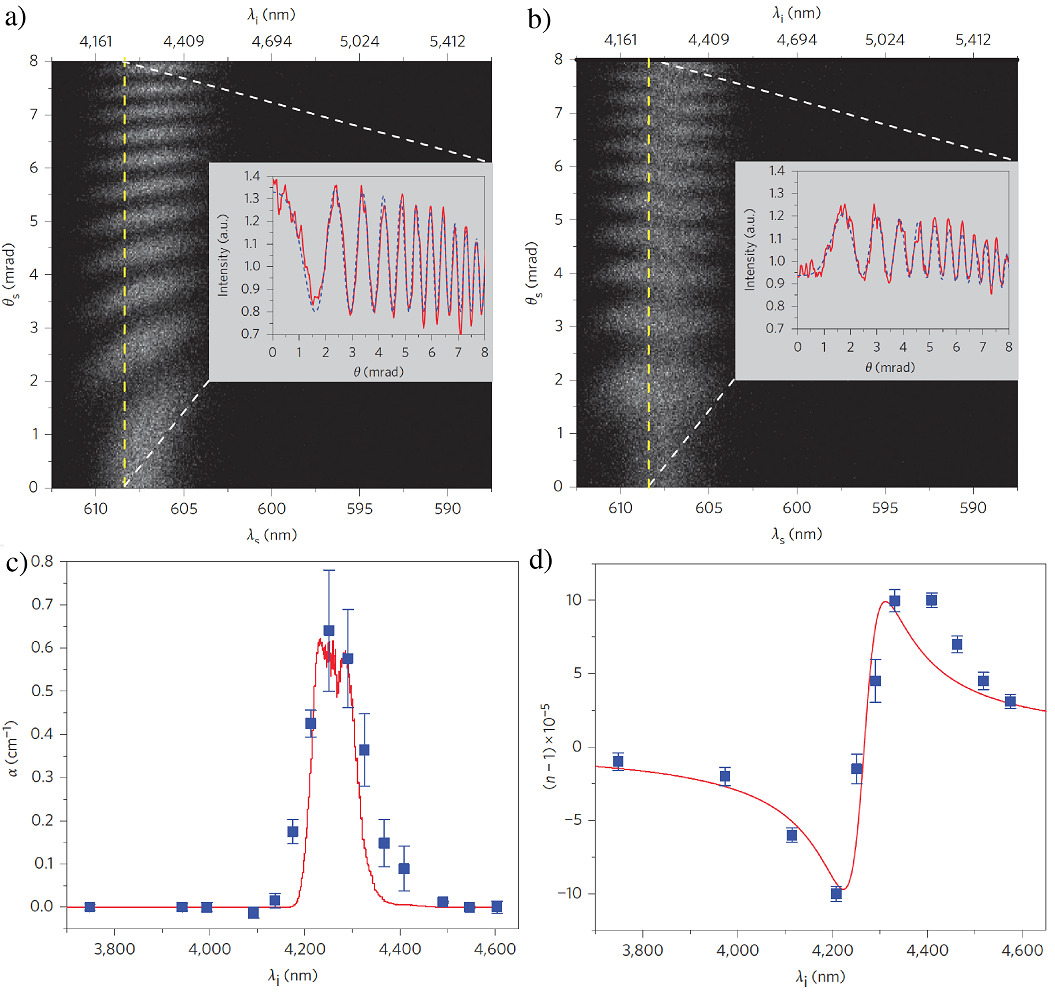}
    \caption{Experimental results. Inset a) shows the calibration measurement in a quasi-vacuum regime. Selecting a specific wavelength cut and fitting eq.\ref{eq:spectroscopy-main} yields the absorption coefficient for vacuum and a phase reference. In b) the spectrograph with the filled carbon-dioxide chamber is displayed. The decrease of visibility and the phase shift is due to the absorption and refractive index change between vacuum in a) and the carbon-dioxide in b). Data displayed in c) and d) shows the wavelength dependence of the absorption and refractive index coefficient at the vicinity of the resonance for carbon-dioxide at a pressure of 10.5 torr. Blue squares represent experimental measurements, and the red curve is a theoretical calculation using HITRAN data (http://hitran.iao.ru) for c) and a Kramer-Kronig relation for d). Data figures are taken from Kalashnikov et al.~\cite{kalashnikov2016infrared}.}
    \label{fig:spectroscopy-results}
\end{figure*}

\begin{align}
\label{eq:spectroscopy-main}
    I(\lambda_s,\theta_s)&\propto \frac{1}{2}\text{sinc}^2\Big(\frac{\delta}{2}\Big)\Big[1+|\tau_{i,m}|\cdot\text{cos}(\delta+\delta_m)\Big]\nonumber\\
    \delta(\lambda_s,\theta_s)&=\frac{L\cdot(k_p-k_i-k_s)}{\text{cos}(\theta_s)}\\\nonumber
        \delta_m(\lambda_s,\theta_s)&=\frac{L_m\cdot(k_p-k_i-k_s)}{\text{cos}(\theta_s)}
\end{align}
where $\theta_s$ describes the emission angle, the phase shifts depending on the wavelengths and emission angles resulting from the non-linear crystals $\delta$ and the medium to be studied $\delta_m$. The wavevectors $k_j$ are related to the wavelength and their respective refractive index $n_j$ via $k_j=2\pi n_j/\lambda_j$. Furthermore, the transmittivity amplitude $\tau_{i,m}$ is connected by $\text{exp}(\text{-}\alpha_m L_m)$ to the amplitude absorption coefficient $\alpha_m$. For vanishing transmittivities $\alpha_m\rightarrow0$ the visibility $V$ also vanishes. Here the visibility is defined as $V=(I_\text{max}-I_\text{min})/(I_\text{max}+I_\text{min})$. Light scattering, especially in bio-tissues can be taken into account, for details see Supplementary information of~\cite{kalashnikov2016infrared}.

Imaging the SPDC radiation onto a slit in front of the spectrograph results in a two-dimensional wavelength-angular spectrogram recorded with an electronically multiplied charge-coupled device (EMCCD) camera. The small angular spread of the SPDC of about 10~mrad introduces the necessary phase shift between signal and idler to measure the visibility. Using a spectrograph allows to precisely control the detected visible spectrum around 610~nm. Energy conservation within the down-conversion process combined with knowledge of the pump and signal wavelengths allows inferring the idler wavelength interacting with the medium to be studied. Fig.\ref{fig:spectroscopy-results} shows a typical experimental measurement of a two-dimensional wavelength-angular spectrogram. First, a calibration measurement under ideal near-vacuum conditions is performed, see Fig.\ref{fig:spectroscopy-results}a). Next, the medium of interest, in this case, $\text{CO}_2$ is placed between the two non-linear crystals. Due to absorption and refractive index differences of $\text{CO}_2$ to vacuum, the observed visibility drops and the relative phase changes respectively, as shown in Fig.\ref{fig:spectroscopy-results}a) and b). Selecting one specific wavelength cut and fitting eq.~\ref{eq:spectroscopy-main} to the experimentally measured data allows to determine the absorption coefficient $\alpha_m$ as well as the refractive index $n_m$, depicted in Fig.\ref{fig:spectroscopy-results}d). 

%Kramer-Kronig relation between absorption and refractive index
The complex refractive index describes both the absorption and the refractive index in a single complex number. The Kramer-Kronig relations connect the real and imaginary part of the complex refractive index. If one of the two parts of the complex number is known, the corresponding other part is uniquely determined. Using these relations, the experimental data can be subjected to a consistency check. Indeed, the experimentally observed absorption spectrum and refractive index changes nicely reproduce the Kramer-Kronig relations, as depicted in Fig.\ref{fig:spectroscopy-results}c) and d).

%Limitations Outlook and potential use (Frequency correlations TeraHertz spectroscopy)
The suggested method of spectroscopy with undetected photons allows the direct measurement of the complete
imaginary refractive index of a medium \cite{paterova2018measurement}. The strength of this technique lies in the possibility of probing the
medium at far-infrared to even terahertz frequencies\cite{tonouchi2007cutting} and detect the signal at visible wavelength. 

A combination of the imaging and spectroscopy techniques allow performing spectroscopy without using a spectrometer by detecting photons only on a camera \cite{paterova2017nonlinear}.

\subsection{Optical Coherence Tomography with undetected Photons}
\begin{figure}
    \centering
    \includegraphics[width=0.8\linewidth]{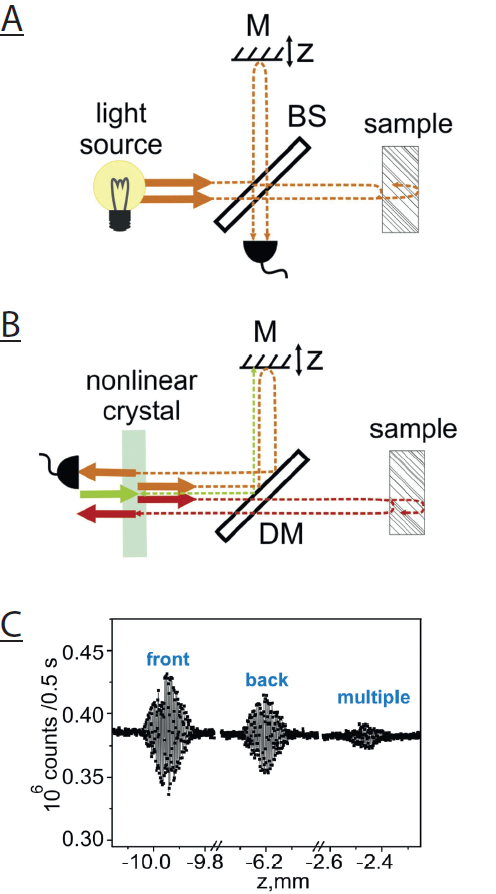}
    \caption{Classical Optical Coherence Tomography and Optical Coherence Tomography with Undetected Photons. Classical OCT (A) is typically implemented with light of a short coherence length in a Michelson interferometer. The reflection from different layers within the object produces an interference pattern if recombined with light that traveled the corresponding distance in the second path. The sample can be laterally resolved by tuning this distance using the mirror M. The mentioned implementation of OCT with undetected photons (B) uses non-degenerate photon pairs produced by a non-linear crystal. The signal (idler) beam emitted towards the right is reflected back through the crystal via the mirror M (via the sample). Consequently, interference in the rate of detected photon pairs can be observed. Only the signal beam is detected towards the left of the source. The observed interference is affected by the properties of the sample if the coherence length requirements are met. This is the case if the idler photon is reflected at a particular depth of the object, that corresponds to the length of the signal path. It is thus possible to probe different depths of the sample. As an example, (C) shows the detected rate of signal photons (at 582 nm) as the signal path length is scanned by translating the mirror M in the z-direction. The sample in the idler beam (at 3011 nm) in this case is a Si-window. Reflections at both the front and the back surfaces, as well as multiple reflections result in interfering signal photon rates at the corresponding mirror positions. Figure adapted from \cite{paterova2018tunable}.}
    \label{fig:oct}
\end{figure}

Optical coherence tomography (OCT) is a method in classical optics to determine optical properties of a sample at specific depths within it \cite{huang1991optical,fercher2003optical}. The technique is typically implemented in a Michelson interferometer (Fig. \ref{fig:oct}A). A thick sample is placed in one path of the interferometer. Light is partially reflected from the sample's front surface and partially penetrates into the sample before it is reflected from various inside layers. The reflected light is then recombined with light from the second path of the interferometer and interference is observed.
By employing light with a short coherence length, the path length requirements to observe interference are met only if the reflection occurs at a particular depth of the sample. Thus, properties determined from the interferogram (e.g. reflectivity) correspond to a particular depth of the sample. An adjustment of the path length difference by moving the mirror in the second path allows to tune at what depth the object is probed. In this way, the method provides a way of imaging different ``depth sections".
Among other fields, the technique is frequently used in life sciences and medicine (e.g. \cite{puliafito1995imaging,fujimoto1995optical,spaide2018optical}).

The technique can be extended using the concept of path identity. In this case, the interference resulting from indistinguishability of the origin of a photon pair is used to probe sample properties in a laterally resolved way. 
Instead of employing a classical interferometer, a Zou-Wang-Mandel \cite{valles2018optical} or a Herzog-Rarity-Weinfurter-Zeilinger \cite{paterova2018tunable} arrangement have been used to probe a sample placed in the undetected beam between the two photon-pair sources. %Recent experiments demonstrated laterally resolved reflectivity measurements.

The schemes based on path identity make use of the coherence length requirements of the non-linear interferometer (see e.g. \cite{jha2008temporal}). In particular, they exploit the fact that interference is only observed if the path length difference between signal and idler photons is tuned in a way that does not allow to recover the source of a photon pair by a difference in the detection times of the two photons.

This allows to perform the following procedure. If the object is thick and partially transparent, multiple possibilities exist for the idler beam to be reflected between the two SPDC processes. Reflections at different depths of the sample correspond to different path lengths of the idler beam between the two possible SPDC processes. Typically, the idler beam can be reflected by layers at different depths of the sample that are separated by a lateral distance that exceeds the coherence length. In this case, it is possible to tune the path length of the signal beam to observe interference corresponding to a particular path length of the idler beam (Fig. \ref{fig:oct}B). Thus, the transmission and phase shift at different depth sections of the sample (for which the path length requirements are met) affect the observed interference pattern in the signal beam.

As an example, we show a result of the experiment \cite{paterova2018tunable}. There, the method was used to determine the positions of reflective layers in different samples, including the reflections off the front and back surfaces of a Si-window (Fig. \ref{fig:oct}B,C).
Notably, the object is probed using IR light, while the wavelength of the detected light lies in the visible range.

This allows probing different depth sections of the object that reflects only undetected light while the detection and the scanning of the depth is performed in another light beam, which typically is at a different wavelength.

Recently a further adaptation of this scheme has been demonstrated that featured the detection in the Fourier domain, i.e. replacing the detector by a spectrometer \cite{vanselow2019mid}.
This modification allows to reconstruct the properties of the sample at different depths via Fourier transform (see \cite{fercher2003optical} for the classical technique) of the spectroscopic data. It thus eliminates the requirement to physically move a mirror in order to scan the path length.

\subsection{Dual-Wavelength properties}\label{sec:two-wvlength}
\par
The imaging, spectroscopic, and tomographic schemes discussed above have one important common feature: the wavelength of the detected light can be different from the wavelength of the light that probes the sample. This fact opens up a new era in these fields because now one can probe samples at a frequency (wavelength) for which efficient and cheap detectors are not available. 
\par
\begin{figure}[htbp]
    \centering
    \includegraphics[width=9cm]{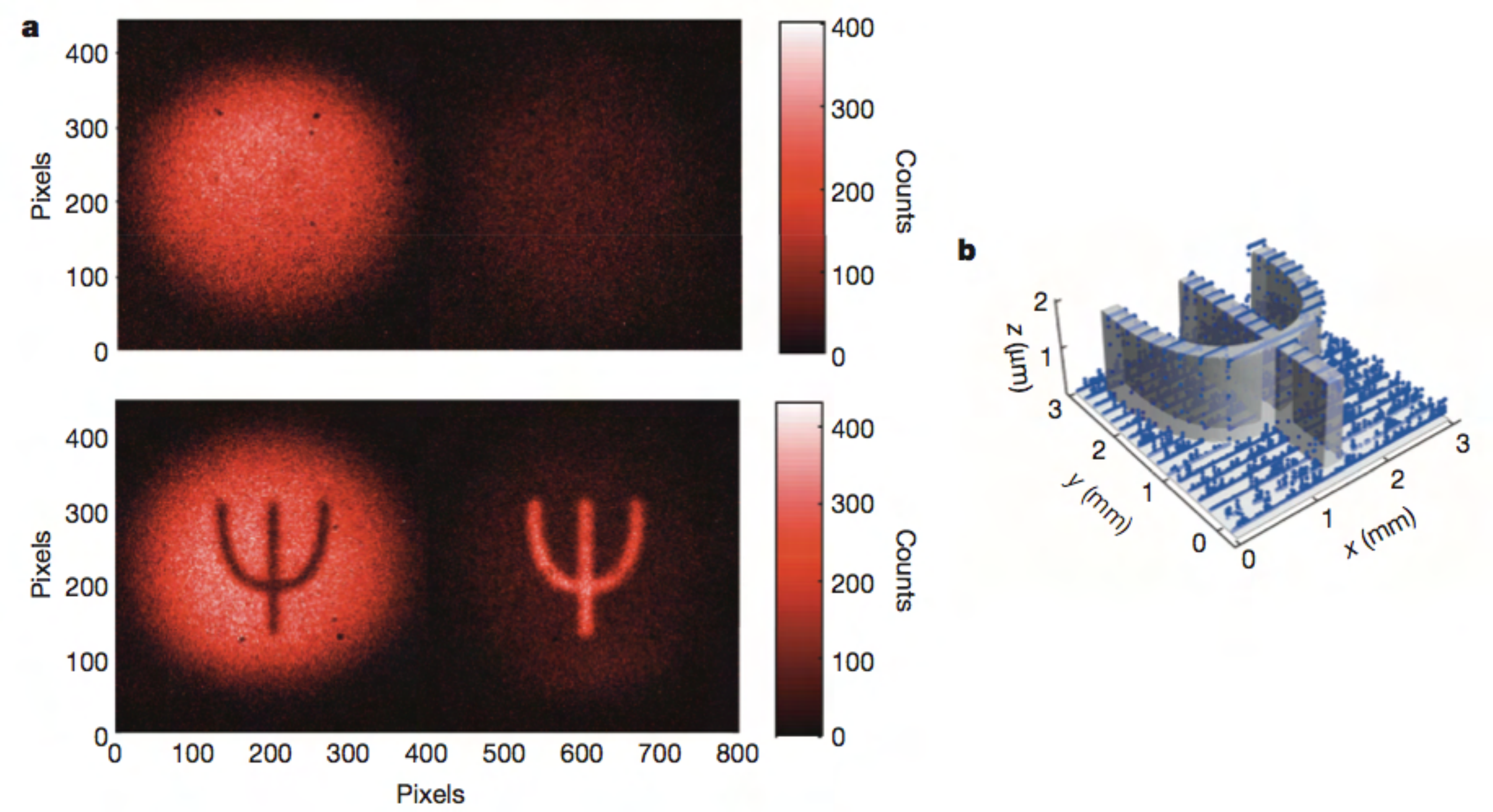}
    \caption{(adapted from \cite{lemos2014quantum})
        a) Top: no image is obtained when the object is imaged by 810 nm light. Bottom:
        Both outputs of the beam splitter contain the image when the object is placed in the idler beam (1550 nm). b) The phase
        object that is invisible to 810 nm light.}\label{fig:Undetected3}
\end{figure}

As an example, we consider the imaging experiment. Here, the wavelength of the detected light is 810 nm, and the wavelength of the light probing the object is 1550 nm. Figure \ref{fig:Undetected3} shows images of a phase object that is invisible to the detected light. This is because of the chosen object [Fig. \ref{fig:Undetected3}(b)] introduces a relative phase shift of approximately $2\pi$ for the detected light. However, the same object produces approximately $\pi$ phase shift for 1550 nm light and is therefore clearly imaged in this scheme [Fig. \ref{fig:Undetected3}(a)].
It is, therefore, impossible to realize transmission imaging by illuminating it with the detected light. However, one can obtain the image of this object in the imaging scheme with undetected photons. These two-wavelength effects not only have potential for commercial applications in THz or deep-UV spectroscopy but also lead to interesting questions on whether interference properties scale with $\lambda_S$ or $\lambda_I$. The surprising answer is, with neither, as we show now.

\subsection{Wavelength dependence of interference fringes with undetected photons}
\label{eq-wl}
It has been shown that the setup for quantum imaging with undetected photons (Fig. \ref{fig:Undetected2}) can be used for interferometry with undetected photons \cite{hochrainer2017interference}. The associated phenomena exhibit several interesting features compared to standard classical interferometry.

In a traditional two-path interferometer, spatial fringes appear if the interfering beams are misaligned relative to one another. A tilt of one of the beams results in a striped interference pattern, whereas a further propagation or defocused lens system causes circular fringes to appear.

\begin{figure}
    \centering
    \includegraphics[width=\linewidth]{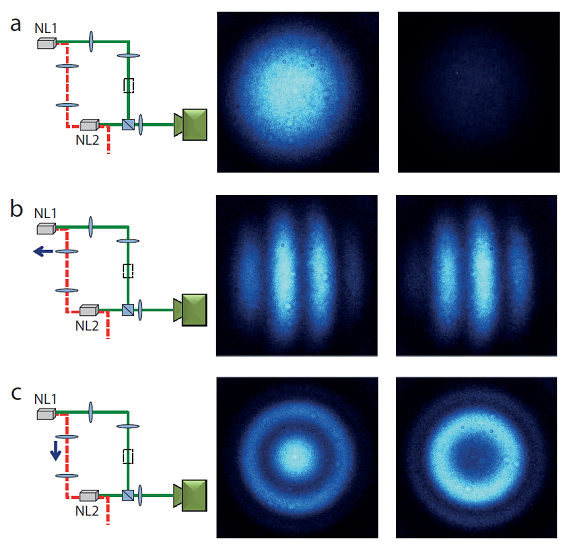}
    \caption{Interference fringes produced with undetected photons. The spatial interference pattern observed in the superposed beam can be controlled by manipulating a third beam, which is not detected. The appearance of the pattern is analogous to that obtained if the same manipulations were performed on one of the interfering beams in a traditional two-path interferometer. Figure from \cite{hochrainer2017interference}.}
    \label{fig:fringeformation}
\end{figure}

In a Zou-Wang-Mandel experiment, it is possible to observe analogous interference fringes if the undetected idler beam from one source is slightly misaligned with respect to the other \cite{hochrainer2017interference}. Without directly interacting with the interfering signal beams, spatial interference patterns are formed as depicted in Fig. \ref{fig:fringeformation}. 
{}

While such patterns in classical interferometry are often explained by the wavefronts of the two interfering beams being tilted or curved with respect to one another, it is not possible to attribute a specific phase structure to one photon beam of a down-converted pair. Thus, a classical explanation clearly fails to explain the observed interference phenomenon, which can be understood in a similar way as the phase imaging with undetected photons (see Sec. \ref{subsec:q-img}). It relies on the fact that a single phase factor is attributed to the photon pair as a whole and not to the individual photons of a pair.

The observation of a spatial structure in ZWM interference can explain earlier results obtained in \cite{grayson1994spatial}, in which a reduced visibility was observed due to a slight misalignment of the undetected beams. If the total intensity over the entire spatially structured interference pattern is determined using a bucket detector, the fringes amount for a lower measured visibility. % is reduced upon slightly misaligning the idler beam, as it integrates over the fringe pattern.

In classical interferometry, the scaling of the spatial structure of the interference patterns depends on the wavelength of the interfering light.
Due to the different wavelengths of signal and idler beams, the question arises, which wavelength characterizes the pattern of interference fringes that are controlled with undetected photons. As the fringes are manipulated in the idler beam and observed in the signal beam, two wavelengths are involved in their formation. 

The above question was analyzed by comparing circular fringes obtained after defocusing the lens system in the idler beam to circular fringes obtained if the same lens manipulation were performed on one of the interfering beams in an analogous classical interferometer  \cite{hochrainer2017interference}.
In the classical case, the radius $\rho_n$ of the $n$th minimum/maximum obeys the relation\footnote{integer $n=0,1,2,..$ correspond to maxima, and half-integer $n=\frac{1}{2},\frac{3}{2},\frac{5}{2},....$} (cf. \cite{born_wolf})

\begin{equation}
\frac{d}{2f_C^2}\rho_n^2+\varphi_0=n\lambda,
\label{classminmaxcondition}
\end{equation}

where $d$ denotes the effective additional propagation distance that is caused by the lens shift, $f_C$ represents the focal length of the lens in front of the camera (see Fig. \ref{fig:fringeformation}C), $\varphi_0$ subsumes all spatially independent phase factors, and $\lambda$ is the wavelength of the interfering light.

\begin{figure}
    \centering
    \includegraphics[width=\linewidth]{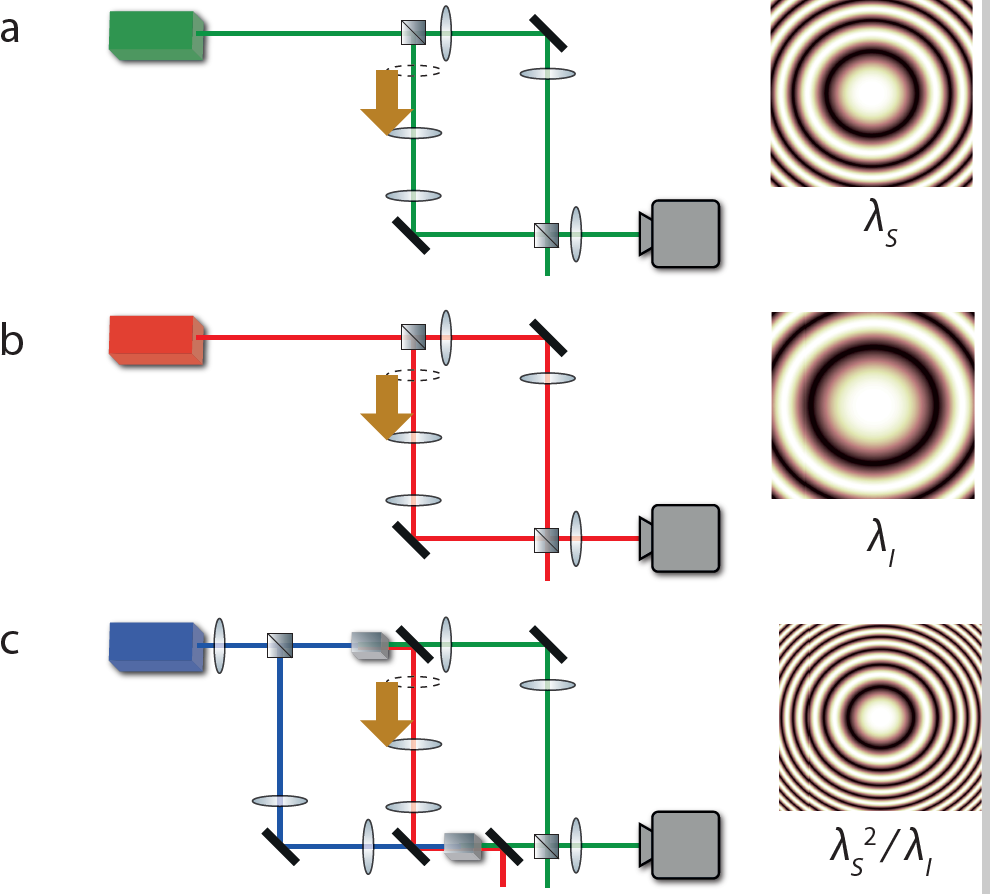}
    \caption{Wavelength dependence of circular interference fringes. The first two lines depict classical interferometers and circular fringes obtained by shifting the lens by a fixed distance. In (a) the light is at the signal wavelength (810 nm), whereas in (b), it is at the idler wavelength (1550 nm). (c) shows the interference fringes obtained in the ZWM setup, where the spacing of the fringes corresponds to the equivalent wavelength $\lambda_{eq}=\lambda_S^2/\lambda_I$ Picture taken from \cite{meinediss}.}
    \label{fig:wlcompare}
\end{figure}

If Eq. \ref{classminmaxcondition} is applied to circular fringes produced in the ZWM interferometer, the wavelength characterizing the pattern was determined to (see Fig. \ref{fig:wlcompare})
\begin{equation}
\lambda=\frac{\lambda_S^2}{\lambda_I}.
\label{eqWL-result-theo}
\end{equation}
%with an ``equivalent wavelength" $\lambda_{eq}$.

{}

Note that this observed "equivalent wavelength" ($420 \pm 7 \text{ nm}$) is significantly smaller than any of the involved wavelengths (signal 810 nm, idler 1550 nm), even the pump (532 nm).

This peculiar feature is understood as a combined effect of the phase shifts being introduced at the wavelength of the undetected idler photons and the detected photons at the signal wavelength \cite{hochrainer2017interference}. Due to momentum conservation in non-degenerate SPDC, the photon at the longer wavelength is emitted at a larger angle with respect to the optical axis than its partner photon at the shorter wavelength. The difference in emission angles is determined by the wavelength ratio. This results in a wavelength-dependent scaling of the interference pattern, which, in our case, leads to fringes with smaller spacing than expected considering only the idler wavelength. The same effect gives rise to the wavelength-dependent magnification (or demagnification in this case) in quantum imaging with undetected photons, see Sec. \ref{subsec:q-img}.

\section{Reconstruction of the Quantum State of Undetected Photons}\label{sec:two-ph-corr}
%\subsection{General Merging Paragraph}
The experiments described in Sec. \ref{section:objectrecon} make use of induced coherence without induced emission in a multi-mode setting. The results show that it is possible to determine properties of a sample placed in one of the beams by performing measurements only on the other beam. Instead of employing photon pairs in a known quantum state in order to determine unknown properties of an object, it is possible to use a known object and determine properties of the employed photon pairs, or to obtain information about one photon by detecting the other. 

A second branch of applications of the concept of path identity exploits the possibility of measuring correlations between two photons by detecting only one of them.
Correlation measurements of photons are used ubiquitously in today's quantum optics. Traditionally, such measurements are performed using the method of coincidence detection, which is a powerful experimental tool in both fundamental science and technology \cite{burnham1970observation,pan2012multiphoton}.

%As demonstrated in the experiments described below, the application of path identity allows correlation measurements to be performed without the use of coincidence detection. 
These adaptation of these techniques to interference by path identity extend the reach of optical experiments to regimes in which only one of the photons can actually be detected (e.g. if one of two correlated photons is at a wavelength for which no suitable efficient detector exists currently).

From a fundamental perspective, the experiments show that information stored in the correlation of a photon pair can be accessed by measurements on only one of its constituents. In other words, a second order property of light can be transferred to a first order property and can be determined using a single detector.

\subsection{Quantifying the momentum correlation between two photons by detecting one}

%The analysis of a single photon interference pattern in a multimode ZWM interferometer was used to determine the strength of the transverse momentum correlation between two photons \cite{hochrainer2017quantifying}.

\begin{figure*}
	\centering
	\includegraphics[width=1\linewidth]{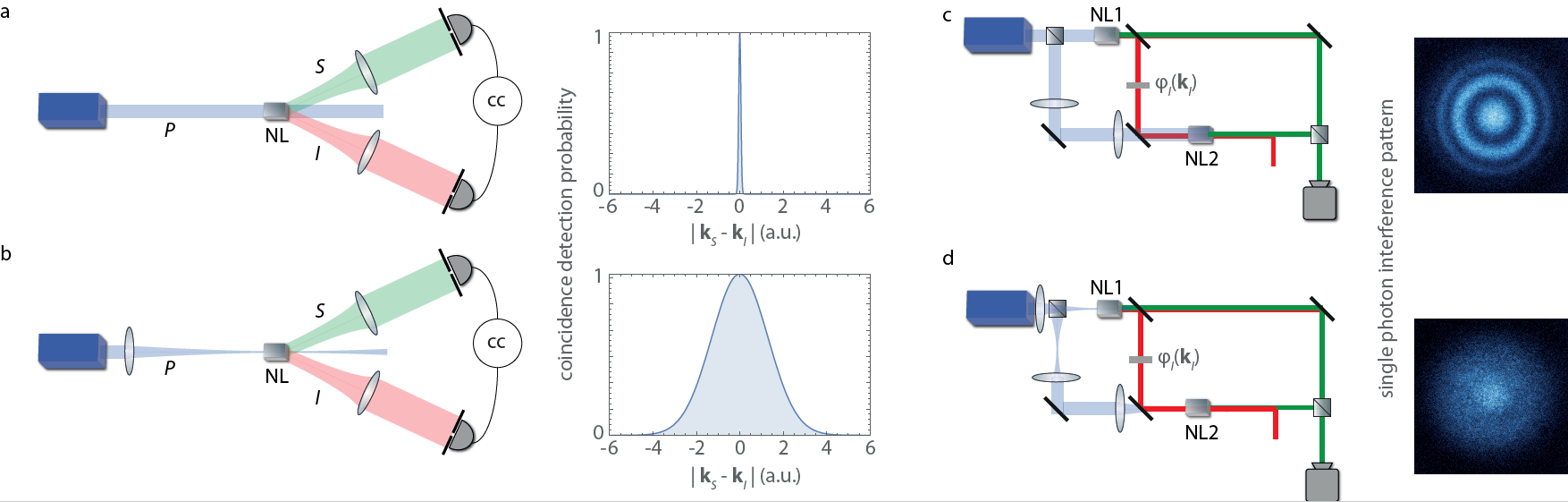}
	\caption{Measuring the transverse momentum correlation of a photon pair with and without coincidence detection. In the traditional method (a,b), each of the two photons is registered by a detector in a way that its individual momentum can be inferred. By comparing the coincidence detection events at different pairs of momenta, the correlation is quantified. A sharp correlation (implemented by a large pump focus) leads to a narrow coincidence peak at a particular relative momentum (a), whereas weakly correlated photons (small pump waist) lead to a similar amount of coincidence counts in a wider range of momenta (b) \cite{monken1998transfer}. Using the concept of path identity, two copies of the photon pair source are arranged in a ZWM configuration. By introducing a spatially dependent phase shift on the undetected idler beam between the two sources, a single-photon interference pattern is observed in the superposed signal beam. The visibility of the pattern depends on how strongly signal and idler photons are correlated. In the case of a loosely focused pump beam, the photons are highly correlated in momentum, and spatial features of an interference pattern are clearly visible (c). The opposite is true for weakly correlated photons (d). The visibility is determined from measurements on only on one of the photons and can be used to quantitatively reconstruct the correlation between the two photons. Figure adapted from \cite{meinediss}.}
	\label{fig:standard-mom-corr-measurement}
\end{figure*}

Traditionally, the transverse momentum correlation between two photons is measured by the coincident detection of both photons. One way to implement such a measurement is depicted in Fig. \ref{fig:standard-mom-corr-measurement}a,b. The two photons are detected with individual detectors that resolve the transverse momentum of the respective photon.
The transverse momenta are proportional to the transverse component of a wave vector of signal and idler photon (denoted by $\qq_S$ and $\qq_I$).
By scanning the relative positions of the detectors, it is possible to determine the joint probability distribution, e.g. $P(\qq_I|\qq_S)$ of detecting one photon with a particular momentum, given the momentum of the other photon is known. % the idler photon at a particular momentum $\qq_I$, given the momentum of the other (signal) photon has been determined to be $\qq_S$.  %the probability of detecting a photon pair with any pair of momenta is determined. %The coincident detection of the two photons at selected pairs of momenta allows to determine the momentum correlation between them.
The variance $\sigma^2(\qq_I|\qq_S)$ of this conditional probability distribution serves as a quantitative measure for the strength of the momentum correlation. Such measurements have been implemented in a variety of different ways, e.g.  \cite{howell2004realization,edgar2012imaging}. However, all of these traditional methods require the joint detection of both correlated photons. %, the variance of this conditional probabiltiy 
Using the concept of path identity, it is possible to determine the correlation strength without this requirement. 

A quantitative measurement of the momentum correlation between two photons by detecting only one of them has been demonstrated experimentally \cite{hochrainer2017quantifying} (see \cite{lahiri2017twin} for a theoretical description).
In this experiment, a correlated photon pair is produced in a superposition of two sources NL1 and NL2. One of the photon beams from each source is aligned to be indistinguishable (Fig. \ref{fig:standard-mom-corr-measurement}c,d).
%\paragraph{Principle of the Experiment}
%A recent experiment \cite{hochrainer2017quantifying} exploited the features of the concept of path identity in order to determine the quantity $\sigma^2(\qq_I|\qq_S)$. In contrast to the traditional method, this measurement did not use coincident detection. 
%This measurement can be performed without coincidence detection, based the following principle. 
% of using one source and two detectors, the experiment used two identical copies (NL1 and NL2) of the source for correlated photon pairs.  to be analyzed are placed in a ZWM interferometer (Fig. \ref{fig:standard-mom-corr-measurement}c,d). 
The other beams from each source are superposed on a beam splitter and subsequently detected on a camera. The detection is performed in a way that the momentum of a detected photon can be inferred, although it is unknown, which source it was initially emitted from.  %The second beam from NL1 is overlapped and aligned to be identical to the second beam emerging from NL2. It remains undetected.
A momentum dependent phase shift $\varphi_I(\qq_I)$ is introduced on the undetected (idler) beam between the sources. The phase shift was experimentally implemented by defocusing the lens system between the two sources. As in Sec. \ref{eq-wl}, it results in the formation of a circular interference pattern on the camera.

\begin{figure}
	\centering
	\includegraphics[width=1\linewidth]{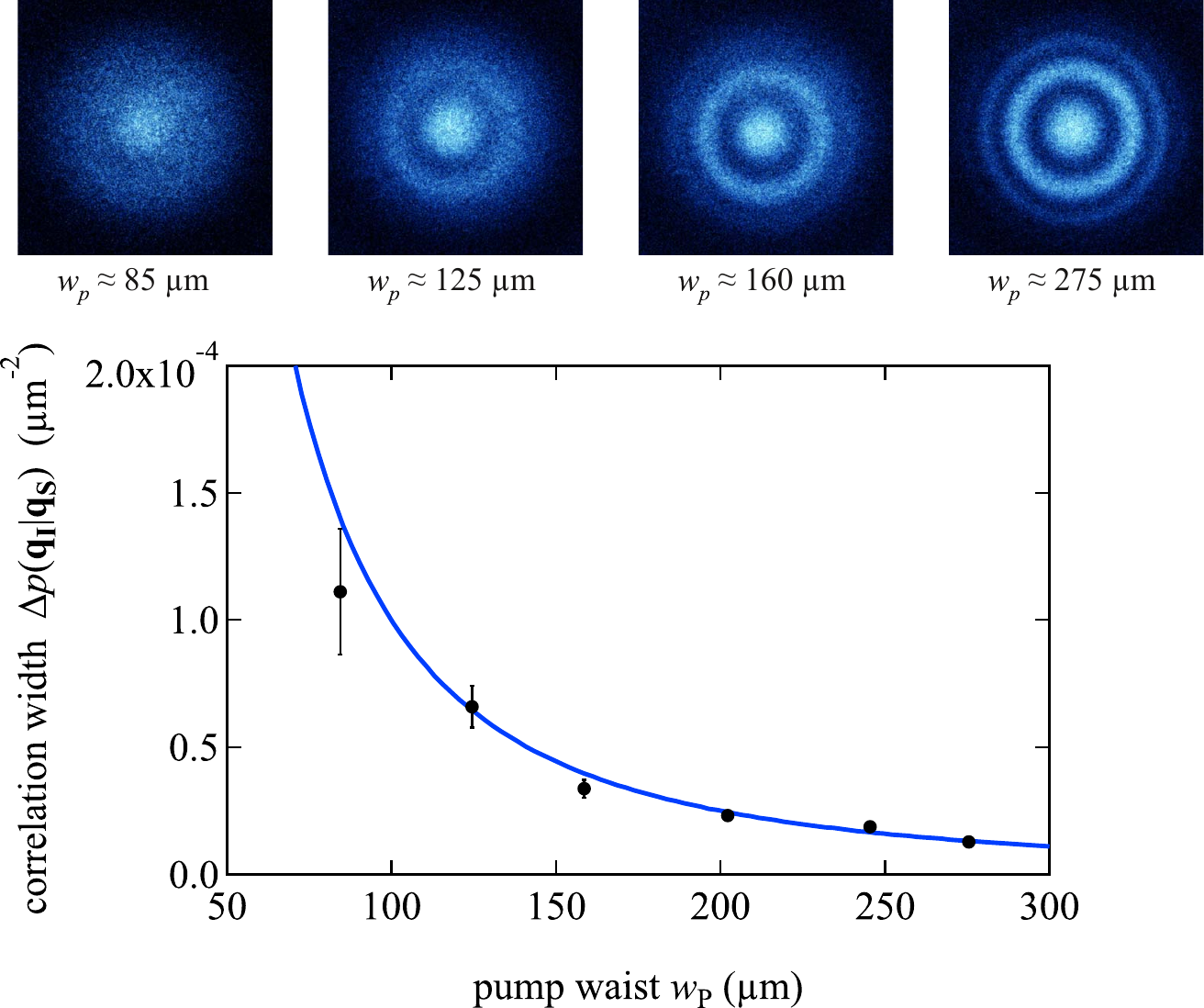}
	\caption{Experimental results of the correlation measurement between two photons by detecting only one of them. The correlation between signal and idler photons was tuned by changing the waist ($w_P$) of the pump beam in both crystals simultaneously. The fringe pattern observed in the interference of the two signal beams gradually exhibits a higher visibility, as the correlation between signal and idler photons is increased. This allows to numerically reconstruct the variance of the conditional probability distribution of the momentum of an idler photon, given the momentum of a signal photon, without relying on coincidence detection. The results are shown compared to the theoretical prediction. Figure adapted from \cite{hochrainer2017quantifying}.
		%The accuracy by which a phase shift in the idler beam is resolved in the pattern, allows to determine the correlation between signal and idler photons. a,b,c...*replace w fringe pics and correlation graph
	}
	\label{fig:fig-quantifying-mom-corr}
\end{figure}

By varying the momentum correlation between signal and idler photons, the appearance of the pattern changes (Fig. \ref{fig:standard-mom-corr-measurement}c,d).
This fact allows to reconstruct, how strongly the momenta of two photons of a pair are correlated, although the measurement is performed by observing only one of them.
%In the experiment, the momentum correlation between signal and idler photons was tuned by controlling the pump waists at the two crystals simultaneously. A small pump waist leads to a weak correlation between signal and idler photons, whereas a large pump spot causes the emitted photon pairs to be strongly momentum correlated \cite{monken1998transfer,hochrainer2017quantifying}. 
In the experiment, the spatially dependent visibility of the resulting interference patterns was used to numerically evaluate the strength of the momentum correlation $\sigma^2(\qq_I|\qq_S)$ \cite{hochrainer2017quantifying}, see Fig. \ref{fig:fig-quantifying-mom-corr}.

%In the experiment, the phase shift $\varphi_I(\qq_I)$ was implemented by defocusing the lens system in the idler beam between the two sources.
The reason for the dependence of a single photon interference pattern on the correlation between two photons becomes apparent considering how the pattern is formed (compare Sec. \ref{subsec:q-img}).
The camera is located behind a lens system, which maps one transverse momentum of the superposed signal beam $\qq_S$ to one point on the camera.
The interference fringe at a selected point on the camera is observed by detecting signal photons and is influenced by the phase shift introduced on the corresponding partner idler photons. % of a signal photon. % (One point on the camera corresponds to one $\qq_S$).
%One point on the camera corresponds to one $\qq_S$. 
In the case of perfect momentum correlation between the detected signal photon and the undetected idler photon, this phase is controlled by phase shifts of a particular momentum of the undetected beam. %one $\qq_S$ corresponds to exactly one $\qq_I$(***explain explicitly?). The phase shift $\varphi_I(\qq_I)$ imparted on the corresponding idler mode controls the interferometric phase detected in the signal interference at this point on the camera.
On the other hand, if the momentum correlation between the two photons is imperfect, the detection of a signal photon at a particular point on the camera does not allow to precisely infer the momentum of its partner idler photon. %Instead, it can have a range of different momenta. %be found in a range of $\qq_I$ modes. 
Therefore, the phase shift $\varphi_I(\qq_I)$ that determines the interference at that point on the camera can vary within a range that is determined by the momentum correlation. The observed phase in this case is not uniquely determined, but is given by a weighted average over the possible phase shifts corresponding to the possible $\qq_I$. %The idler photons corresponding to the same momentum of the signal photon (the same point on the camera) have different momenta $\qq_I$. Thus, 
%The interference observed at that point on the camera is determined by a weighted average over the possible phase shifts corresponding to the possible $\qq_I$. 
As a result, the visibility of the observed interference pattern is reduced for a weak momentum correlation.% with a higher variance of $P(\qq_I|\qq_S)$.%the interference pattern ``blurs out"

%The experiment used the knowledge about $\varphi_I(\qq_I)$ to relate the observed spatially dependent interferometric visibility with the strength of the momentum correlation. This allows to reconstruct the correlation strength between signal and idler photons form the observed interference pattern in the signal beam only.

%\paragraph{What we can learn from it}
The experiment shows that it is possible to access properties of a photon pair by measuring only a part of it. A measurement that traditionally requires coincidence detection of both photons can be performed with only one detector. In that sense, the ``path identification" of the undetected beams plays the role of coincidence detection and can be used to access higher order properties of the photon pair with a single detector.%using the detection of a single photon interference pattern. 
%This shows that using the idea of path identity, the information stored in a pair of photons can be transferred to the interference/coherence properties of one of its constituents.

The question arises, to what extent the presented technique can be generalized in order to perform measurements of other properties that usually require coincidence detection.
It would be interesting to adapt the method to determine not only the momentum correlation, but also the position correlation of a photon pair -- or, in general, correlations in any conjugate or mutually unbiased measurement bases. This would allow an experimental test of quantum entanglement that relies solely on measurements of one photon.
%\clearpage

%\subsection{Conclusions/Outlook}
%From a fundamental perspective, the presented experiments show that it is possible to access information stored in the correlation of a photon pair on one of the photons alone. This stimulates the investigation of this effect from a quantum information point of view. 
%The results further highlight the potential of the concept of path identity as a tool in quantum science, as the idea can in principle be generalized to perform other measurements with single photon detection, which to date require coincidence detection and which could in the future be performed without it. %**DIscussion of what else one could do (tomography, USD, HOM?, HBT?....).

\section{Single and Entangled Photon Sources using Path Identity}\label{section:EbPI}
In this chapter, we will explore the connection between the path identity principle and photonic quantum information experiments. Photonic experiments which exploit quantum mechanical effects, for example, Bell violations or BosonSampling, require either single-photon sources or entangled photon-pair sources.
We start by explaining how to create single-photon sources using path identity. Furthermore, we describe how to generate entangled photons in higher-dimensions using path identity. Finally, we present general schemes based on path identity to create genuine multi-photon and high-dimensionally entangled quantum states.

Since introducing entanglement in detail is beyond the scope of this review, we refer the interested reader to the following references\cite{bruss2002characterizing,horodecki2009quantum,plenio2014introduction}. A thorough review of how to experimentally create and detect genuine two- and multi-photon entanglement in two- and higher-dimensions can be found here\cite{pan2012multiphoton,friis2019entanglement,erhard2019advances,slussarenko2019photonic}. 

\subsection{Single Photons by Path Identity}\label{section:SPbPI}
\begin{figure}[ht]
  \includegraphics[width=.5\textwidth]{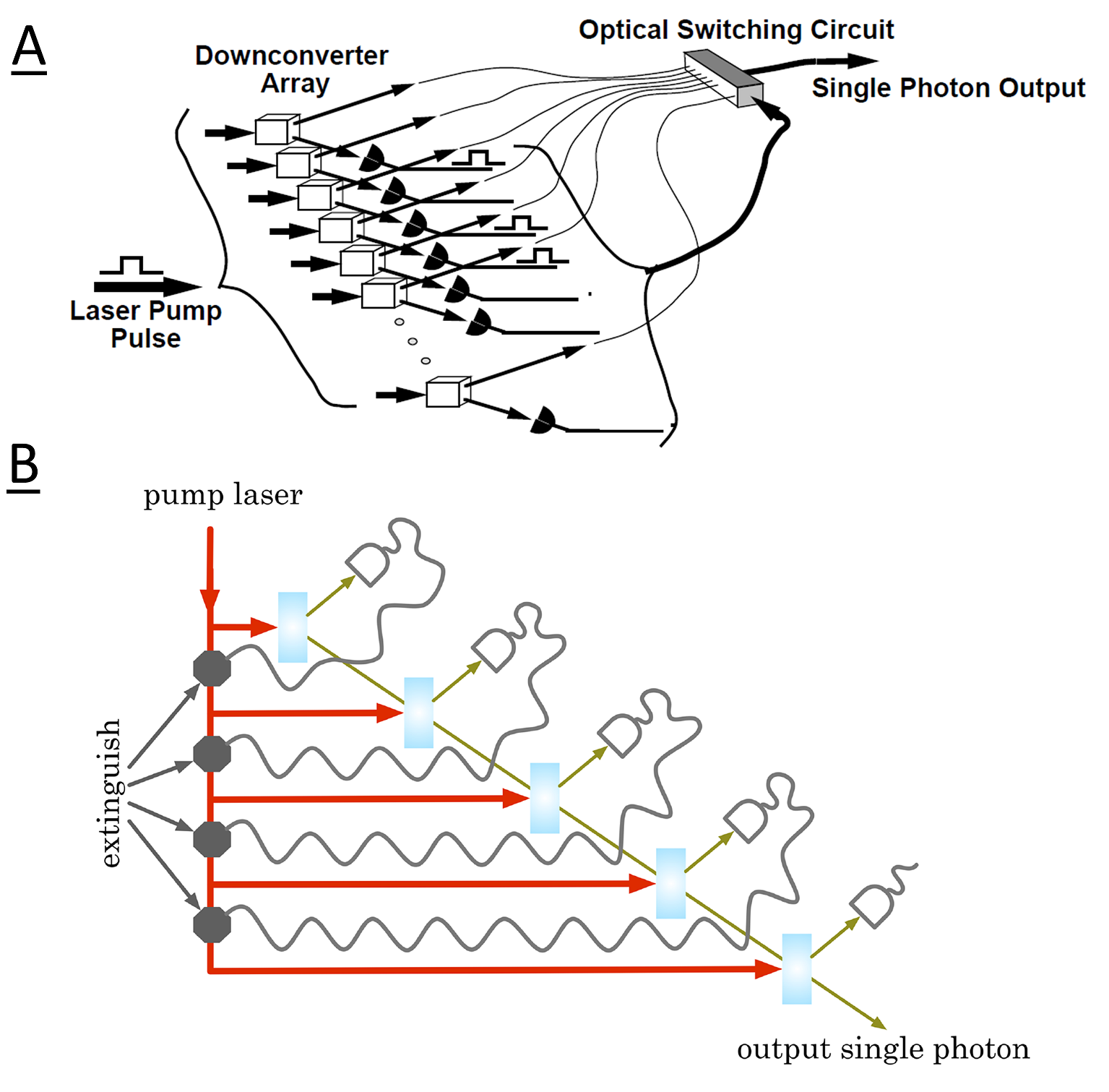}
    \caption{Two different ways to generate high-quality single-photon sources from SPDC crystals. \underline{A:} An array of SPDC crystals are pumped simultaneously, each idler photon is detected, while the signal photon goes to an optical switch. The optical switch navigates the signal photon into the one output path (Image from \cite{migdall2002tailoring}). \underline{B:} An alternative approach, presented by Rudolph \cite{rudolph2017optimistic} (Image from there), avoids the active influence of the single photons. Instead, the scheme denoted as \textit{Dump-the-pump} identifies the paths of the idler photon and measures the signal photon. If one of the detector fires, the pump is actively blocked. }
    \label{fig:rudolfSinglePhotonSource}
\end{figure}

For scalable photonic quantum technology, sources for pure, indistinguishable photons are of utmost importance. While SPDC has been used for nearly 20 years as a workhorse in quantum optics, it is intrinsically probabilistic. Due to the probabilistic nature, it cannot be guaranteed that only one pair of photons is generated at a time. In particular, if the pump power is increased to improve the photon pair rate, also accidental multi-pair event increase. As a consequence, increasing the photon pair rate reduces the quality of the fidelity of the produces multi-photon state \footnote{Photon-Number sensitive detectors could avoid this problem, but they are not standard equipment of quantum optics laboratories yet. Furthermore, increasing the timing resolution of the detection could further improve the quality, but due to physical limits, the timing jitters cannot be reduced arbitrarily.}. One possible route towards true single-photon emitters, which has seen impressive advances in recent years are quantum dots (see, for instance, \cite{wang2017high, wang2019demand}). 

A different method is denoted as \textit{active multiplexing}, see Figure \ref{fig:rudolfSinglePhotonSource}A. The main idea is to use $N$ crystals instead of one, and pump each of them with small pump power, such that the probability of producing a multi-pair emission is neglectable. One of the photons of the pair is detected with one of the $N$ detectors. The information which crystal produced a photon pair is transmitted to an optical switch, which routes the partner photon in one single output path. This scheme has been explained in \cite{migdall2002tailoring} and demonstrated in \cite{ma2011experimental}. One potential challenge is the active interaction with the path of the single-photon, which is itself brittle. That interaction might introduce additional loss, noise and distinguishability.

As an alternative approach, appropriately called \textit{dump-the-pump}, Rudolph has proposed a method which exploits path identification \cite{rudolph2017optimistic}. The clever idea is to use an array of SPDC crystals, where each of them is pump weakly such that multi-pair emissions in a single crystal are unlikely. Then the signal photon of each crystal is detected, while the idler photon paths are identified. The detection of a signal photon then triggers the blocking of the pump laser, such that no further crystal can emit a photon, thus only one photon is emitted into the output mode. Rudolf estimates that the maximal emission rate of 1/5 for a single crystal could be boosted up to 2/3 or even 3/4 using six cascaded SPDC crystals. So far, no experimental demonstration of this potential, influential method has been shown.

\subsection{High-Dimensional Entanglement by Path Identity}\label{sec:two-photon-high-dim-entanglement}
\begin{figure}[ht]
  \includegraphics[width=.45\textwidth]{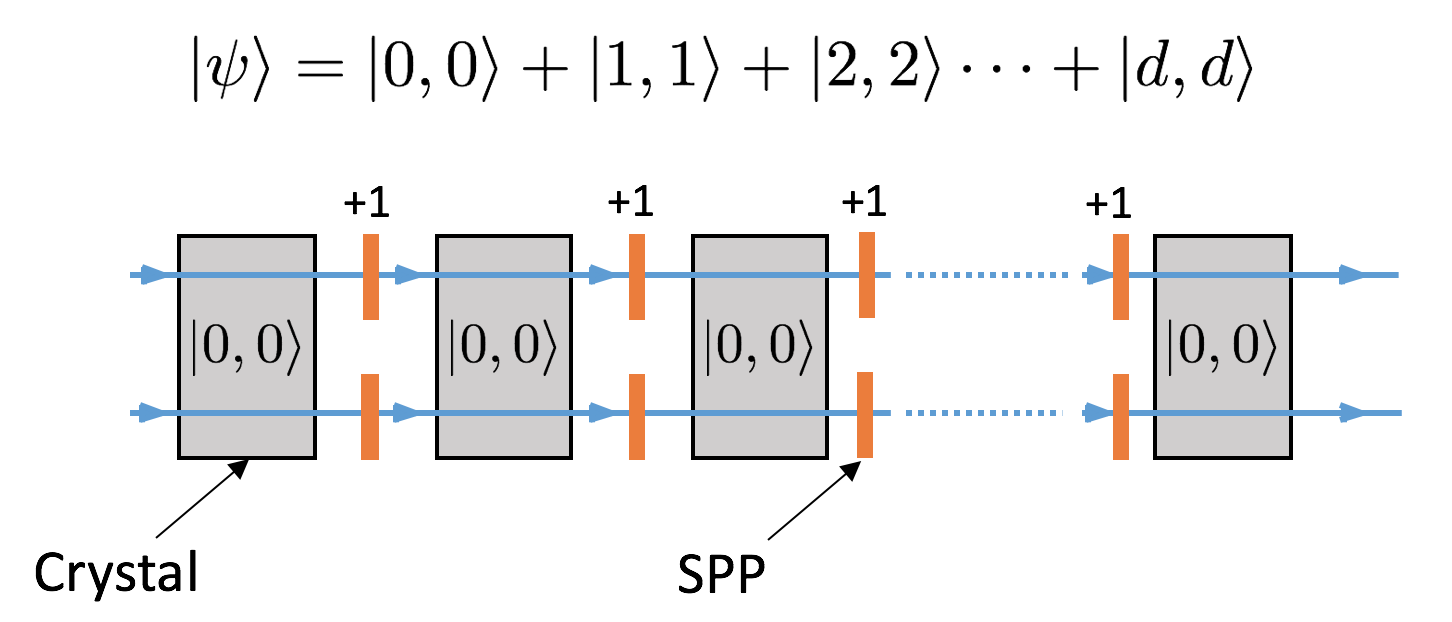}
    \caption{\label{fig:entanglement-sources-1}A new scheme to create two-photon high-dimensionally entangled photon pairs using path identity -- denoted \textit{Entanglement by Path Identity}\cite{krenn2017entanglement}. Consecutively stacked indistinguishable photon-pair sources (non-linear-crystals -- NLC) coherently emit photon pairs. Identifying their paths leads to a coherent superposition of possible origins of a photon pair. Placing spiral-phase-plates in between two NLCs that add one quantum of orbital-angular-momentum to the incoming photons results in a $d$-dimensionally entangled quantum state $|\psi\rangle$.}
\end{figure}

We have seen in Section \ref{Sec:EbPIIntro}, that Hardy proposed a source of polarization-entangled photons by overlapping the output of two crystals, and modifying the polarisation between the crystals. While this two-photon polarization-entangled source is a standard workhorse in quantum optics, it took 25 years until it was understood that the concept is much more generally applicable \cite{krenn2017entanglement}. Interestingly, the generalization was only found in the solutions of a computer program for designing quantum experiments \cite{krenn2016automated}.

To encode high-dimensional quantum information, we employ a multi-level physical degree of freedom. For photons there exist several degrees of freedom capable of encoding quantum information beyond qubits. For example the frequency-bin\cite{olislager2010frequency,bernhard2013shaping,reimer2016generation}, time-bin~\cite{franson1989bell,tittel1998violation}, path~\cite{reck1994experimental,schaeff2012scalable,wang2018multidimensional} or the spatial modes~\cite{allen1992orbital,mair2001entanglement} form such multi-level encoding degree of freedom. In the following, we briefly introduce the spatial degree of freedom, in particular, the orbital-angular-momentum (OAM) of photons\cite{yao2011orbital, erhard2018twisted}.

The OAM of photons spans an in principle unbounded state space and is thus ideally suited to encode high-dimensional quantum states. Physically, the OAM essentially stems from a spatially varying phase distribution which helically wraps around the axis of propagation according to $exp(i\ell\phi)$, with $\phi$ describing the azimuthal angle and the integer $\ell$ defining the amount of OAM in units of $\hbar$. A photon with non-zero OAM exploits one or more phase-singularities\footnote{At phase-singularities, the phase is undefined.} where the amplitude is zero. These phase-singularities lead to the typical doughnut-shaped intensity distributions for light beams carrying OAM. The OAM forms an ideal testbed for proof-of-principle experiments quantum experiments since many techniques to create, manipulate and measure OAM on a single photon level exist in the laboratory\cite{heckenberg1992generation, marrucci2006optical, leach2002measuring, berkhout2010efficient, babazadeh2017high, morizur2010programmable, fontaine2019laguerre, brandt2019high}.

Historically, creating high-dimensional entanglement relied on conservation laws of the utilized photon creation process\cite{mair2001entanglement,vaziri2002experimental,krenn2014generation}. For example, in SPDC, the OAM is conserved. Therefore, the OAM of the down-converted photons sums up to the OAM of the pump photon $\ell_p=\ell_s+\ell_i$. Using a pump beam with zero OAM ($\ell_p=0$) yields perfect anti-correlation $\ell_s=-\ell_i$ within the entangled quantum state. The desired coherent overlap is guaranteed since, in principle, no information is available on which combination of OAM modes is realized. However, the probability that a certain OAM correlation occurs is different for all OAM combinations. It is more likely that the two down-converted photons are found in a lower order OAM mode than in higher orders~\cite{miatto2012bounds}. This, in turn, results in an inherently non-maximally, though high-dimensionally entangled quantum states. There exist methods to pre-~\cite{kovlakov2018quantum,liu2018coherent} or post-compensate~\cite{vaziri2003concentration,dada2011experimental} for this naturally occurring unbalance, which results in lower creation efficiencies or limitations in terms of versatility. 

\begin{figure*}[ht]
    \centering
    \includegraphics[width=0.9\linewidth]{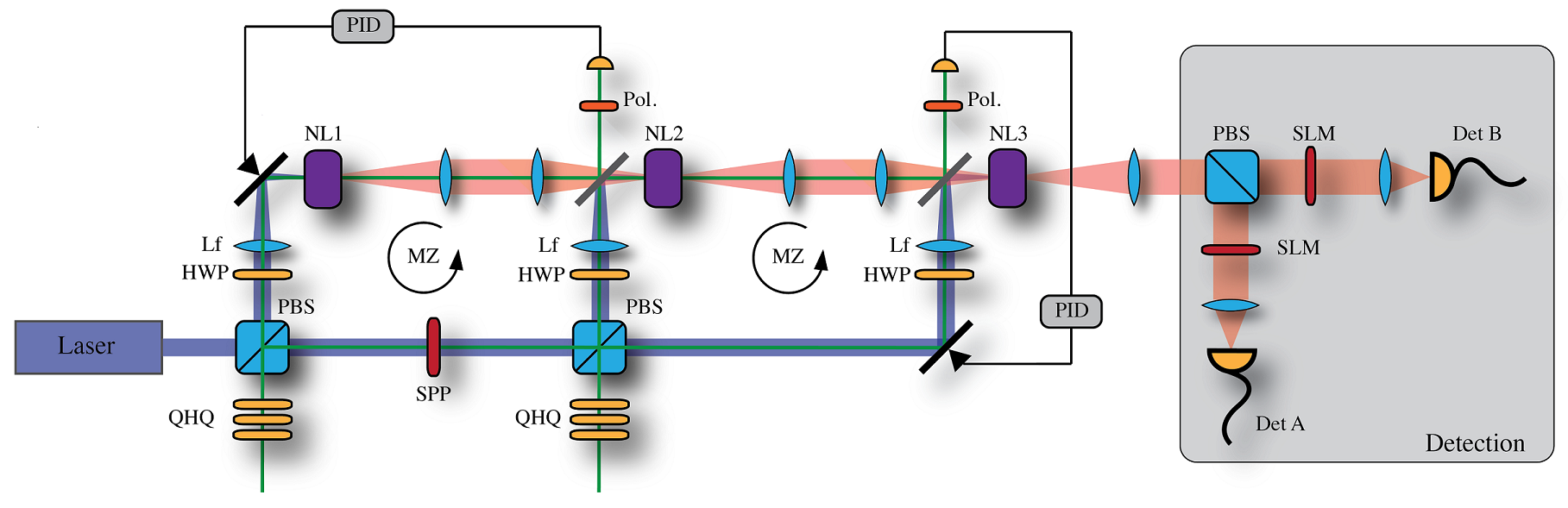}
    \caption{Detailed experimental setup for creating three-dimensionally entangled photon pairs using path identity \cite{kysela2020path}. A continuous-wave laser centered at 405nm is split with polarizing beam splitters (PBS) and is guided to three non-linear-crystals(NLC) made from periodically poled KTP. For experimental simplicity, the spiral-phase-plate (SPP) is placed within the pump beam instead of the down-converted photons as depicted in Fig.\ref{fig:entanglement-sources-1}. A lens (Lf) is used to focus the pump beam onto the NLC. Using a 4f optical imaging system between two consecutive NLCs ensures spatial indistinguishability of the down-converted photon pairs. Actively stabilized Mach-Zehnder interferometer using a PID-controller ensure the interferometric stability between two NLCs. The phase between two NLCs can be set using the quarter-half-quarter (QHQ) waveplates of the stabilization laser depicted in green. The photon pairs are deterministically separated using a PBS at the detection part. Arbitrary superposition projections can be measured with a spatial-light-modulator (SLM) in combination with a single-mode fiber.}
    \label{fig:EBPI_Experiment}
\end{figure*}

%concept of creating high-dim entanglement using path-identity
Creating high-dimensionally entangled photon pairs using path-identity represents a paradigm shift to previous schemes. Conceptionally, creating high-dimensionally entangled photon pairs using path-identity relies on indistinguishable, probabilistic and coherently emitting photon-pair sources. As shown in Fig.~\ref{fig:entanglement-sources-1}, each non-linear crystal probabilistically emits identical\footnote{in all degrees of freedom} photon pairs in the lowest order Gaussian spatial mode, which is denoted as $|0,0\rangle$. Inserting a mode shifting element between two successive non-linear crystals, that is capable of performing a $+1$ operation on the quantum state $|0,0\rangle\rightarrow|1,1\rangle$ yields a $d$-dimensionally entangled quantum state of the form
\begin{equation}
    |\psi\rangle=1/\sqrt{d}(|0,0\rangle+e^{i\phi_1}|1,1\rangle+\cdots+e^{i\phi_d}|d-1,d-1\rangle).
\end{equation}
Adding additional control on the emission rate of each non-linear crystal results in the ability to create arbitrary high-dimensionally entangled quantum states $\sum_{i=0}^{d-1}\alpha_i|i,i\rangle$, with $\alpha\in\mathbb{C}$ and $\sum_i|\alpha_i|^2=1$. 

To realize this scheme experimentally, one needs indistinguishable photon-pair sources in all degrees of freedom. This means their joint-spectral-amplitude, polarization and paths are perfectly identical. Furthermore, to ensure coherent emission of two or non-linear crystals poses two main constraints~\cite{zou1991induced,herzog1994frustrated,jha2008temporal, kulkarni2017transfer}:

First, analogous to eq.(\ref{coh-length-cond-1}), the optical path-length difference of the pump beam and the two down-converted photons must be smaller than the coherence length of the pump laser, e.g.
\begin{equation}
    |L_P-L^a_{DC}-L^b_{DC}|\leq L_P^{coh}
    \label{eq:coherence-condition-1}
\end{equation}
with $L_P$ denoting the optical path length of the pump beam, $L_{DC}$ the optical path length of the respective down-conversion photon and $L_P^{coh}$ the coherence length of the pump laser.

The second condition is given by the optical path length difference of the down-conversion photons and their coherence length:
\begin{equation}
    |L^a_{DC}-L^b_{DC}|\leq L_{DC}^{coh}
    \label{eq:coherence-condition-2}
\end{equation}
with $L_{DC}^{coh}$ describing the coherence length of the down-conversion photons.

The first condition is given by eq. (\ref{eq:coherence-condition-1}) can be fulfilled by choosing a narrowband pump laser with a coherence length of several centimetres. The second condition is more difficult to meet. Typically, photons created via SPDC have a spectral bandwidth on the order of nanometres. In turn, this yields a coherence length of approximately tens of micrometres. Matching the optical path length of the two down-conversion photons can be difficult. For example, birefringence in the non-linear crystals can lead to a substantial mismatch of the temporal overlap and thus to indistinguishability. These effects can be avoided altogether using a type-0, or type-I phase-matched SPDC source\footnote{both create down-conversion photons with identical polarisation} at the cost of the possibility to deterministically split the photon-pairs in a collinear arrangement using polarisation (as for type-II sources). 

%experimental realization:
\begin{figure}[b]
    \centering
    \includegraphics[width=0.45\textwidth]{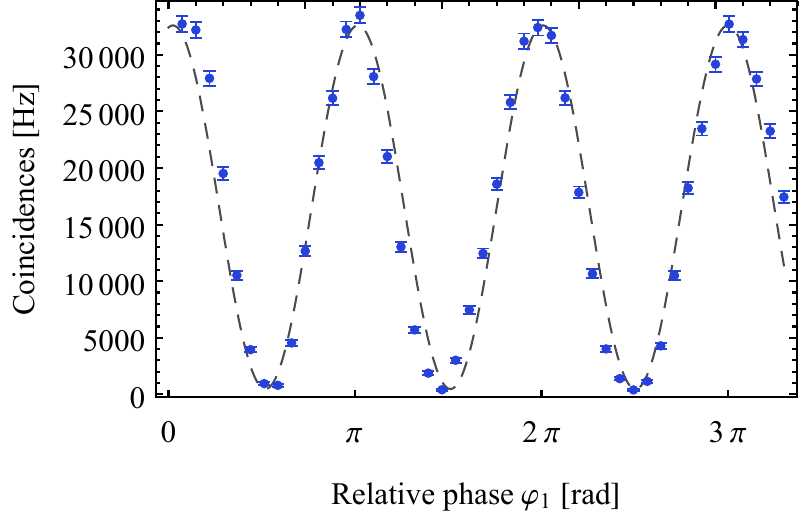}
    \caption{Coincidence interference fringes of two consecutive non-linear-crystals. In this plot, the relative phase $\varphi_1$ between NL1 and NL2 is altered. The coincidences are measured without an SPP, thus the interference is occurring between the fundamental Gaussian modes (denoted with 0) and the corresponding quantum state reads $|0,0\rangle_\text{NL1}+e^{i\varphi1}|0,0\rangle_\text{NL2}$. If the observed visibility reaches 1, then the two photon-pair sources are identical, meaning NL1=NL2. In this experiment, visibility of $0.971\pm0.005$ is observed.}
    \label{fig:entanglement-path-identity-visibility-plot}
\end{figure}
%create indistinguishable photon pairs
A quantitative measure for the achieved indistinguishability is given by the interference visibility between the two crystals. Experimentally the phase is adjusted by splitting the pump from the down-conversion photons and shifting the optical path-length between the two. Using an additional single-mode-fibre and detecting simultaneous two-fold photon counts, visibility of $97.1\pm0.5\%$ has been demonstrated, see Fig.\ref{fig:entanglement-path-identity-visibility-plot}. 
%what determines the indistinguishability and the coherence? -- IMO too detailed (MK)
%The spectral indistinguishability is ensured by precisely adjusting the temperature of the two consecutive ppKTP\footnote{periodically poled potassium titanyl} crystals within $0.1^\circ$ to maximize the overlap between the joint-spectral-amplitudes. To compensate for birefringent effects within the 10mm long type-II quasi-phase-matched SPDC processes, the orientation of the two crystals is inverted\footnote{Here, the birefringence introduces a maximal temporal walk-off of roughly 150um, whereas the coherence length of the down-conversion is only about 70um. Without inverting the two ppKTP crystals, the visibility is drastically reduced.}. 

%how does it impact the quality of the entangled quantum state
The indistinguishability of the two photon-pair sources is an important measure because it directly determines the coherence of the entangled quantum state.

Given this basic ingredient, the next step is to introduce a mode-shifting element that is capable of shifting the spatial mode. In this first proof-of-principle experiment \cite{kysela2020path}, the pump and down-conversion photons are split up to access both wavelengths separately. This avoids chromatic aberrations\footnote{without using specially designed optical elements} and allows the simultaneous manipulation of both pump and down-conversion photons. Here, the pump beam is split into three parts that coherently pump all three ppKTP crystals. The pump beam for the second and third crystal is modified with a spiral-phase-plate that adds (subtracts) four-quanta of OAM to (from) the pump beam. According to the conservation of OAM in the SPDC process, the second crystal thus emits photon pairs with two quanta of OAM $|2,2\rangle$ and the third with opposite $|\text{-}2,\text{-}2\rangle$ correlated photon pair. The resulting quantum state reads
\begin{equation}
\ket{\psi} = \alpha \! \! \underbrace{\ket{0,0}}_{\text{crystal 1}} + \ \beta  \! \! \! \underbrace{\ket{2,2}}_{\text{crystal 2}} + \ \gamma  \underbrace{\ket{-2,-2}}_{\text{crystal 3}}.
\label{eq:3dstate}
\end{equation}
where the magnitudes of $\alpha,\beta$ and $\gamma$ are adjusted by controlling the relative pump power and the phases by adjusting the relative phase within the Mach-Zehnder interferometers. To guarantee phase stability, the two Mach-Zehnder interferometers are actively stabilized with an additional phase-locking laser.

%Results
Performing full state tomography yields reported fidelities ranging from $85\%$ to $90\%$ for two- and three-dimensionally entangled quantum states. Thereby, different maximally and non-maximally entangled photon pairs are created in two and three dimensions to demonstrate the versatility in terms of state creation using entanglement by path identity. 

%Potentially upscaling the dimensionality?
The principle of generating entanglement by path identity is suited to scale up the dimensionality of the entangled state. One possible way to scale up the dimensionality is to miniaturize the unit dimensional cell consisting of a non-linear crystal and a phase-shifting element. Smaller distances in combination with integrated fabrication technologies as demonstrated in~\cite{wang201818} could substantially increase the stability and cross-talk quality. Complementarily, for example, a purely linear arrangement without interferometer could be implemented as in the original proposal. This would require an element which, for example, only performs a +1 operation on the down-conversion photons but not on the pump beam. The q-plate~\cite{marrucci2006optical} would be a possible realization. 

%Multi-Photon Entanglement with references to GHZ/ Quantum 
%Multi-Photon Entanglement with references to GHZ/ Quantum 
\subsection{Multi-Photon Entanglement by Path Identity}
%Multi-Photon 2d-GHZ creation using path identity and generalisation to 3d-GHZ
Multi-photonic interference phenomena lie at the heart of many key experiments~\cite{pan2012multiphoton}. Ranging from technological demonstrations such as quantum teleportation~\cite{bouwmeester1997experimental,ren2017ground,wang2015quantum,PhysRevLett.123.070505}, entanglement swapping~\cite{pan1998experimental}, fault-tolerant~\cite{shor1996fault} and blind quantum-computation~\cite{barz2012demonstration} to fundamentally and philosophically appealing experimental demonstrations of rejecting local-realistic theories using genuine multi-photon entanglement~\cite{greenberger1990bell,bouwmeester1999observation,pan2000experimental,zhong201812}.

In general, multi-particle entanglement is still a very active research area due to the vast complexity it involves. Even for small systems consisting of four qubits only, there exist nine different ways to be entangled~\cite{verstraete2002four}, for five there are infinitely many. The interested reader is referred to this small and incomplete list of references~\cite{horodecki2009quantum,plenio2014introduction,bruss2002characterizing}. 

An important class of maximally entangled multi-photon states are the Greenberger-Horne-Zeilinger (GHZ) states~\cite{greenberger1989going}. These states have been investigated in the fundamentally interesting context of local-realistic theories. In contrast to Bell's theorem~\cite{bell1964einstein}, the GHZ theorem allows for a qualitatively different way of refuting local-hidden-variable theories\cite{greenberger1990bell,mermin1990extreme}. But these maximally entangled multi-particle states are not only of fundamental interest. Error-correcting schemes in quantum computers~\cite{shor1996fault} are based on GHZ states, or quantum-secret sharing protocols~\cite{hillery1999quantum} use these strong correlations to exceed classical limitations. These prospects have started technological developments on various physical platforms ranging from trapped-ions~\cite{monz201114}, Rydberg atoms\cite{Omran570}, super-conducting qubits\cite{kelly2015state,song201710} and photons\cite{zhong201812,wang201818}. 

Since this review is focused on photonic systems, we will discuss the experimental principles behind the GHZ state and introduce the concept of entanglement by path identity to the multi-photonic regime. Furthermore, we also discuss the generalization to higher-dimensional and multi-photonic systems that have been developed and realized recently~\cite{malik2016multi,erhard2018experimental}.

\SkipTocEntry\subsubsection*{GHZ entanglement for QuBits}\label{ch:2d-GHZ}
\begin{figure}[htbp]
    \centering
    \includegraphics[width=0.45\textwidth]{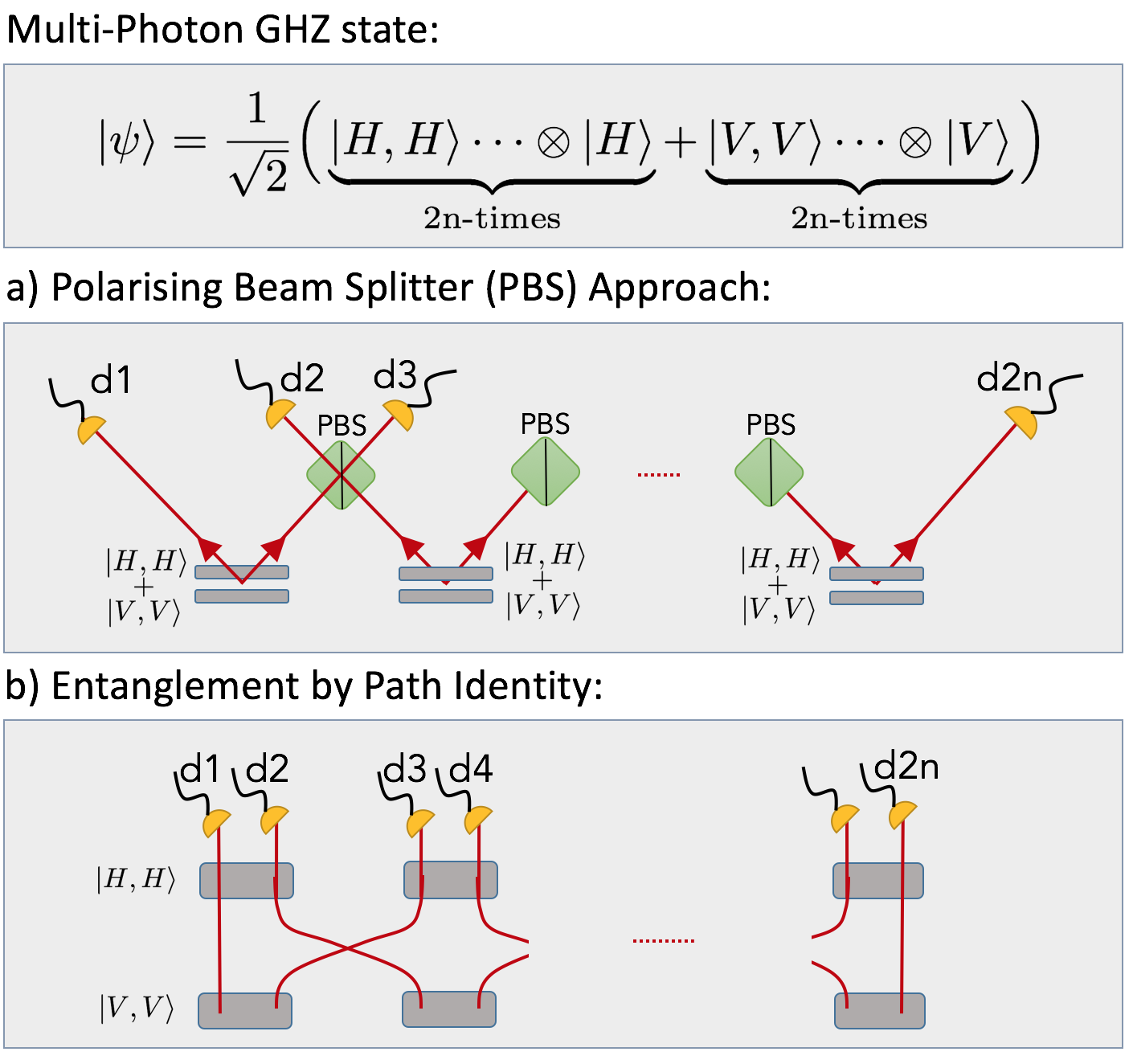}
    \caption{Mutli-Photon Greenberger-Horne-Zeilinger (GHZ) state creation. \textbf{a)} The commonly used technique\cite{pan2012multiphoton} to create GHZ states is to overlap two maximally entangled two photons in the state $|\phi^+\rangle=(|H,H\rangle+|V,V\rangle)/\sqrt{2}$, with $H,V$ denoting the polarisation state of the photon. Since the polarizing beam splitter (PBS) only transmits horizontally (H) polarised photons and reflects only vertically (V) polarized photons, simultaneous 2n-fold detection of photons at detectors $\{d1,...,d2n\}$ results in a maximally and genuine multi-photon (2n) GHZ state. \textbf{b)} Creating multi-photon GHZ states using path identity. Two rows of n crystals, where all 2n crystals can coherently emit n-photon pairs. The lower row creates vertically polarised photons, while the upper row solely creates horizontally polarized photon pairs. By identifying the paths of the photons in the lower row with the ones in the upper row (as indicated above) and conditioning upon 2n-fold photon detection results in a GHZ state. Because of the path identification, a 2n-fold photon detection event can only occur if all crystals of the upper row emit photon pairs or if all crystals of the lower row emit photon pairs.  }
    \label{fig:2d-ghz-scheme}
\end{figure}
For photons there exists a particularly simple scheme to create an arbitrary number of entangled photons in principle.
As shown in Fig.\ref{fig:2d-ghz-scheme}, we use two two-dimensionally entangled photon-pair sources $|\psi\rangle=(|H,H\rangle+|V,V\rangle)/\sqrt{2}$ entangled in their polarisation degree-of-freedom. Combining one photon of one source with another photon of the other source at a polarising beam splitter. A polarizing beam splitter reflects vertically and transmits horizontally polarised photons, thus the resulting quantum state reads 
\begin{align}\label{eq:2d-ghz-after-pbs}
    (|H,H,H,H\rangle_{ABCD}+|H,H,V,V\rangle_{ABBD}\\\nonumber
    +|V,V,H,H\rangle_{ACCD}+|V,V,V,V\rangle_{ABCD})/2,
\end{align}
where $H,V$ describes the polarisation state (horizontal and vertical) and the indices $d1, d2, d3$ and $d4$ label the detectors. Post-selecting on simultaneous four-photon detection events on all four detectors $d1, d2, d3, d4$ results in the desired GHZ type entangled four photon state
\begin{equation}\label{eq:2d-ghz}
    (|H,H,H,H\rangle_{ABCD}+|V,V,V,V\rangle_{ABCD})/\sqrt{2}.
\end{equation}
This scheme can now be generalised to any photon number by adding more two-dimensionally entangled photon pair sources and combining them at polarising beam splitters, as depicted in Fig.\ref{fig:2d-ghz-scheme}. 

In contrast, using path identity one can create GHZ type entangled quantum states without using PBSs as described above. Instead, two rows of 2n non-linear-crystals (NLC) are coherently pumped, such that they can simultaneously emit n photon pairs, see Fig.\ref{fig:2d-ghz-scheme}b). The upper row of NLCs solely emits horizontally polarised photon pairs $|H,H\rangle$, while the lower row only creates vertically polarized photon pairs $|V,V\rangle$. By using the path identity principle, we now cross the paths of the photons between next neighbouring NLCs, as depicted in Fig.\ref{fig:2d-ghz-scheme}b). Conditioning onto 2n-fold photon events, meaning that all 2n photodetectors simultaneously detect a photon, results in a genuinely multi-photon entangled GHZ state. The reason for this is, that the only two possibilities for such a 2n-fold photon detection exists. First, either all crystals from the upper row simultaneously emit one photon pair, or all crystals from the lower row simultaneously emit one photon pair. Whenever only a single crystal of the lower (upper) row emits a photon pair instead of the upper (lower) row, then no 2n-fold detection can occur because at least one detector will be empty (not detecting a photon). Since all crystals are pumped coherently and the upper and lower row have different polarisations, we can write these two creation possibilities in a coherent superposition as stated by $|\psi\rangle$ in Fig.\ref{fig:2d-ghz-scheme}.

%efficiency comparison of both protocols (PBS vs. PathId)?
The creation efficiency of the protocol can be calculated from eqs.~\ref{eq:2d-ghz-after-pbs} and \ref{eq:2d-ghz}. For n-photon pairs there exist in general $2^n$ terms after the PBS, see eq.\ref{eq:2d-ghz-after-pbs}. However, we are only interested in a maximally entangled GHZ state, which always only consists of two terms. Hence, the efficiency of the generation protocol using entangled photon pairs in combination with PBSs is given by the ratio between the two expected terms and all possible terms, which evaluates to $2^{-n+1}$.

%experimental efficiency?
Despite this simple principle scheme, the largest number of photons entangled in a GHZ manner is twelve~\cite{zhong201812}. To go beyond this number, several challenges of inherently probabilistic photon pairs sources need to be overcome. For example, the probability that one photon pair is created is $p$ then the probability that six photon pairs are emitted simultaneously is $p^6$. Usually, the probability of creating one photon pair in a single laser pulse is $p\approx 10^{-6}-10^{-2}$, which leads to a 12-photon detection rate of approximately one event in 10 hours. In addition, SPDC sources produce with probability $p$ one pair and with probability $p^n$ n-photon pairs. Multi-pair emissions reduce the fidelity of the entangled quantum state if no number resolving detectors are utilized. Among others, faster triggering rates of detectors, higher photon detection efficiencies, photon-number-resolving detectors are currently investigated and optimized\cite{rudolph2017optimistic,slussarenko2019photonic}.

\SkipTocEntry\subsubsection*{GHZ entanglement beyond QuBits}
%what makes it interesting to create such states?
Generalizing the GHZ theorem to higher-dimensional QuDits only recently succeeded~\cite{ryu2014multisetting,ryu2013greenberger,lawrence2014rotational,tang2017multisetting}. Also the first experimental implementation of a genuinely higher-dimensional GHZ state (3-dimensional) was just performed shortly later~\cite{erhard2018experimental}. In contrast to the two-dimensionally entangled GHZ state, the experimental implementation is not as simple and has only been found using computational algorithms~\cite{krenn2016automated,krenn2020computer}. However, despite the unintuitive experimental creation of the three-dimensionally entangled GHZ state using linear-optics, there is an intuitive way of creating such states using entanglement by path identity.

\begin{figure}[hb]
    \centering
    \includegraphics[width=0.45\textwidth]{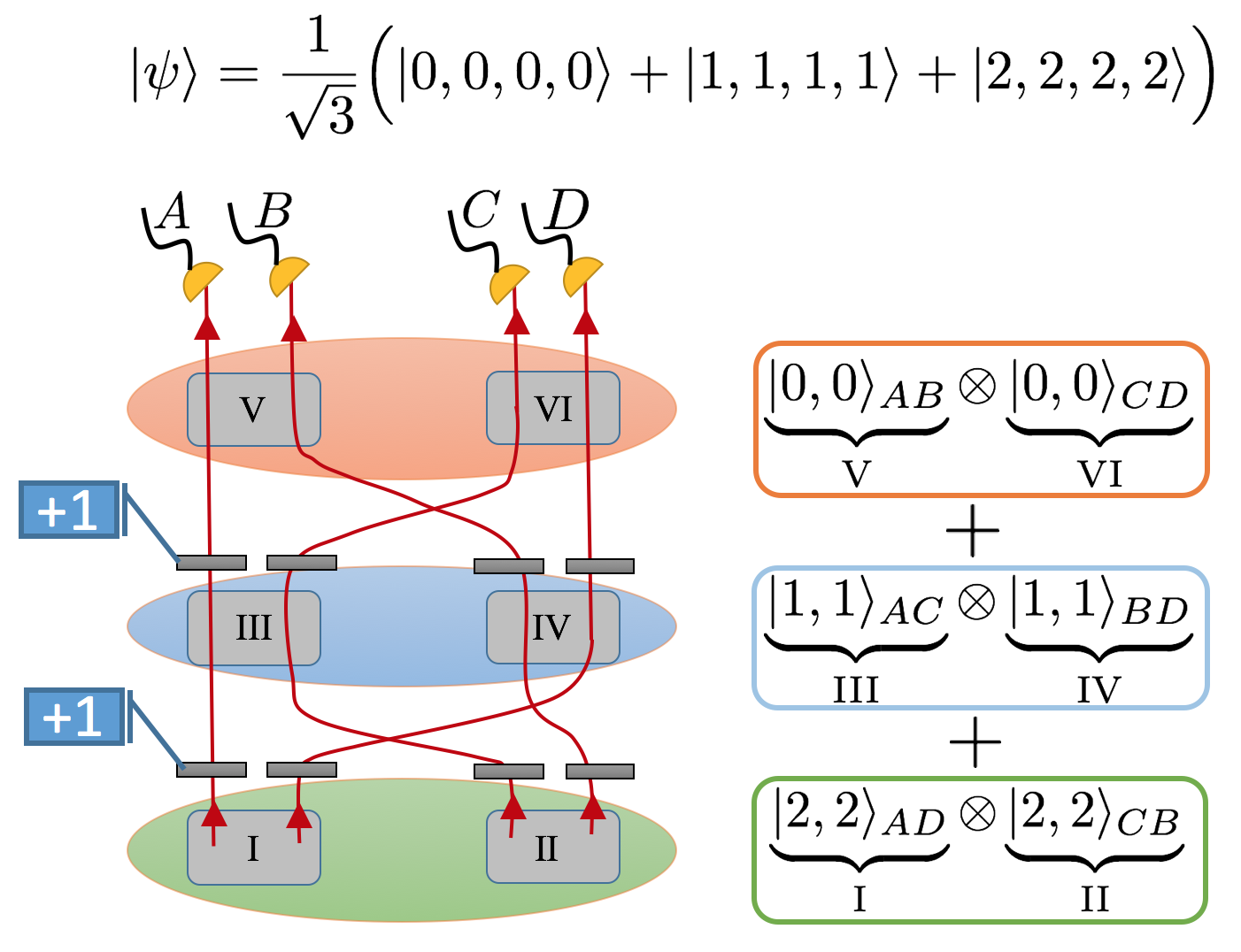}
    \caption{Creating genuine multi-photon and high-dimensional quantum entanglement using path identity. Six non-linear-crystals (NLC) labelled with roman numbers where each NLC can emit photon pairs in the transverse spatial Gaussian-mode (denoted by $|0,0\rangle$) are utilized. Each photon pair emitted by one of the six crystals is connected to a pair of detectors $\{A,B,C,D\}$ as indicated on the right in the figure above in such a way that overlapping paths of different crystals are identified. Simultaneous four-photon events where each detector registers a photon can occur in three possible ways: Either the lowest, the middle or the top row emits two pairs simultaneously. All other combinations cannot occur because of the specific path identification and routing of the photon paths. Inserting spiral-phase plates between the different layers of crystals add one quantum ($+1\hbar$) of orbital-angular-momentum (OAM) to incoming photons. Since all photon pair emissions occur coherently, the three possibilities can be written in a coherent superposition yielding a genuinely four-photon and three-dimensionally entangled GHZ state $|\psi\rangle$.}
    \label{fig:3d-GHZ_path-identity}
\end{figure}

In Fig.\ref{fig:3d-GHZ_path-identity}, the principle of generating a four-photon three-dimensionally entangled GHZ state is shown. Similar to the two-photon high-dimensional case, we employ the OAM of photons to illustrate entanglement by path identity for multiple photons. All six crystals $\text{(I, II, III, IV, V, VI)}$ are pumped coherently and emit with the same probability amplitude one photon pair in the fundamental Gaussian mode denoted as $|0,0\rangle$. Furthermore, we post-select onto events where four-photons are detected simultaneously in all detectors $A, B, C$ and $D$. We can now identify the photon pairs paths such, that there exist exactly three possible ways how such a four-photon detection event can appear: Either the two crystals in the first, second or third row in Fig.\ref{fig:3d-GHZ_path-identity} emit simultaneously. If the two-photon pairs have been produced in the first row, they propagate twice through a spiral-phase plate that adds one quantum of OAM to the photons. Thus at the detector, these two-photon pairs are described by the probability amplitude $|2,2\rangle_\text{AD}\otimes|2,2\rangle_\text{CB}$. Similarly, for the second and third row, as depicted in Fig.\ref{fig:3d-GHZ_path-identity}. Since the photon pair emission events occur such that there is in principle no information about their origin, we have to write all three possibilities in a coherent superposition
\begin{equation}
    |\psi\rangle=\frac{1}{\sqrt{3}}\Big(|0,0,0,0\rangle+|1,1,1,1\rangle+|2,2,2,2\rangle\Big),
\end{equation}
which is exactly the desired three-dimensionally entangled four-photon GHZ state.

%protocol efficiency?
For the two-dimensional GHZ state creation with entanglement by path identity, no quantitative difference in terms of efficiency or achievable fidelity was found. However, for the three-dimensional GHZ state, there is indeed an advantage in terms of creating efficiency. The linear-optical approach realized in~\cite{erhard2018experimental} only succeeds in $\approx 5\%$ of all four-photon emission events. In the entanglement by path identity approach, every detectable four-photon event succeeds. Thus using this new method results in a stunning 20-fold improvement and hence reduces the estimated measurement time from roughly two weeks to less than a day. 

%experimental constraints, coherent (pulsed lasers->short coh-length) & indistinguishalble photon pairs (JSA)?
From an experimental point of view, such an experiment puts stringent constraints in terms of coherent and indistinguishable photon pair emission. As discussed in section~\ref{sec:two-photon-high-dim-entanglement}, the path length of the down-converted photons are restricted to the pump and down-conversion coherence length. While in the two-photon case a continuous-wave laser with a long coherence length can be employed, the multi-photon scenario usually requires pulsed pump lasers to identify simultaneous two-pair events. Femtosecond pulsed lasers are routinely used in such experiments and have a Fourier-limited coherence length on the order of micrometer. Also, the joint-spectral-amplitude of the down-converted photon pairs not only need to be identical but also separable\cite{pan2012multiphoton}. 

%can path-identity be generalize to the d-dim case?
Lastly, the scheme of creating genuinely three-dimensional entangled GHZ states seems to be generalizable to any $d$-dimensional GHZ state. It would require to identify different paths in a non-linear crystal network such that exactly $d$ possible ways exist to create a four-photon detection event. This question can be answered using graph theory and has a surprising answer, which will be analyzed in detail in the next chapter.

\subsection{Manipulating Entangled States without Direct Interaction}\label{subsec:ent-control}
After showing how to generate entangled states using path identity, it was reported how the concept could be generalized to multi-photon emitters. Interestingly, these generalizations allow for manipulations of quantum states without ever interacting with the involved photons. The method also emphasizes the deep connection between entanglement and interference, a connection that has fascinated scientists for a long time \cite{greenberger1989going,greenberger1990bell,horne1986einstein,zukowski1988bell,horne1989two,rarity1990experimental,pan2012multiphoton}.
\par
\begin{figure}[htbp]  \centering
    \includegraphics[width=0.9\linewidth]{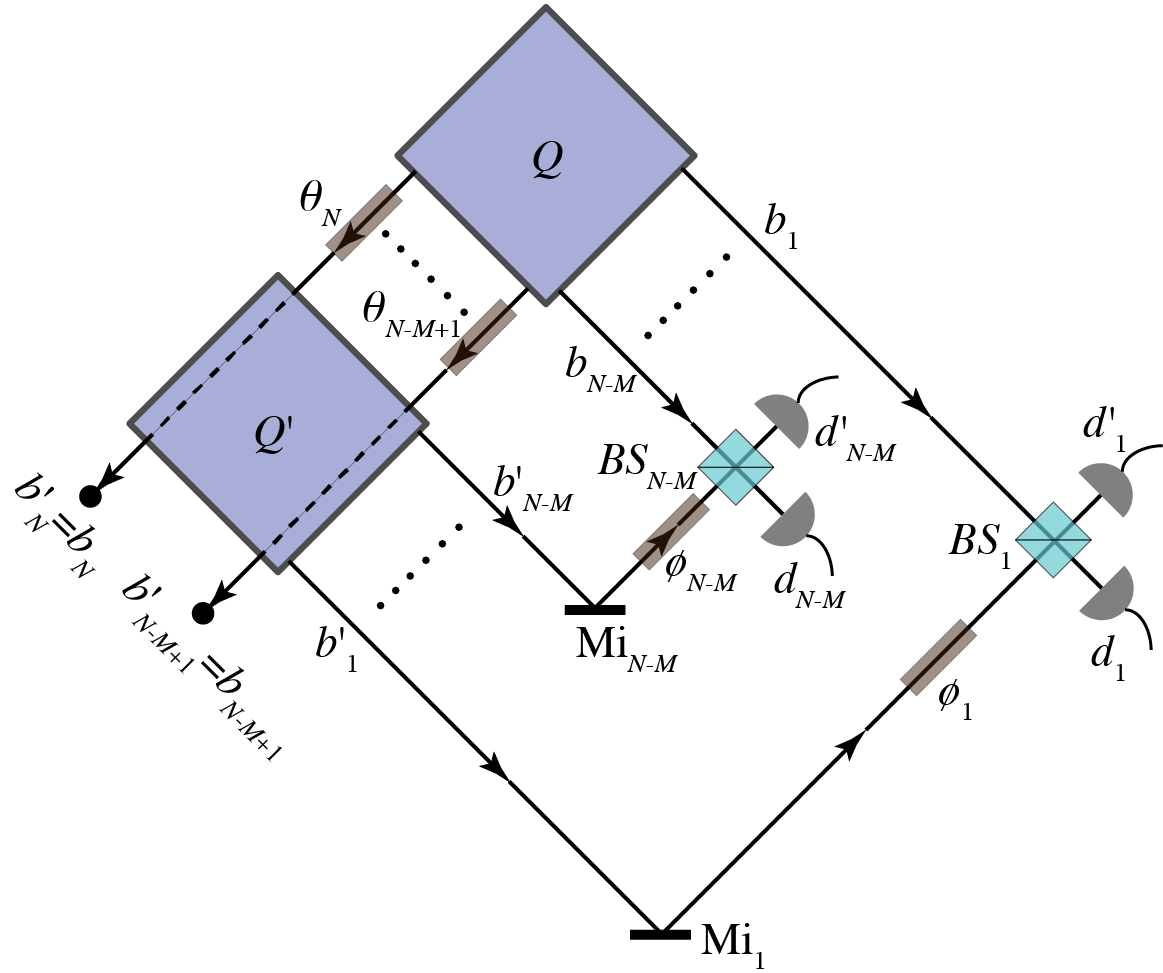}
    \qquad \caption{Scheme of generating and controlling many-particle entangled states. Two identical $N$-particle sources emit particles
        ($1,2,\dots, N$) into paths ($b_1,b_2, \dots, b_N$) and ($b_1',b_2',
        \dots, b_N'$) respectively. Path identity is applied for $M$ particles ($N-M+1,\dots,
        N$). Rest of the particles ($1,2,\dots, N-M$)
        produce many-particle interference patterns and many-particle entangled states
        when their paths are superposed by beam splitters. (Adapted from \cite{lahiri2018many}.)} \label{fig:gen-setup}
\end{figure}

The scheme is illustrated in Fig. \ref{fig:gen-setup}. 
There are two identical sources, $Q$ and $Q'$, each of which can emit $N$ particles. We label the particles by numbers $1,2, \dots, N$. Suppose that $Q$ emits these particles in paths $b_1,b_2, \dots, b_N$. Similarly, $Q'$ emits them in paths $b_1',b_2', \dots, b_N'$. If the sources emit in quantum superposition, they generate the following $N$-particle path-entangled GHZ state:
\begin{align}\label{state-N-part}
\ket{GHZ}_N=\frac{1}{\sqrt{2}}\left(\prod_{j=1}^N
\ket{b_j}+e^{i\phi_0}\prod_{j=1}^N \ket{b_j'}\right).
\end{align}
We make the paths of particles $N-M+1,\dots,N$ identical by sending
the paths $b_{N-M+1},\dots,b_{N}$ through $Q'$ and perfectly
aligning them with $b_{N-M+1}',\dots,b_{N}'$. The corresponding transformations of kets are given by $\ket{b_l}_l \to \exp[i\theta_l]\ket{b_l'}$, where $l=N-M+1,\dots,N$ and  
$\theta_l$ is the phase introduced in path $b_l$ between $Q$ and $Q'$.
If the remaining pairs of beams ($b_1,b_1'$), ($b_2,b_2'$),
$\dots$, ($b_{N-M},b_{N-M}'$) are superposed by beam
splitters, entangled states are produced at the outputs of these beam splitters. 
\par
Suppose that the outputs of the beam splitters are collected by pairs of detectors ($d_1,d_1'$),
($d_2,d_2'$), $\dots$, ($d_{N-M},d_{N-M}'$) and we measure $(N-M)$-fold coincidences with a set of
detectors, each placed at an output of a distinct beam
splitter. These $N-M$ fold coincidence counts vary with the phase of the undetected particles in a sinusoidal way and generate a $N-M$-particle interference pattern.
\par
The generation of the interference patterns and the entangled states can be mathematically understood as follows. If we apply transformations (\ref{BS-1}) and the path-identity conditions
to Eq. (\ref{state-N-part}), we find that the quantum state becomes
\begin{align}\label{state-N-part-final}
&\ket{\psi_N} \nonumber
\\ &=\left(\frac{1}{\sqrt{2}}\right)^{N-M+1}\left[\sum_{r=0}^{N-M}
(i^r+i^{N-M-r}e^{i\xi^{(N)}_M}) \ket{D_r}^{N-M}\right]
\nonumber
\\ & \qquad \qquad \qquad \qquad \otimes \prod_{j=1}^M \ket{b_{N-M+j}'},
\end{align}
where
$\xi^{(N)}_M=\phi_0+\sum_{k=1}^{N-M}\phi_k-\sum_{j=1}^{M}\theta_{N-M+j}$
and $\ket{D_r}^{N-M}$, is a ($N-M$)-particle Dicke state
\cite{dicke1954coherence,toth2007detection}, i.e. a sum of
$\binom{N-M}{r}$ terms (states), each being a product of $r$ primed
states ($\ket{d_k'}$) and $N-M-r$ unprimed states ($\ket{d_k}$). We note that the
kets $\ket{b_{N-M+j}'}_{N-M+j}$ factor out. This fact implies
that one does not need to detect
the $M$ particles used for path identity in order to observe the patterns. The entangled state representing $N-M$ particles emerging from the outputs of the beam splitters
depends on the value of the phase
$\xi^{(N)}_M$. Since $\xi^{(N)}_M$ contains the
phases $\theta_{N-M+j}$, it can be varied without interacting with
the entangled particles. Therefore, the scheme allows us to manipulate the entangled states and the interference patterns in an interaction-free way.
\begin{figure}\centering
   \subfigure[] {
    \label{figa:3-part-1-Pi-patterns}
    \includegraphics[width=0.45\linewidth]{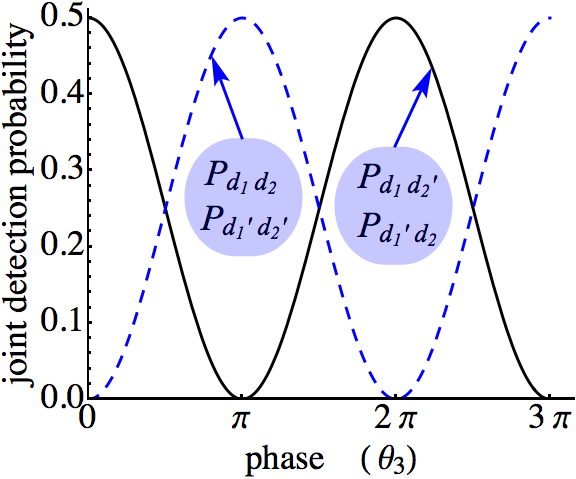}
} \hskip 0.2cm
   \subfigure[] {
    \label{figb:concplot}
    \includegraphics[width=0.45\linewidth]{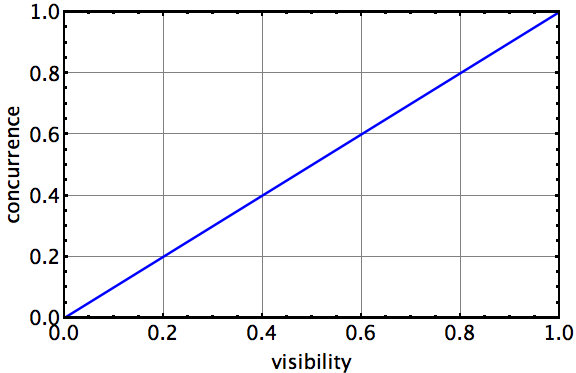}
} 
     \caption{Two-particle interference patterns and entangled states.
interference patterns. (a) Probabilities of joint detection at the pairs of
detectors ($d_1$,$d_2$), ($d_1$,$d_2'$), ($d_1'$,$d_2$), and
($d_1'$,$d_2'$) vary sinusoidally with the phase introduced by one of the undetected particles ($\theta_3$). Interference patterns
$P_{d_1d_2}$ and $P_{d_1'd_2'}$ are in phase (dashed line). They are complementary to the patterns $P_{d_1d_2'}$ and
$P_{d_1'd_2}$ (solid line). Maximum and minimum of any of these patterns correspond to two distinct Bell states. (b) Controlling the amount of entanglement. The concurrence is equal to the
visibility of the two-particle interference pattern. Both
concurrence and visibility are controlled by attenuating the beam(s) of undetected particle(s) between the two sources. (Adapted from \cite{lahiri2018many}.)} \label{fig:patern-ent-control}
\end{figure}
\par
\emph{Example of Entangled States.}\textemdash As an example let us consider the case in which $N-M=2$. In this case, the two-particle interference patterns are given by [Fig. \ref{figa:3-part-1-Pi-patterns}]
\begin{subequations}\label{two-part-pattern}
    \begin{align}
    P_{d_1d_2}=P_{d_1'd_2'}&=\frac{1}{4}[1-\cos(\Phi^{(2)}-\sum_{j=3}^N\theta_j)], \label{two-part-pattern:a} \\
    P_{d_1d_2'}=P_{d_1'd_2}&=\frac{1}{4}[1+\cos(\Phi^{(2)}-\sum_{j=3}^N\theta_j)],
    \label{two-part-pattern:b}
    \end{align}
\end{subequations}
where $\Phi^{(2)}=\phi_0+\phi_1+\phi_2$. When the argument of the cosine take values $2m\pi$ and $(2m+1)\pi$, the pair of particles (1,2) is in the following distinct Bell states respectively ($m=0,\pm 1,\pm 2,\dots$):
\begin{subequations}\label{Bell-st-output}
\begin{align}
\ket{\Psi^{+}}&=\frac{1}{\sqrt{2}}
(\ket{d_1}\ket{d_2'}+\ket{d_1'}\ket{d_2}), \label{Bell-st-output:a} \\
\ket{\Phi^{-}}&=\frac{1}{\sqrt{2}}
(\ket{d_1}\ket{d_2}-\ket{d_1'}\ket{d_2'}).
\label{Bell-st-output:b}
\end{align}
\end{subequations}
Clearly, when the coincidence counts at
($d_1'$,$d_2$) and ($d_1$,$d_2'$) maximize and the coincidence
counts at ($d_1$,$d_2$) and ($d_1'$,$d_2'$) minimize, the state $\ket{\Psi^{+}}$ is obtained. Similarly, the state
$\ket{\Phi^{-}}$ is obtained when coincidence counts maximize at
($d_1$,$d_2$) and ($d_1'$,$d_2'$), and minimize at ($d_1'$,$d_2$) and
($d_1$,$d_2'$).We can therefore switch
between the two Bell states without interacting with the pair of
particles.
\par
We consider as another example
the case in which $N-M=3$ and $\xi^{(N)}_M=(2m+1/2)\pi$. It follows from Eq.
(\ref{state-N-part-final}) that the entangled state has the form (replacing the unprimed states by $0$ and primed states by 1)
\begin{align}\label{GHZ-output}
\frac{1}{2}(&\ket{0}\ket{0}\ket{0}-\ket{1}\ket{1}\ket{0}-
\ket{1}\ket{0}\ket{1}\nonumber
\\&-\ket{0}\ket{1}\ket{1}).
\end{align}
This state is a three-particle Greenberger-Horne-Zeilinger-class
state (see, for example, \cite{rubens2009tripartite}). It has highest
(unit) ``three-tangle'' or ``residual entanglement'' \cite{coffman2000distributed}: the
concurrence \cite{hill1997entanglement,wootters1998entanglement} of each qubit with
the rest of the system is $1$, and all the pairwise concurrences are
$0$.
\par
\emph{Controlling the Amount of Entanglement}.\textemdash As we mentioned in Sec. \ref{subsec:theory-ZWM}, path identity can be controlled by
inserting attenuator(s) in the
path(s) of aligned particle(s) between the two sources. In this case the visibility of the interference patterns and the amount of entanglement change.
\par
As an example we consider the case $N-M=2$. We insert an attenuator with the amplitude transmission coefficient $T$ in the path of one of the undetected particles between $Q$ and $Q'$. (We assume $T$ to be real without any loss of generality.) In this case, the two-particle interference patterns are given by 
\begin{subequations}\label{two-part-pattern-loss}
\begin{align}
P_{d_1d_2}=P_{d_1'd_2'}&=\frac{1}{4}[1-T\cos(\Phi^{(2)}-\sum_{j=3}^N\theta_j)], \label{two-part-pattern-loss:a} \\
P_{d_1d_2'}=P_{d_1'd_2}&=\frac{1}{4}[1+T\cos(\Phi^{(2)}-\sum_{j=3}^N\theta_j)].
\label{two-part-pattern-loss:b}
\end{align}
\end{subequations}
The visibility of the interference patterns is given by $\mathcal{V}=T$. If the argument of cosine takes the values $2m\pi$ and $(2m+1)\pi$, the two-particle entangled state takes the following forms respectively:
\begin{subequations}\label{do-form-sp}
\begin{align}
&\widehat{\rho}_{+}=\frac{1}{2}\left(1-T \right)
\ket{\Phi^{-}}\bra{\Phi^{-}}+\frac{1}{2}
\left(1+T\right)\ket{\Psi^{+}}\bra{\Psi^{+}},\label{do-form-sp-1} \\
&\widehat{\rho}_{-}=\frac{1}{2}\left(1+T \right)
\ket{\Phi^{-}}\bra{\Phi^{-}}+\frac{1}{2}
\left(1-T\right)\ket{\Psi^{+}}\bra{\Psi^{+}}.\label{do-form-sp-2}
\end{align}
\end{subequations}
The concurrence
\cite{wootters1998entanglement} of both states is given by $\mathcal{C}(\widehat{\rho}_{+})=\mathcal{C}(\widehat{\rho}_{-})=T=\mathcal{V}$. Therefore, in this case the concurrence is equal to the visibility of the two-particle interference pattern [Fig. \ref{figb:concplot}]. Clearly, we
can change the concurrence by
varying $T$, i.e. thereby without interacting with the entangled particles. A similar argument applies when
the number of entangled particles is
more than two. This is because the
placement of attenuators would result in the conversion of a pure
output state to a mixed state for any number of particles.

\newpage
\section{Quantum Experiments described by Graph Theory}\label{ChapterGraphs}
Multi-photonic quantum entanglement experiments based on path identity, such as those in the previous section, can ideally be described using graph theory \cite{krenn2017quantum}. This different point of view allows for more systematic manual \cite{gu2019quantum3, krenn2019questions} and algorithmic \cite{krenn2020conceptual} design methods for quantum experiments, insights into new quantum interference effects and connections to quantum computation \cite{gu2019quantum2}.
\begin{figure}[!ht]
\centering
\includegraphics[width=0.5\textwidth]{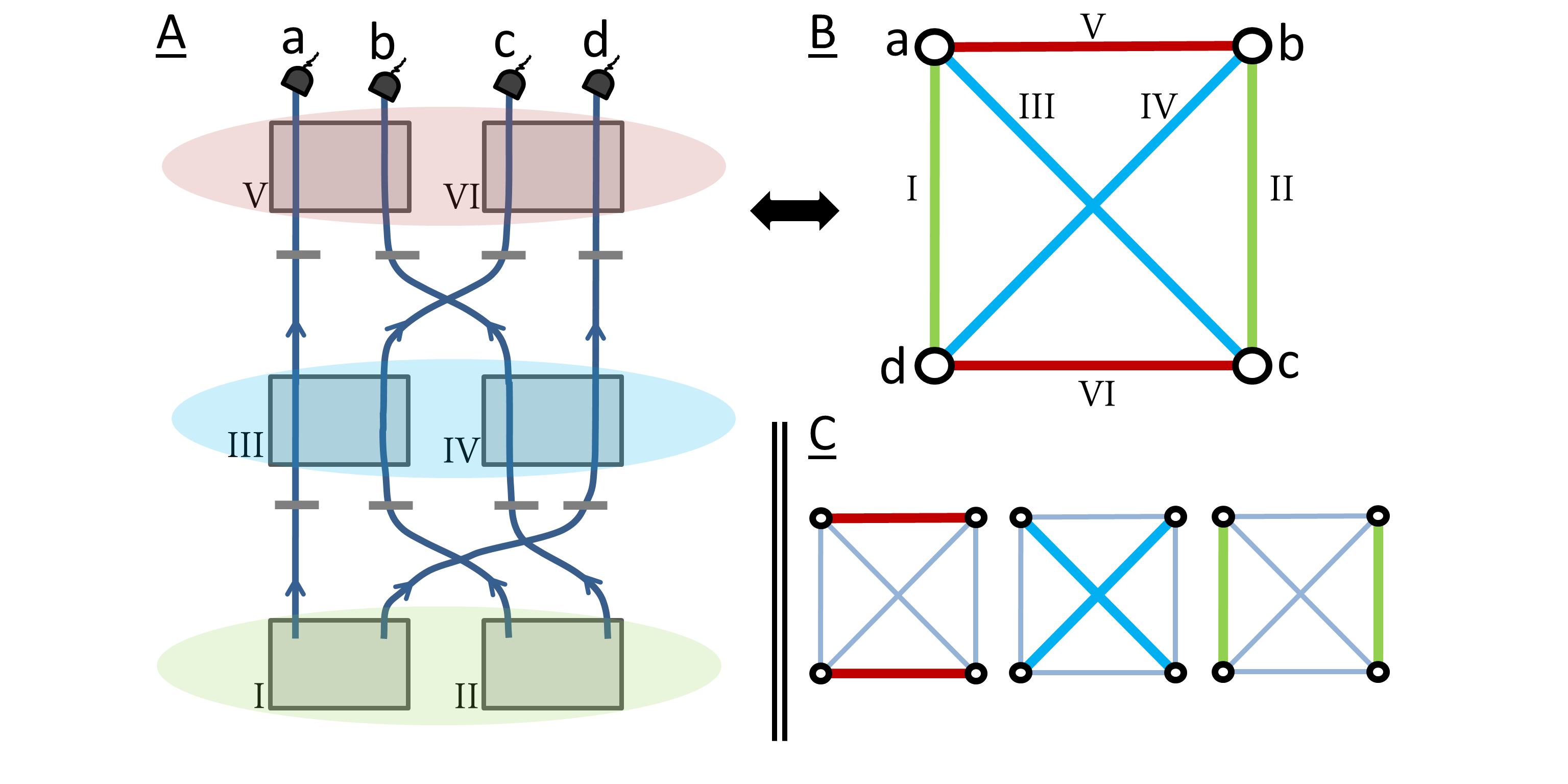}
\caption{A quantum optical experiment for the generation of a 3-dimensional GHZ-State, and its correspondence to a Graph. Every vertex is a photon path, and every edge corresponds to a non-linear crystal. Colours stand for the mode numbers. A resulting state, which arises conditioned on four-fold coincidence clicks, corresponds to perfect matchings of the graph.}
\label{fig:3dGHZ_graph}
\end{figure}

We describe the connection using an example: Figure \ref{fig:3dGHZ_graph}a depicts the setup of a 3-dimensional GHZ state, which was already discussed in Figure \ref{fig:3d-GHZ_path-identity}. The corresponding graph is shown in \ref{fig:3dGHZ_graph}b. Every vertex of the graph corresponds to a photon path, every edge between two vertices corresponds to a non-linear SPDC crystal which can produce two photons in two photonic paths. Photons produced in different layers of the crystal lead to different mode numbers. This is represented by the colour of the crystals and their corresponding edges. Conditioning the outcome of the experiment on a four-fold coincidence count leads to a 3-dimensional GHZ state. A four-fold coincidence count happens when every detector fires exactly once. In the corresponding graph, a four-fold coincidence count can be identified for a subset of edges, which contain every vertex exactly once. This property is denoted as \textit{perfect matching} in graph theory. The results of a quantum optical experiments can, therefore, be interpreted as coherent superpositions of perfect matchings of a graph. The detailed link between quantum experiments and graphs can be seen in Table \ref{tab:compare}.

\begin{table}[b]
  \centering
    \begin{tabular}{ | l | l |}
    \hline
    \textbf{Quantum Experiment} & \textbf{Graph Theory}\\ \hline \hline
    Optical Setup with Crystals & undirected Graph $G(V,E)$ \\ \hline
    Crystals & Edges $E$ \\ \hline
    Optical Paths & Vertices $V$ \\ \hline
    n-fold coincidence & perfect matching \\ \hline
    layers of crystals & disjoint perfect matchings \\ \hline
    \#(terms in quantum state) & \#(perfect matchings) \\ \hline
    maximal dimension of photon & degree of vertex \\ \hline           
    \hline
    \end{tabular}
  \caption{The analogies between Quantum Experiments involving multiple crystals and Graph Theory, adapted from \cite{krenn2017quantum}.}
  \label{tab:compare}
\end{table}

\subsection{Application to Designing Experiments for Quantum States}
\begin{figure*}[t]
\centering
\includegraphics[width=\textwidth]{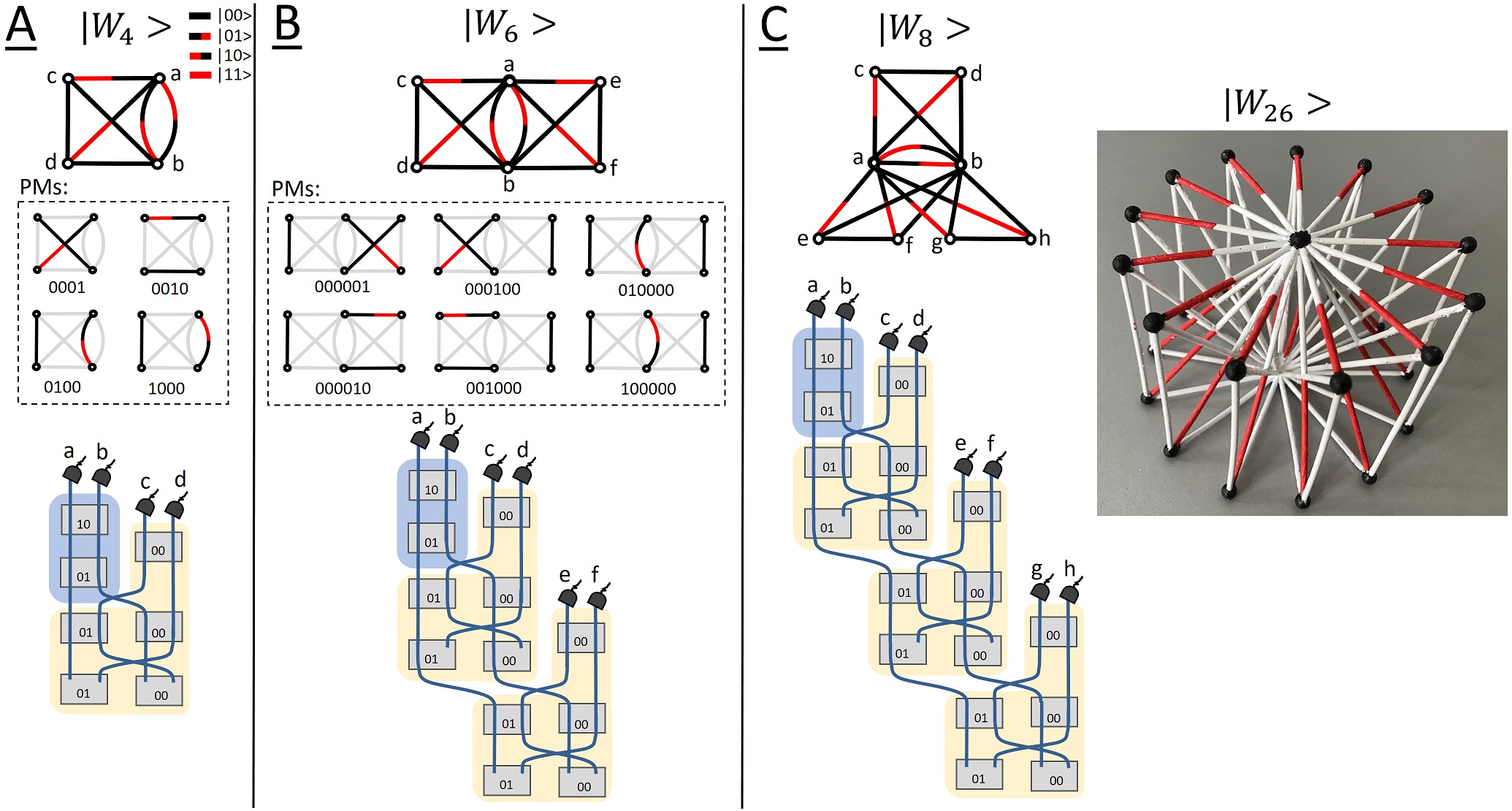}
\caption{Design of Quantum Experiments using Graph Theory. The connection to graph theory is an ideal descriptive tool of quantum experiments, and can be used to design experiments in order to design specific quantum states. In this example, presented in \cite{gu2019quantum3}, a general recipe for the generation of W-states is given.}
\label{fig:Wstate_graph}
\end{figure*}

Designing quantum experiments is very challenging because universal rules for multi-photonic systems do not exist, and multi-party quantum effects and interference are difficult to grasp intuitively. The connection between quantum experiments and graphs allows for a good descriptive tool where structures of the quantum states are encoded into the structures of the graph, and subsequently, the resulting graph directly corresponds to an experimental setup.

In the previous chapter, the (2-dimensional) GHZ-state was introduced as an important class of multi-particle entangled states. It was discovered in 2000, that three qubits could be entangled in two inequivalent ways \cite{dur2000three}. That means that there are two classes of three-qubit states which cannot be transformed into each other using only local operations and classical communications. One of them is the GHZ state, and the other one is the so-called W-states \cite{zeilinger1992higher, bourennane2004experimental}.

An intuitive understanding is that GHZ states are the strongest entangled states, while W-states encode the most robust entanglement. A three-particle W-state is defined as
\begin{align}
\ket{W_3}=\frac{1}{\sqrt{3}}\left(\ket{1,0,0}+\ket{0,1,0}+\ket{0,0,1} \right).
\label{Wstaten3}
\end{align}
It is a coherent superposition of one excitation (indicated by $\ket{1}$) being delocalised over all three particles. In the $n$-particle generalisation, it is a delocalisation of one excitation over all $n$ photons.

An experimental configuration for a 4-particle W-state using entanglement by path identity has been shown in \cite{krenn2017entanglement}, its $n$-party generalization was discovered in \cite{gu2019quantum3}, by exploiting the descriptive nature of the corresponding graphs, see Fig.\ref{fig:Wstate_graph}. There, similar techniques have been exploited to generate setups using Path Identity for much more general high-dimensional and multipartite quantum states. Examples involve Dicke states (which generalise W-state to multi-excitations) \cite{dicke1954coherence}, Schmidt-Rank Vector states (which classify quantum entanglement in a high-dimensional multipartite scenario) \cite{huber2013structure, huber2013entropy}, or absolutely maximally entangled states \cite{goyeneche2015absolutely, huber2018bounds, cervera2019quantum}. The design principle using entanglement by path identity has recently been generalized to multi-photon emitters, which involve hypergraphs as an descriptive tool \cite{gu2020quantum}.

The bridge between graph theory and quantum experiments can also be used to show which quantum state can not be produced with probabilistic photon-pair sources. The key idea is to translate a question in quantum physics into an equivalent question in graph theory, solve the question with the tools of graph theory and translate it back \cite{krenn2017quantum, gu2019quantum3}. A specific example is the following: The question \textit{Which $d$-dimensional $n$-photon GHZ state can be created with probabilistic photon-pair sources?} can be translated into \textit{Which graph with $n$ vertices exist, that has $d$ perfect matchings which are all disjoint?}. One can show that the only graphs which can fulfil this requirement are $n>$2,$d$=2 and $n$=4, $d$=3 \cite{krenn2017quantum, bogdanov267013}, which restricts the generations of GHZ states (without the employment of additional tools such as ancillary states). Many similar quantum physics questions are translated into the language of graph theory and can be solved \cite{gu2019quantum3} or yet have to be solved \cite{krenn2019questions}.

\subsection{Application in Quantum Random Networks}\label{ChapterRandomNetwork}

\begin{figure}[t]
\centering
\includegraphics[width=0.5\textwidth]{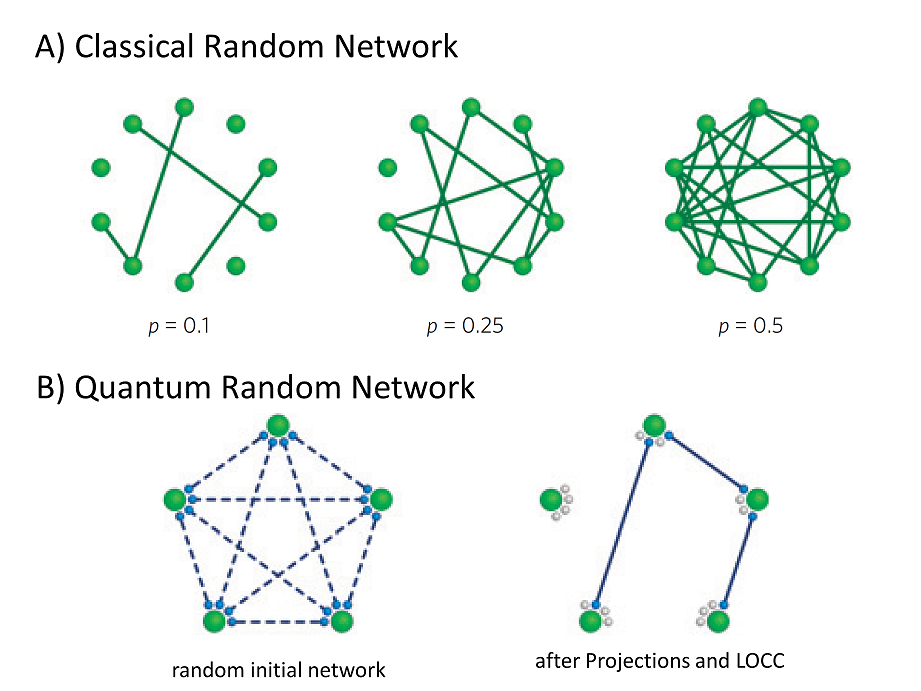}
\caption{Random Networks. \textbf{A}: In a classical random network, edges have a probability of $p$ to be formed between two vertices. \textbf{B}: A generalization to quantum networks \cite{perseguers2010quantum} introduces coherent superpositions of edges being formed or not -- with probability $p$. In quantum random networks, the critical behaviour of emergence of subgraphs happens at a much smaller probability, if projective measurements, local operations and classical communications are employed.}
\label{fig:randomNetwork}
\end{figure}

Classical random networks were introduced by Erd\H{o}s and R\'enyi in 1959 to describe many real-world features of networks, such as the small-world problem\cite{erdos1959random, erdos1960evolution}. These graphs are described by $G=(V,E)$, where $V$ are the vertices and $E$ are edges between nodes and another parameter $p$ which describes the probability that an edge forms between two nodes. An example of a classical random graph can be seen in Figure \ref{fig:randomNetwork}A.

A fascinating result is that in classical networks, many properties appear suddenly. For example, as N goes to infinity, the probability that a certain subgraph exists in the network is zero for $p<p_c(N)$ and is one for $p>p_c(N)$, where $p_c(N)$ is a critical probability. The critical probability scales as $N^z$, with the critical exponent $z \in (\infty, 0]$. A concrete example is the emergence of a fully connected graph of four vertices, denoted as $K_4$, which happens at $z=-\frac{2}{3}$. 

\begin{figure}[t]
\centering
\includegraphics[width=0.5\textwidth]{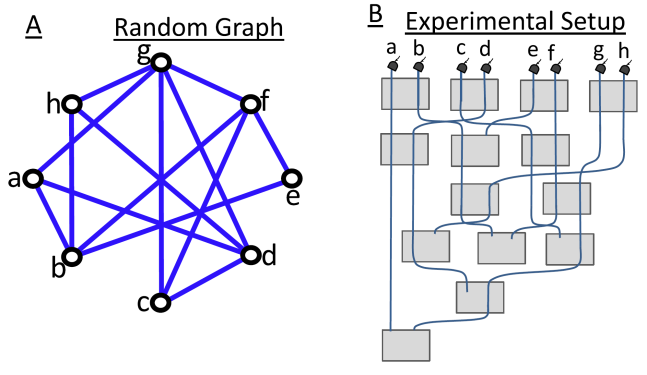}
\caption{Quantum Random Networks \cite{perseguers2010quantum} have interesting, critical properties -- such as the emergence of certain quantum states when the edge probability $p_c(N)>N^{-2}$, where N are the vertices. \underline{A:} A specific random graph with 14 edges connecting 8 vertices is a Quantum Random Network. \underline{B:} The corresponding setup, which is described by the graph, consists of 14 crystals which are coherently pumped and output into $N$=8 paths. The SPDC probability $p$ corresponds to the edge probability. Figure from \cite{krenn2017quantum}.}
\label{fig:randomGraph}
\end{figure}

Quantum Random Networks \cite{perseguers2010quantum} have been invented as generalizations of random networks in the quantum regime. The graph is built as a coherent superposition of all edges being inserted (with probability $p$) and not being inserted (with probability $\sqrt(1-p^2)$). The authors show, that exploiting projective measurements, local operations and classical communications (LOCC), that arbitrary quantum states of finite subgraphs can be obtained with a critical exponent of $z=-2$, which is much smaller than for classical random networks. 

Entanglement by Path Identity can be used to generate an arbitrary undirected graph, which creates quantum networks in the form as introduced in \cite{perseguers2010quantum}, as shown in Figure \ref{fig:randomGraph}. A single SPDC crystal produces quantum a quantum state that can be approximated by

\begin{align}
\ket{\psi_{a,b}}&=\Big(1+p\left(\hat{a}_{a}^{\dagger} \hat{a}_{b}^{\dagger} - \hat{a}_{a} \hat{a}_{b}\right)+ \nonumber\\
&+ \frac{p^2}{2} \left(\hat{a}_{a}^{\dagger} \hat{a}_{b}^{\dagger} - \hat{a}_{a} \hat{a}_{b}\right)^2 + ...\Big)\ket{0}.
\label{SPDC}
\end{align}
where $p$ is the SPDC amplitude. The complete quantum (random) network is a combination of all crystals being pumped coherently, which is a tensor product over all existing edges in the form of
\begin{align}
\ket{\psi_{\textnormal{network}}}=\bigotimes_{e(i,j)\in E}\ket{\psi_{i,j}}
\label{SPDC}
\end{align} 
where $i$ and $j$ are the vertices which are connected by the edge $e \in E$. These setups can be used to simulate striking phenomena of Quantum Random Networks, such as critical exponents, in a natural and inexpensive way.

\subsection{Generalization to general linear optical experiments}
The graph theoretical description has been generalized to arbitrary linear-optical systems, by realizing that every linear transformation is related to a certain graph transformation -- thus linear-optical elements cannot go beyond the graph-theoretical picture \cite{gu2019quantum2}. As a consequence, all conclusions about the construction of multi-photonic quantum states hold for linear-optical setups. It allows for the explanation of multi-photonic protocol such as quantum teleportation \cite{bennett1993teleporting, bouwmeester1997experimental} or entanglement swapping \cite{zukowski1993event, pan1998experimental} using simple pictorial diagrams. 

\begin{figure}[t]
\centering
\includegraphics[width=0.5\textwidth]{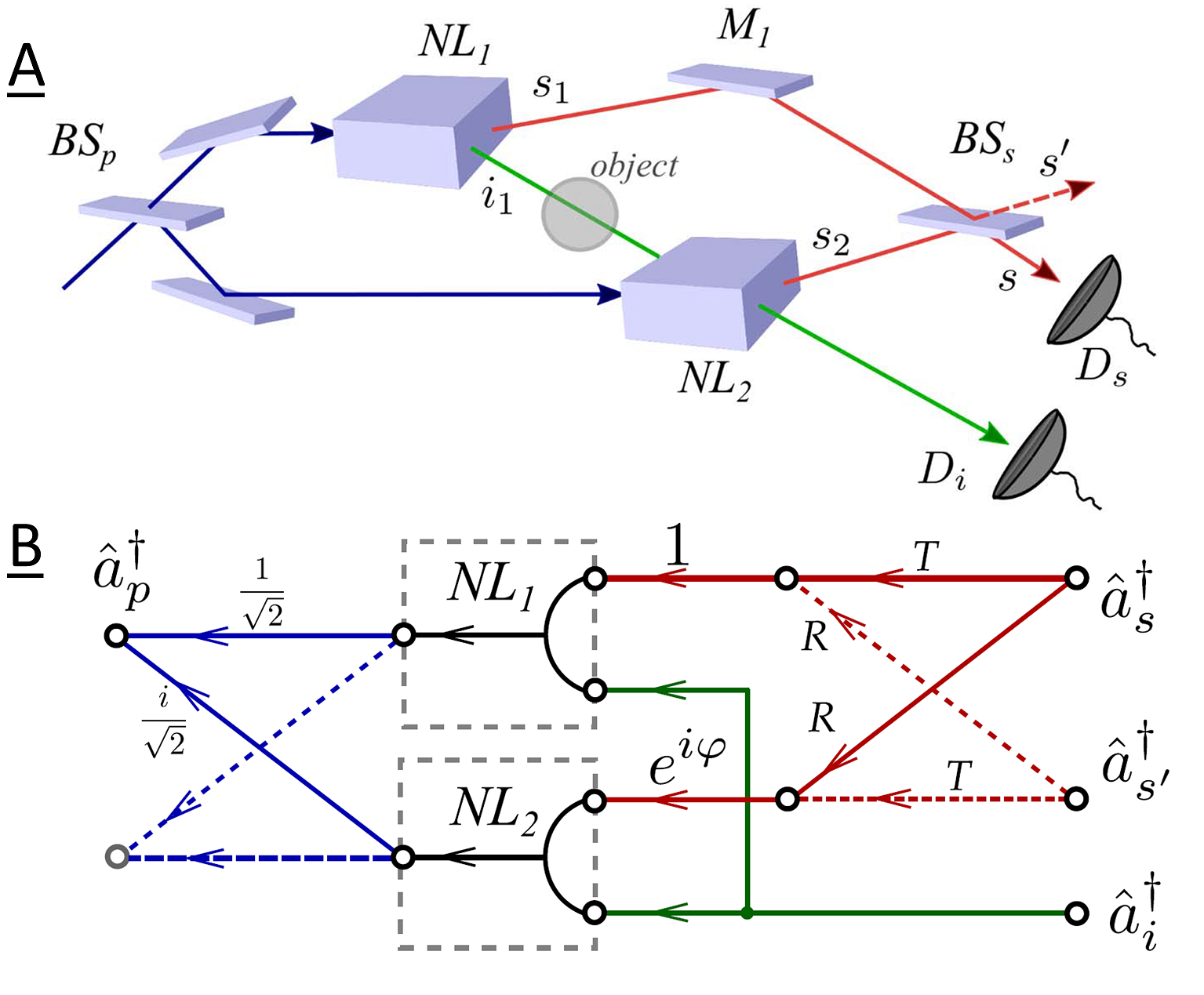}
\caption{Graphical Approach to linear-optical quantum experiments developed by Ataman \cite{ataman2015quantum, ataman2018graphical}. In \underline{A}, the ZWM experiment is depicted, including the pump beam. The translation to a graphical model is shown in \underline{B}. The idea is to represent photons as creation operators. The transformation of every linear-optical element as well as non-linear crystals for the creation of photon pairs can be understood in photon transformations of the photon's paths.}
\label{fig:AtamanGraph}
\end{figure}

A different approach to investigating photonic experiments has been shown by Ataman \cite{ataman2015quantum, ataman2018graphical}. The main idea is to translate creation operator rules, which define linear operators, into rules for photon paths. With that, various quantum experiments (at least for photon pairs) can be described. An example is the ZWM experiment shown in Figure \ref{fig:AtamanGraph}. Extending Ataman's description to multi-photonic experiments could be achieved by extending the graph-theoretical background, in particular, introducing the concept in (perfect) matchings of graphs.

\section{Quantum Interference in general photon creation Processes}\label{section:multiphotonInterf}
Here we describe interesting extensions of Herzog's experiment (\cite{herzog1994frustrated}, see Chapter \ref{Sec:HerzogIntro}). In the original experiment, two SPDC processes were organized in such a way that the resulting photon pairs were destructively or constructively interfered. Here we describe three extensions of this concept. The first one shows that the photon pairs, which interfere, do not necessarily come from the same source. The second extensions show a generalization to multiphotonic systems and a link to quantum computing. The third one demonstrates nonlinear interference in a four-wave mixing process in integrated photonics.

\subsection{Weak coherent laser + SPDC}
\begin{figure}[!ht]
\centering
\includegraphics[width=0.4\textwidth]{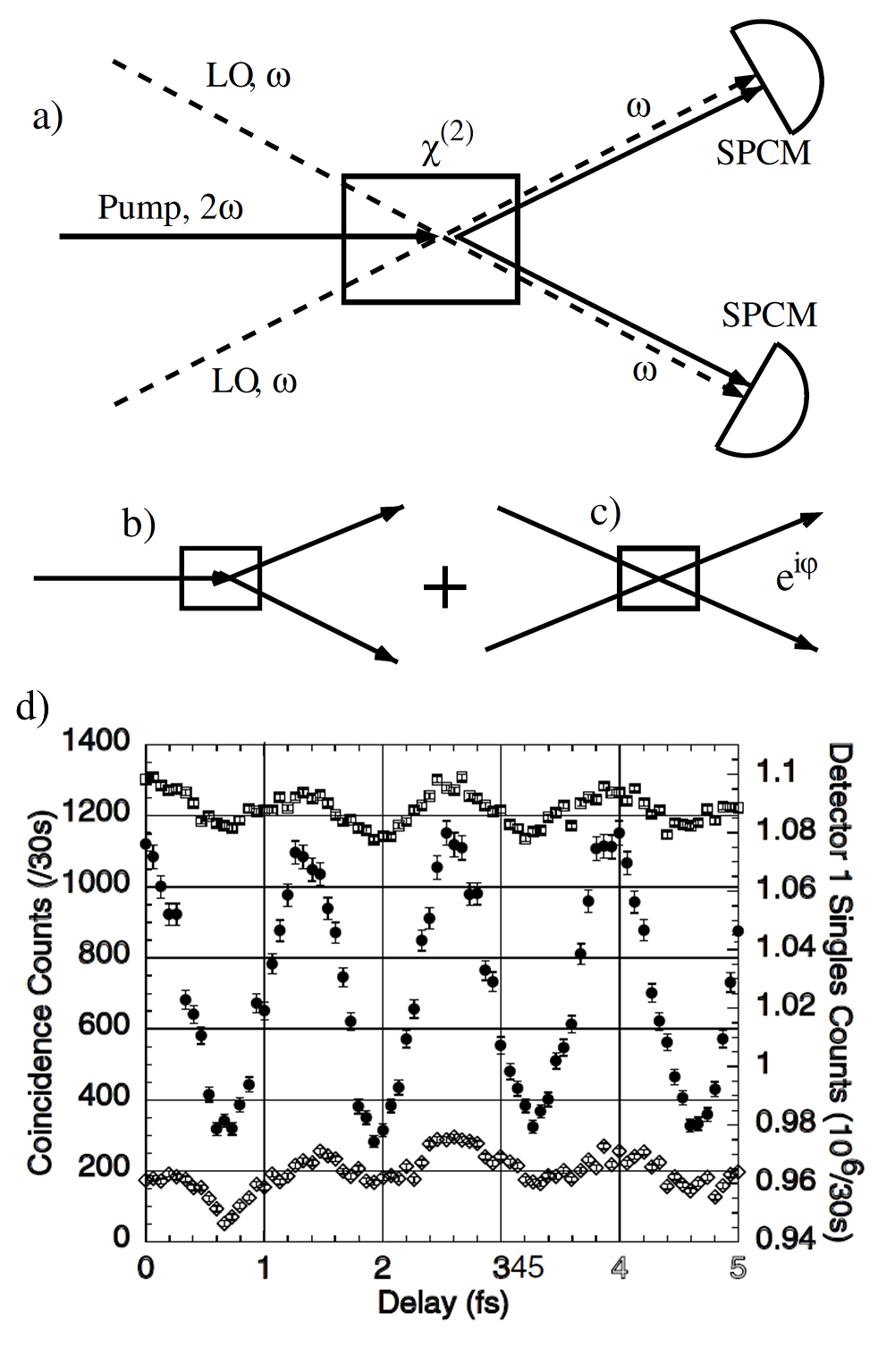}
\caption{Interference of photons from different sources (figures adapted from \cite{resch2001nonlinear}). A beam of a local oscillator (a laser beam with a defined coherence relation to the pump beam of the SPDC) is overlaped with the output of an SPDC crystal (a). There are two different possibilities how a pair of photons could be generated: Either via SPDC (b) or from the weak local oscillator (c). The authors were able to make these two possibilities indistinguishable, and therefore observe interference between them. In (d), solid circles stand for coincidence counts, with a fringe visibility of more than 50\%. The open squares stand for the single counts, in which one can also see statistically significant modulation as the phase between the two possibilities varies.}
\label{fig:SPDCLaserResch}
\end{figure}

The quantum interference is agnostic to the source of the photon pairs -- it is only essential that the two possibilities are fundamentally indistinguishable and share a well-defined phase relation. A remarkable experiment which has shown an analogue to the frustrated down-conversion interference, but from two different types of sources, was presented in \cite{resch2001nonlinear}. The sketch of their idea is shown in Figure \ref{fig:SPDCLaserResch}. The idea is to use a weak local oscillator (i.e. a part of a laser which is upconverted and acts as a pump for the SPDC process) and overlap it with the output of an SPDC process. The authors indeed see high-quality interference fringes.

Their experiment indicates that any generation process that can be performed in a coherent way allows for quantum interference. It again shows the significance of information -- as long as there is no information anywhere in the universe, that could help distinguishing in which of those processes the pair is created, quantum interference may occur.

\subsection{Multiphotonic Quantum Interference}
The objective of the multiphotonic setups in chapter \ref{section:EbPI} was the production of complex novel entangled quantum states - in a way generalizing Hardy's entanglement source \cite{hardy1992source} to higher dimensions and a larger number of particles. Hardy's entanglement source also has a close relation to Herzog's interference experiment \cite{herzog1994frustrated}. While in Hardy's experiment, the interference can be observed by measuring entanglement, in Herzog's experiment, interference can be observed by measuring direct photon count rates.

\begin{figure}[t]
\centering
\includegraphics[width=0.5\textwidth]{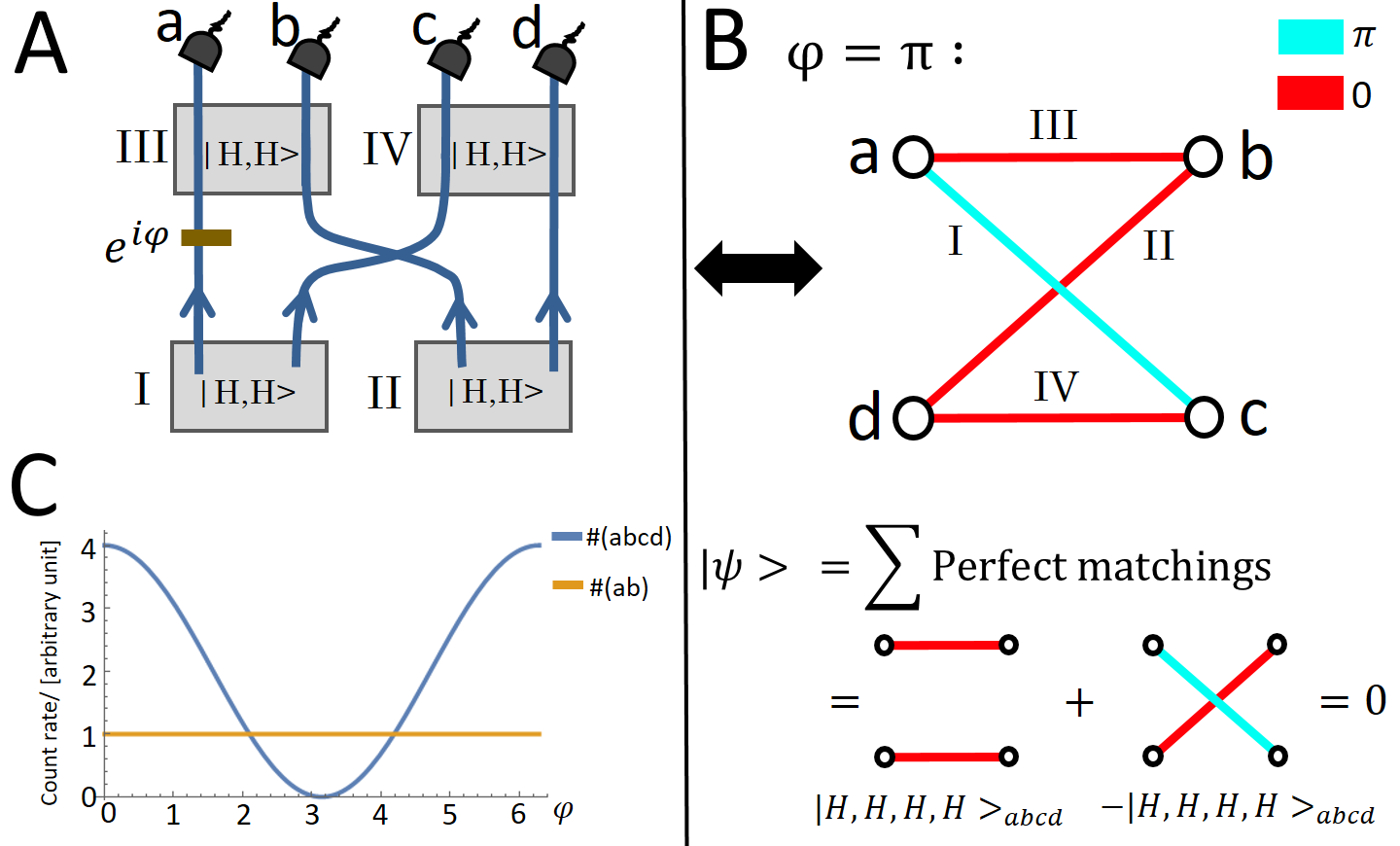}
\caption{Constructive and destructive quantum interference based on Path Identity \cite{gu2019quantum2}. In \textbf{A}, four crystals are aligned such that the emission of four-fold coincidence clicks in all four detectors $a$,$b$,$c$ and $d$ can only happen when crystal I and II emit a pair of photons, or crystal III and IV emit a pair of photons each. Here, all photons are indistinguishable. These two possibilities lead to two terms in the quantum state, which are coherently superposed. A phase shifter in one of the arms changes the relative phase between these terms, thus lead either to increased or decreased rates of four-fold counts. In \textbf{C}, the rate of two-fold counts in detector $a$ and $b$, and four-fold counts in all four detectors is shown when the phase $\phi$ is rotated. While the pair counts do not change, the four-fold counts can vanish. Figure \textbf{B} shows an interpretation in using graph theory, where weighted edges lead to phases between the perfect matchings which can cancel each other. This interpretation will help to find and understand follow-up applications.}
\label{fig:EbPIInterference}
\end{figure}

With the Hardy/Herzog analogy in mind, and with the possibility to generalize the Hardy entanglement source to multiple particles, one can ask whether also Herzog's experiment can be generalized to multiple particles. Indeed, it has been shown in \cite{gu2019quantum2} that experimental setups that can generate entanglement by path identity can be modified to show interference. In contrast to entanglement, this interference can be observed directly in the rate of emitted multi-photon states. The concept is shown in Figure \ref{fig:EbPIInterference}. There, mode shifters are replaced by phase shifters. The four-fold coincidence which, which arises from crystal I and II has thereby a relative phase between the four-photon term from crystal III and IV.

The down-conversion process can be approximated as a series expansion in the form of 
\begin{align}
\hat{U}_{a,b}=1&+g\left(\hat{a}_{a}^{\dagger} \hat{a}_{b}^{\dagger} - \hat{a}_{a} \hat{a}_{b}\right) \nonumber\\
&+ \frac{g^2}{2} \left(\hat{a}_{a}^{\dagger} \hat{a}_{b}^{\dagger} - \hat{a}_{a} \hat{a}_{b}\right)^2 + \mathcal{O}(g^3)
\label{eq:SPDC}
\end{align}
where $\hat{a}_{a}^{\dagger}$ and $\hat{a}_{a}$ are creation and annihilation operators for a photon in the mode $a$, respectively, and $g$ is proportional to the SPDC rate and the pump power. For simplicity, we restrict ourselves to single-mode analysis. In the 4-photon interference setup, four crystals are used, therefore the state can be expressed (taking only cases into account with one photon in each detector) as
\begin{align}
\ket{\psi}&=\hat{U}_{c,d}\hat{U}_{a,b}\hat{P}_{a}\hat{U}_{b,d}\hat{U}_{a,c}\ket{0,0,0,0}\nonumber\\
&=g^2(1+e^{i\varphi})\ket{1,1,1,1}+O(g^3)
\label{eq:4photonInterference1}
\end{align}
where $\hat{P}_{a}$ introduces a phase $\phi$ in path $a$, and $\ket{1,1,1,1}$ stands for a state with one photon in each path. The complete state up to second order SPDC contains exactly one term (depicted in red), which stands for interferences (i.e. its amplitude changes when the phase $\phi$ changes). No other terms, in particular, no other two-photon terms, show that behaviour. Thus, this phenomenon is a genuine multiphotonic interference effect.

This interference effect has an interesting interpretation: If the phase is set to $\phi=\pi$, one will never observe four-fold coincidences in the four detectors. Interestingly, in this case, when one sees a photon pair in detector $a$ and $b$, one can be sure to not observe a photon-pair in detector $c$ and $d$. This is surprising because all crystals produce photons pairs spontaneously (i.e. not deterministically). Furthermore, crystal IV, which would produce photon pairs in $c$ and $d$, can be far away from crystal III and the phase shifter. Thus the setting information of the phase shifter needs to travel to crystal IV. This reasoning indicates that one could construct fascinating experiments investigating time delays of the interference effects \cite{ma2016delayed}.

\subsection{On-Chip Quantum Interference by Path Identity}
\begin{figure*}[t]
\centering
\includegraphics[width=0.9\textwidth]{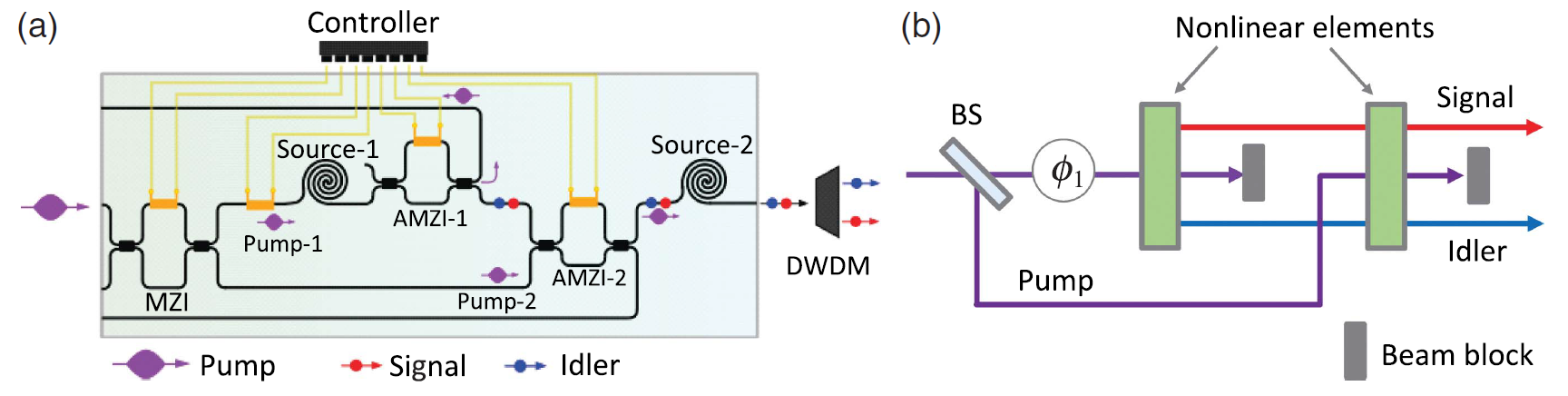}
\caption{Frustrated pair-creation has been observed on an integrated silicon photonic chip by Ono et al. \cite{ono2019observation} in the group of Rarity. The source of photon pairs is a spontaneous four-wave mixing process of a $\chi^3$ nonlinearity. Two sources can each create a pair of photons. As the origin of the pair is undefined, it is in a coherent superposition of being created in either. Thus the authors observe constructive and destructive interference of the resulting photon pair.}
\label{fig:HerzogOnChip}
\end{figure*}
The experiments in the previous chapter require many photon-sources which are phase-stable among each other. One way to guarantee stability is the integration of the whole setup into a photonic chip. Integrated sources of photon pairs have been demonstrated in several experiments over the last years \cite{jin2014chip, silverstone2014chip, silverstone2015qubit, krapick2016chip, wang2018multidimensional, santagati2018witnessing, adcock2019programmable,feng2019chip,lu2020three}.

However, it was only in 2019 that the first non-linear interference experiment has been demonstrated \cite{ono2019observation}. The authors use two sources of spontaneous four-wave mixing and overlap their outputs, as shown in Figure \ref{fig:HerzogOnChip}. They observe very high interference visibility of 96.8\%. 

This experimental demonstration could open the door for using path-identity based interference effects as an additional powerful building block in integrated photonics. Furthermore, it paves the way to observe new interference phenomena described in the previous section.

\subsection{Application in Quantum Computation}
\SkipTocEntry\subsubsection*{Special Purpose Quantum Computations via Sampling}
The setup in Figure \ref{fig:EbPIInterference}A can be generalized to a random network -- similar to chapter \ref{ChapterRandomNetwork}, with random phases between all paths. Let's consider the situation where the experimental setup has $m$ output modes and $N$ crystals, and $n/2<m$ photon pairs ($n$ photons) are generated. To calculate the distribution of the possible output results, one has to find all combinations of crystals which could lead to this particular result, and sum their amplitudes in a coherent way. As $m$ and $n$ increase, this cannot be done efficiently anymore on a classical computer.
\begin{figure}[b]
\centering
\includegraphics[width=0.5\textwidth]{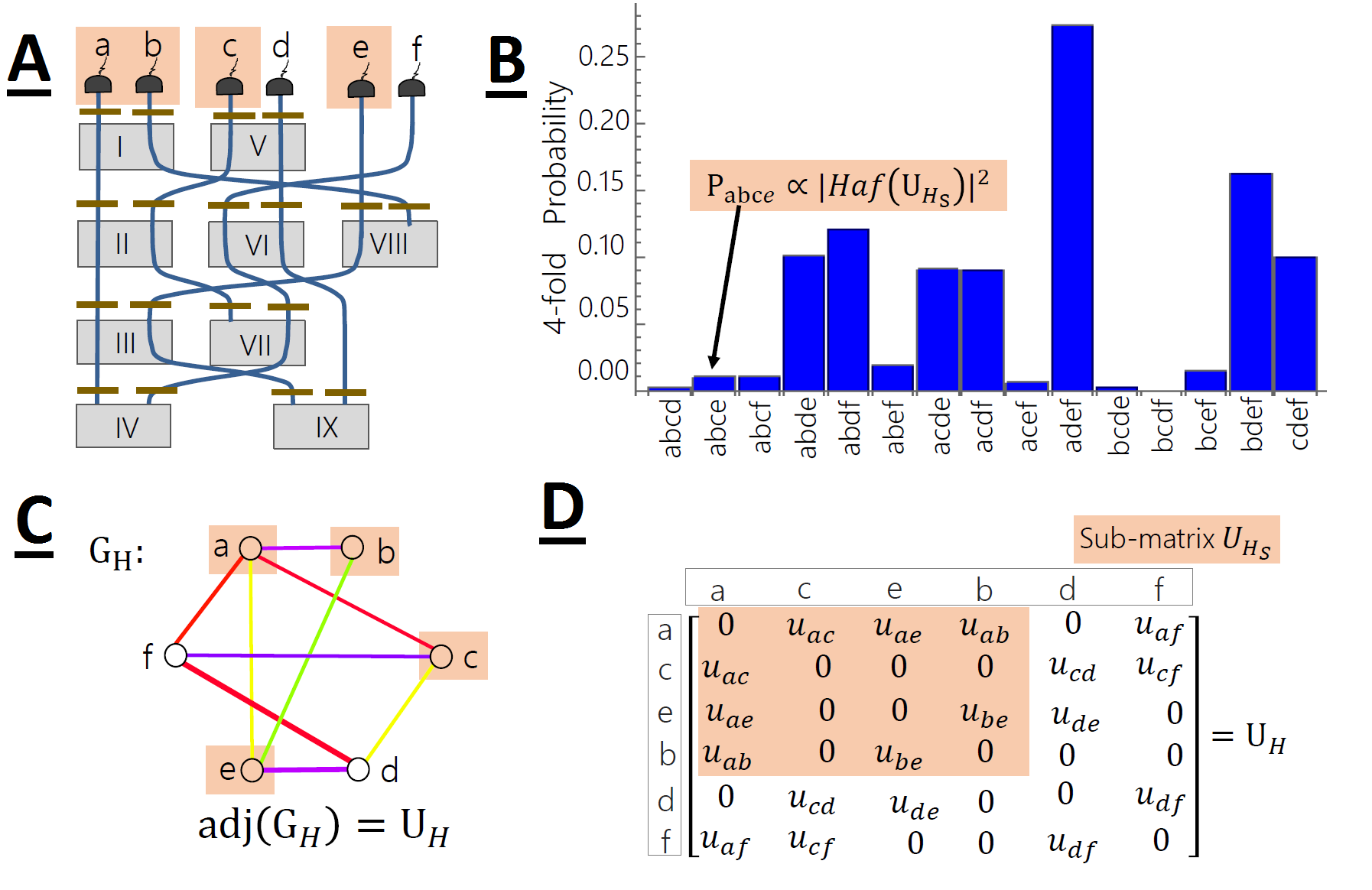}
\caption{Multi-photonic interference can be exploited for special-purpose quantum computation \cite{gu2019quantum2}. Figure \textbf{A} shows path identified photons, which are produced in a random network of crystals. The network has six output modes and is pumped in such a way that events of more than four photons can be ignored. In \textbf{B}, the four-photon count rates are shown. Every experimental setup corresponds to a weighted undirected graph $G_H$. A four-fold coincidence count (for example, in output $a$,$b$,$c$ and $e$) corresponds to a subgraph $G_{H_S}$ of those vertices, in \textbf{C}. The count rate in these output modes can be calculated as the coherent sum of all weighted perfect matchings of $G_{H_S}$. This is equivalent as calculating the matrix function \textit{Hafnian} applied on the adjacency matrix of the graph $G_{H_S}$, seen in \textbf{D} -- a problem known to be extremely difficult to calculate.}
\label{fig:Wstate_graph}
\end{figure}

To understand this better, it is useful to translate the experimental setup into its corresponding graph, as shown in chapter \ref{ChapterGraphs}. The probability for a given combination of $n$ detectors clicking is provided by the sum of the weights of all perfect matchings of the particular subgraph.

The problem is the following: It is easy to verify that a given set of edges form a perfect matching (as a reminder, it is set of edges where every vertex is contained exactly once) in a graph. However, there is no known algorithm that can find a perfect matching for arbitrary graphs in polynomial time. In the words of complexity theory, the question of finding a perfect matching is in the complexity class NP-complete.

Now to calculate the measurement results for a given combination of detectors needs to find \textit{all} perfect matching in the graph, and each of them has a complex amplitude associated with it. Therefore, this problem is even more difficult and lays in the complexity class \#P \cite{valiant1979complexity, aaronson2011linear}.

For bipartite graphs (these are graphs with two sets of vertices, where an edge only contains vertices between from the two sets), calculating the number of perfect matchings corresponds to calculating the matrix function \textit{Permanent} of the adjacency matrix of the graph. For general graph, the generalized matrix function called \textit{Hafnian} \cite{caianiello1953quantum} can be used.

The scenario just described is experimentally entirely different, but mathematically closely related to an idea proposed in 2011 denoted as \textit{Boson Sampling} \cite{aaronson2011computational}. There, $n$ single photons propagate through a random network of beam splitters and phase shifters, and are detected in a combination of $m$ output detectors. The situation can be described as a bipartite graph (a set of $n$ input modes connect to a set of $m$ output modes). This idea was further generalized to general graphs (using Gaussian Boson Sampling \cite{lund2014boson, hamilton2017gaussian, bradler2018gaussian}).

The situation has sparked a lot of excitement because it would allow -- for large enough number of photons and modes -- to directly observe experimental measurement results which cannot be calculated efficiently on a classical computer. On the one hand, it is considered as a method to demonstrate the first \textit{quantum supremacy} -- a calculation that is faster on a quantum device than on any classical available computer \cite{harrow2017quantum}. The fastest quantum Boson Sampling device can outperform the first electronic universal computer ENIAC (1942) and the first transistor-based electronic computer TRADIC (1954) \cite{wang2017high}.

Estimating the output distribution of systems described above also has essential implication in science and technology, and could lead to real-world use-cases of quantum hardware as special-purpose computers. One example is the spectra of vibronic (interactions between electronic and vibrational modes) modes in molecules \cite{huh2015boson}, which are essential in chemistry. Other algorithms involve graph theory applications such as the Dense Subgraph problem \cite{arrazola2018using} or the graph isomorphism problem \cite{bradler2018graph}.

Detailed comparisons or efficiency, error-tolerance or experimental feasibility between the traditional methods of Boson Sampling and the Sampling using Path Identity are important open questions for the future.

\SkipTocEntry\subsubsection*{Application in Gate-based Quantum Computation}
Universal Quantum Computers in the gate-based model have the following scheme: An array of $N$ qubits are initialized in a state $\ket{0}$, followed by a sequence of single- and two-qubit quantum gates which execute the quantum algorithm and subsequent measurements.
\begin{figure}[t]
\centering
\includegraphics[width=0.5\textwidth]{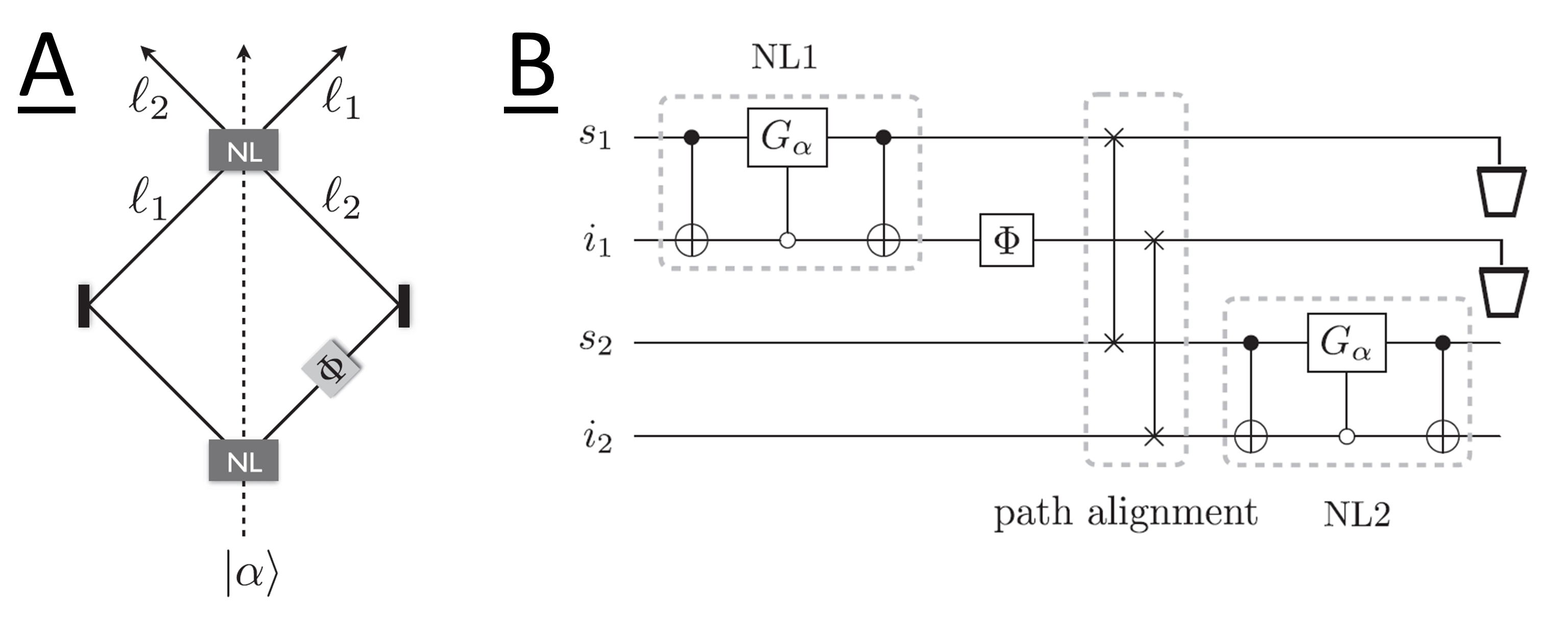}
\caption{A quantum circuit to describe the frustrated SPDC \cite{herzog1994frustrated}, in \textbf{A}. \textbf{B}: Each mode is interpreted as a three-level state. The first level is the state of a non-existent photon, while state two and three encode the usual computational qubit. The pair-creation of NL1 and NL2 (non-linear crystals 1 and 2) are described by two CNOT gates and one controlled-unitary gate. If no qubit exists in the mode, it creates one in the computational state $\ket{1}$. If a qubit with state $\ket{1}$ exists, it will annihilate the qubit. Formally, the mode can be considered as a qutrit (three-level system). The path alignment between the crystals is governed by two SWAP gates. Figure from \cite{alipour2017quantum}.}
\label{fig:QGate}
\end{figure}
These models assume that the qubits already exist at the initialization, and that they always exist during the execution. However, using a frustrated generation of qubit pairs \cite{herzog1994frustrated}, one has the additional potential of exploiting the existence or non-existence of the qubit itself. This potential has been largely unexplored so far in the realm of quantum algorithms.

An initial attempt to describe such type of interference effects in the language of quantum gates has been shown in \cite{alipour2017quantum}. There, the qubits have been extended to qutrits (three-level systems), to carry the additional information of whether the mode is mode is occupied by a photon or not. An example to explain frustrated SPDC is shown in \ref{fig:QGate}. This would allow for a natural way of encoding quantum information for ternary quantum computers (where instead of qubits, quantum systems with three levels are used) \cite{bocharov2017factoring, bocharov2015improved}. It is an open question of how all generalized approaches presented here can be translated to the language of quantum gates and quantum circuits, and whether it can inspire new ideas in the design of quantum algorithms.

\section{Concepts and Ideas related to Path Identity}

\subsection{Non-Demolition Modulation of Interference}
\begin{figure}[t]
\centering
\includegraphics[width=0.5\textwidth]{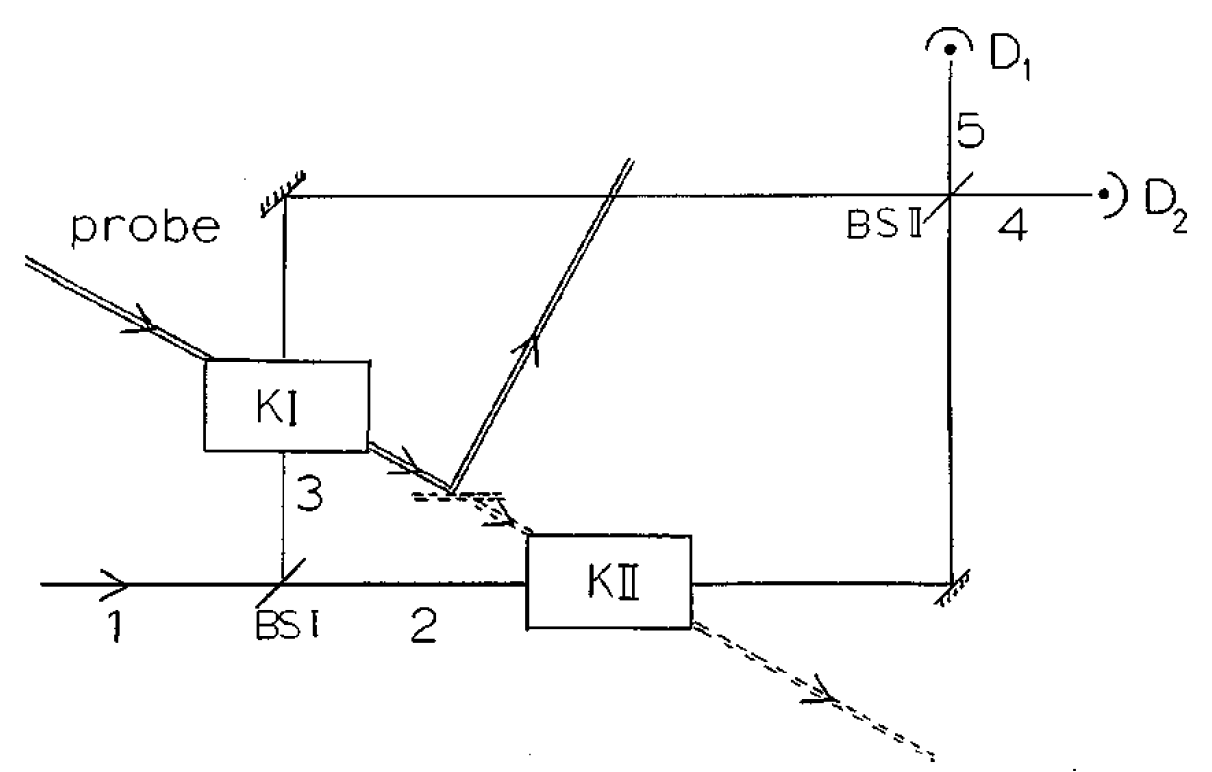}
\caption{Path Identity of a Probe Beam. Kerr cells K1 and K2 are introduced in the two arms of a Mach-Zehnder interferometer. The Kerr cells can identify in which path the photon has propagated, and are read out with a probe beam. When the path of the probe goes through both cells, no information is obtained, and full interference contrast is restored. Partial information introduced by a BS reveals partial information of the photon's path, thus reduced the interference visibility. Image from \cite{genovese2000quantum}.}
\label{fig:KerrCellMZI}
\end{figure}

In all of the examples so far, path identity has been considered in the context of creation processes. However, identifying paths is a general concept and is not tied to a generation process.

An example has been demonstrated in an original proposal in \cite{genovese2000quantum}. There, the authors consider a single photon entering a Mach-Zehnder interferometer with one Kerr cell in each arm of the interferometer, as shown in Figure \ref{fig:KerrCellMZI}. The photon changes the state of the Kerr cell such that it could act as a witness whether the photon has passed through or not. The state of the Kerr cell can be read with a probe beam, which would ultimately reveal the information in which path the photon propagated, thus destroy the interference at the output of the interferometer. Now if the path of the probe beam goes through both Kerr cells, one does not get the information of the photon path, and therefore interference is maintained. If one introduces a partially reflective object in the probe beam between the two objects, one can receive partial information of the photon path, thus partially reduces the strength of interference.

This remarkable idea shows that path identity is deeply connected with the concept of information, thus of quantum coherence.

    \begin{figure*}
        \centering
        \includegraphics[width=1\linewidth]{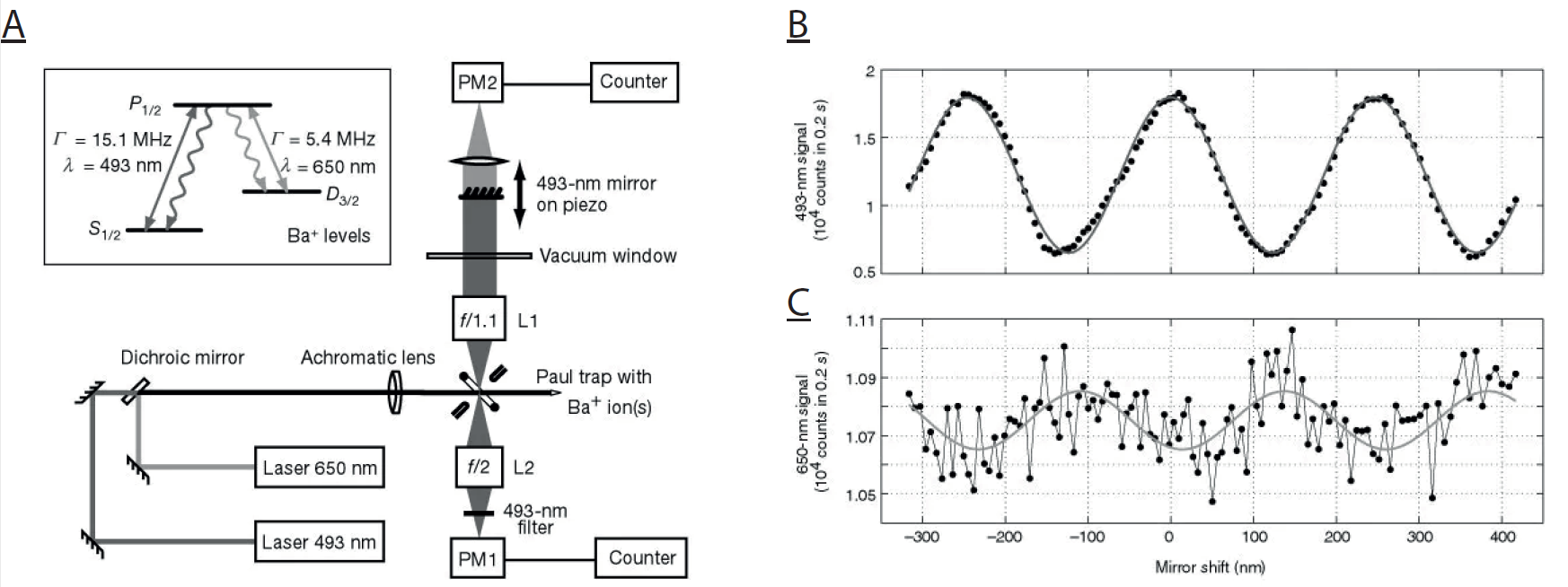}
        \caption{Eschner's experiment. The setup is sketched in A. An ion in a trap can emit in the direction of the detector (PM1), or in the opposite direction, in which case photons are back-reflected by a mirror ans also arrive at detector PM1. The detected spontaneous emission rate (B) exhibits a sinusoidal modulation due to the interference arising from the two indistinguishable emission processes. The fringe in plot C corresponds to a different atomic transition which is directly proportional to the population of the excited state (probed using the different detector PM2). The result shows that at a reduced rate of emission, the ion is held in the excited state for a longer time, thus proving that not only the emitted light takes part in the interference phenomenon, but the entire emission process. Fig. adapted from \cite{eschner2001light}}.
        \label{fig:eschner-setup}
    \end{figure*}

\subsection{Interfering Spontaneous Emission Processes}

The interference caused by aligning photon paths to be indistinguishable has been experimentally demonstrated in other systems than SPDC.
A related effect, namely the modification of the spontaneous emission rate in the presence of mirrors had been studied since the early 70s with molecules and atoms, e.g. \cite{drexhage1974iv,hulet1985inhibited,kleppner1981inhibited,goy1983observation,heinzen1987enhanced,jhe1987suppression}, as well as with a variety of other systems, e.g. \cite{yablonovitch1987inhibited,deppe1990optically}.
    
As an illustrative example of the connection to path identity, we consider an experiment \cite{eschner2001light} that uses a configuration that closely resembles the Herzog-Rarity-Weinfurter-Zeilinger experiment described in Sec. \ref{Sec:HerzogIntro} (Fig. \ref{fig:Intro_3experiments}B). It demonstrates the close analogy of the phenomenon of inhibited spontaneous emission to the concept of path identity.
    
In the experiment \cite{eschner2001light}, a single excited ion is located in a trap and can spontaneously emit photons in any direction (Fig. \ref{fig:eschner-setup}A). A mirror reflects emitted photons back to the ion. A detector is located at the opposite side of the mirror. Thus, a photon can arrive at the detector in two indistinguishable ways: either it is directly emitted by the ion towards the detector, or it is initially emitted towards the mirror and subsequently reflected onto the detector. The experiment is constructed in a way that no information can be obtained, even in principle, as to which direction a detected photon was initially emitted. As a result, interference between the probability amplitudes corresponding to the two processes is observed.
    
    Fig. \ref{fig:eschner-setup}B shows the rate of photons recorded at the detector in the opposite direction of the mirror, as the distance between ion and mirror is varied. The period of the sinusoidal modulation corresponds to the wavelength of the emitted light\footnote{As the mirror is translated by a distance $d$, the optical path increases by 2$d$.} (493 nm). Simultaneously, the population of the excited state of the ion was probed using a different atomic transition. The corresponding interference fringe (Fig. \ref{fig:eschner-setup}c) is proportional to the excitation probability and is anticorrelated to the fringe observed in the spontaneous emission rate. 
    
%    \paragraph{What do we learn from it}
    This result clearly shows that not only the emitted light is involved in the interference phenomenon, but the entire process of spontaneous emission together with the corresponding de-excitation of the ion. The concept of path identity can be used make two alternative ways of the process of spontaneously emitting a photon indistinguishable, which causes not only interference of the emitted light but also changes the internal dynamics of the atom. A theoretical analysis of this experiment can be found in \cite{dorner2002laser}.

\section{Conclusions and Outlook}
We have discussed the concept of path identity and its applications to fundamental and applied physics. Although the concept was in the literature since the early 1990s, it is only the recent developments that show its significance for future directions of research. In addition to its implications for fundamental problems, the concept of path identity has pushed frontiers of imaging, spectroscopy, and quantum information science. In the fields of imaging and spectroscopy, the concept has shown that it is possible to retrieve the object-information without detecting the radiation that illuminates the object. Therefore, the concept of path identity allows us to study the properties of an object at a wavelength for which good detectors are not available and thereby extend our experimental reach. As for quantum information science, the concept has led to distinct avenues of creating, controlling, and measuring entanglement. Furthermore, this concept of path identity has also inspired graph theoretical descriptions of quantum experiments, allowing a much more systematic and efficient way of designing future experiments.  
\par
Although all of the experiments discussed here are performed in the optical domain (i.e. by detecting photons), the concept of path identity is also applicable to other quantum entities. In this context, a very important fact is that all the experiments, discussed in this review, do not require stimulated emission. Therefore, the ideas of such experiments can also be applied to design experiments with fermionic systems. We expect that future experiments with non-photonic quantum systems, based on the concept of path identity, will not only extend our knowledge of fundamental physics but also will result in numerous applications. 

In a similar spirit, path identity could be applied too in atoms. In 2004, Paul Lett argued that an atomic variation of the ZWM experiment could be performed with pairs of atoms emitted from two Bose-Einstein condensates \cite{lett2004correlated}. This proposal has been denoted as \textit{non-trivial}. While this is certainly still true today, the experimental control Bose-Einstein condensate in conjunction with single-atom detection (such as meta-stable Helium \cite{robert2001bose, dos2001bose, vassen2012cold, keller2014bose}) has improved significantly, allowing for quantum optics experiments (such as Hong-Ou-Mandel analogues \cite{lopes2015atomic}). That progress could ultimately also lead to atomic variations of path identity experiments.

Many other vital questions, raised over a span of the last five years, remain open -- both theoretical and experimental. How can we experimentally increase the wavelength difference of signal and idler photons, in order to build highly efficient imaging, spectroscopy and microscopy \cite{kviatkovsky2020microscopy, paterova2020hyperspectral} techniques for deep-UV or the THz regime (Section IV)? Can the effective wavelength, which controls the interference properties, be applied in super-resolution schemes (Section IV.E)? Is it possible to detect quantum entanglement by measuring only one of the photons? Can this idea be generalized to multiphotonic entanglement, to perform GHZ-like paradox (Section V)? Can Rudolph's Single Photon source by Path Identity be experimentally implemented (Section VI.A)? Can one experimentally build a scaleable high-dimensional source based on Path Identity (Section VI.B)? Are (high-dimensional) multiphotonic sources based on Path Identity experimentally more efficient (Section VI.C)? Can the multiphotonic interference, which generalizes the idea of frustrated down-conversion to many photons, be observed in the laboratory (Section VIII)? Can Path Identity be observed for other systems, such as atoms (Section IX)?

Five years ago, Path Identity -- a \textit{sleeping beauty} -- had been woken up, and since then she shows her applicability and inspires new ideas and connections in diverse fields of quantum optics. We are looking forward to the progress in the coming years.

\SkipTocEntry\section*{Acknowledgements}
AZ wishes to thank the late Mike Horne and Danny Greenberger for many discussions of the fundamental questions of the topics, and its consequences over many years. This work was supported by the Austrian Academy of Sciences ({\"O}AW), University of Vienna via the project QUESS and the Austrian Science Fund (FWF) with SFB F40 (FOQUS). ME and AH acknowledge support from FWF project W 1210-N25 (CoQuS). MK acknowledges support from the FWF via the Erwin Schr\"odinger fellowship No. J4309. 

\bibliography{references-rev}

\end{document}